\documentclass{article}[11pt]

\usepackage{custom}

\begin{document}

\title{{Cold Atoms in Space: \\ {\Large Community Workshop Summary and Proposed Road-Map}}}

\abstract{\\
We summarize the discussions at a virtual Community Workshop on Cold Atoms in Space
concerning the status of cold atom technologies, the prospective scientific and societal opportunities
offered by their deployment in space, and the developments needed before cold atoms
could be operated in space. The cold atom technologies discussed include atomic clocks, 
quantum gravimeters and accelerometers, and atom interferometers. Prospective applications
include metrology, geodesy and measurement of terrestrial mass change due to, e.g., climate change,
and fundamental science experiments such as tests of the equivalence principle,
searches for dark matter, measurements of gravitational waves and tests of quantum mechanics.
We review the current status of cold atom technologies and outline the requirements for their space qualification, including the
development paths and the corresponding technical milestones, and identifying possible
pathfinder missions to pave the way for missions to exploit the full potential of
cold atoms in space. Finally, we present a first draft of a possible road-map for
achieving these goals, {that we propose for discussion by the interested  cold  atom,  
Earth Observation, fundamental physics and other prospective scientific user communities, 
together with ESA and national space and research funding agencies.}}
\affiliation[@]{Contact Person}
\affiliation[*]{Section Editor and/or Workshop Organiser}

\author[1]{Iván~Alonso,}  \affiliation[1]{Higher Polytechnic School, University of the Balearic Islands, Valldemossa Road, Palma de Mallorca, 07122, Spain}

\author[2]{Cristiano~Alpigiani,} \affiliation[2]{University of Washington, Physics-Astronomy Building, 15th Ave NE Seattle, WA 98195-1560, USA}

\author[3]{Brett~Altschul,} \affiliation[3]{University of South Carolina, Main Street, Columbia, South Carolina, SC 29208, USA}

\author[4]{Henrique~Ara\'ujo,} \affiliation[4]{ Imperial College London, Prince Consort Road, London, SW7 2AZ, UK}

\author[5]{Gianluigi~Arduini,} \affiliation[5]{CERN, CH-1211 Geneva 23, Switzerland}
\
\author[6]{Jan~Arlt,} \affiliation[6]{Aarhus University, Ny Munkegade, 8000 Aarhus C, Denmark}

\author[7]{Leonardo Badurina,}\affiliation[7]{King's College London, Strand, London, WC2R 2LS, UK}

\author[8]{Antun~Bala\v{z},} \affiliation[8]{University of Belgrade, Pregrevica 118, 11080 Belgrade, Serbia}

\author[9,10]{Satvika~Bandarupally,} \affiliation[9]{Dipartimento di Fisica e Astronomia, Universit\`a di Firenze, Via Sansone 1, 50019 Sesto Fiorentino, Firenze, Italy}
\affiliation[10]{European Laboratory for Non-Linear Spectroscopy (LENS), Universit\`a degli Studi di Firenze, Via Giovanni Sansone, 1, 50019 Sesto Fiorentino, Firenze, Italy}

\author[11]{Barry~C~Barish,} \affiliation[11]{LIGO, California Institute of Technology, 1201 E California, Pasadena, CA 91125, USA}

\author[12]{Michele Barone,} \affiliation[12]{ Institute of  Nuclear and Particle Physics, NCSR Demokritos, Agia Paraskevi 15310, Greece}

\author[13]{Michele~Barsanti,} \affiliation[13]{University of Pisa, Largo Lazzarino, Pisa, 56122, Italy}

\author[14]{Steven~Bass,} \affiliation[14] {Kitzb\"uhel Centre for Physics, Kitzb\"uhel, Austria}

\author[15,16,*]{Angelo~Bassi,} \affiliation[15]{ University of Trieste, Strada Costiera 11, 34151 Trieste, Italy}
\affiliation[16]{Istituto Nazionale di Fisica Nucleare, Trieste Section, Via Valerio 2, 34127 Trieste, Italy}

\author[17]{Baptiste~Battelier,} \affiliation[17]{LP2N, Laboratoire Photonique, Numérique et Nanosciences, Université Bordeaux-IOGS-CNRS:UMR 5298, F-33400, Talence, France}

\author[4]{Charles~F.~A.~Baynham,} 

\author[18]{Quentin Beaufils,} \affiliation[18]{SYRTE, Observatoire de Paris, Université PSL, Sorbonne Université, LNE, 61 avenue de l’Observatoire, 75014 Paris, France}

\author[8]{Aleksandar~Beli\'{c},}

\author[19]{Joel~Berg\'e,} \affiliation[19]{ DPHY, ONERA, Université Paris-Saclay, Chemin de la Hunière - BP80100, F-91123 Palaiseau, France}

\author[20, 21]{Jose~Bernabeu,}  \affiliation[20]{Department of Theoretical Physics, University of Valencia, Joint Centre Univ. Valencia-CSIC, 46100 Burjassot-Valencia, Spain}
\affiliation[21]{IFIC, University of Valencia, Joint Centre Univ. Valencia-CSIC, 46100 Burjassot-Valencia, Spain}

\author[17]{Andrea~Bertoldi,} 

\author[22, 23]{Robert~Bingham,}  \affiliation[22]{ Rutherford Appleton Laboratory, Harwell Campus, Didcot, OX11 0QX, UK}
\affiliation[23]{University of Strathclyde, Glasgow, G4 0NG, UK}

\author[18]{S\'ebastien~Bize,}

\author[24, 25]{Diego~Blas,} \affiliation[24]{Grup de F\'{i}sica Te\`{o}rica, Departament de F\'{i}sica, Universitat Aut\`{o}noma de Barcelona, 08193 Bellaterra, Spain} 
\affiliation[25]{Institut de Fisica d’Altes Energies, The Barcelona Institute of Science and Technology, Campus UAB, 08193 Bellaterra (Barcelona), Spain}

\author[26,*]{Kai~Bongs,} \affiliation[26]{ University of Birmingham, Birmingham, B15 2TT, UK}

\author[17,*]{Philippe~Bouyer,} 

\author[15]{Carla~Braitenberg,}

\author[27]{Christian~Brand,} \affiliation[27]{German Aerospace Center, Institute of Quantum Technologies, Wilhelm-Runge-Stra\ss e 10, 89081 Ulm, Germany}

\author[28, 27]{Claus~Braxmaier,} \affiliation[28]{Ulm University, Institute of Microelectronics, Albert-Einstein-Allee 43 89081 Ulm, Germany} 

\author[19]{Alexandre~Bresson,}

\author[4,29,*,@]{Oliver~Buchmueller,}  \emailAdd{Oliver.Buchmueller@cern.ch}
\affiliation[29]{University of Oxford, Department of Physics, Clarendon Laboratory, South Parks Road, Oxford OX1 3PU, UK}

\author[30, 31]{Dmitry~Budker,}   \affiliation[31]{Johannes Gutenberg-Universit{\"a}t Mainz, Helmholtz-Institut Mainz, GSI Helmholtzzentrum f{\"u}r Schwerionenforschung, 55128 Mainz, Germany} 
\affiliation[31]{Department of Physics, University of California, Berkeley, CA 94720, USA}

\author[32]{Luís~Bugalho,} \affiliation[32]{Instituto de Telecomunicações, Instituto Superior Técnico, Av. Rovisco Pais, Torre Norte, Lisboa, 1049-001, Portugal}

\author[33]{Sergey~Burdin,} \affiliation[33]{University of Liverpool, Liverpool, L69 7ZE, UK}

\author[34,*]{Luigi~Cacciapuoti,} \affiliation[34]{European Space Agency, Keplerlaan 1, PO Box 299, NL-2200 AG Noordwijk, The Netherlands}

\author[35]{Simone~Callegari,} \affiliation[35]{Physikalisch-Technische Bundesanstalt, Bundesallee 100, Braunschweig, 38116, Germany}

\author[36]{Xavier~Calmet,} \affiliation[36]{University of Sussex, Brighton, BN1 9QH, UK}

\author[37]{Davide~Calonico,} \affiliation[37]{Istituto Nazionale di Ricerca Metrologica, Strada delle Cacce 91, Turin, 10035, Italy}

\author[17]{Benjamin~Canuel,}

\author[38]{Laurentiu-Ioan~Caramete,} \affiliation[38]{Institute of Space Science, 409 Atomistilor Street, Bucharest, Magurele, Ilfov, 077125, Romania}

\author[34,*]{Olivier~Carraz,} 

\author[40]{Donatella~Cassettari,}  \affiliation[40]{University of St Andrews, North Haugh, St Andrews, KY16 9SS, UK}

\author[41]{Pratik~Chakraborty,} \affiliation[41]{Leibniz Universität Hannover, Welfengarten 1, Hannover, 30167, Germany} 

\author[42, 43, 31]{Swapan~Chattopadhyay,} \affiliation[42]{Fermilab, Batavia, Illinois, USA}
\affiliation[43]{Stanford University, Stanford, CA 94305, USA}

\author[44]{Upasna~Chauhan,} \affiliation[44]{Indian Institute of Technology, New Delhi, India}

\author[45]{Xuzong~Chen,}  \affiliation[45]{Peking University, Beijing 100871, China}

\author[46, 47, 48]{Yu-Ao~Chen,} \affiliation[46]{Hefei National Laboratory for Physical Sciences at the Microscale and Department of Modern Physics, University of Science and Technology of China, Hefei 230026, China} 
\affiliation[47]{Shanghai Branch, CAS Center for Excellence in Quantum Information and Quantum Physics, University of Science and Technology of China, Shanghai 201315, China} 
\affiliation[48]{Shanghai Research Center for Quantum Sciences, Shanghai 201315, China}

\author[13, 49,*]{Maria~Luisa~Chiofalo,} 
\affiliation[49]{Istituto Nazionale di Fisica Nucleare, Sezione di Pisa, Largo Pontecorvo 3, I-56125 Pisa, Italy}

\author[33]{Jonathon~Coleman,} 

\author[18]{Robin~Corgier,}

\author[4]{J.~P.~Cotter,}

\author[26,*]{A. Michael Cruise,}

\author[50]{Yanou~Cui,}  \affiliation[50]{Department of Physics and Astronomy, University of California, Riverside, CA 92521, USA}

\author[4]{Gavin~Davies,}

\author[51, 5,*]{Albert~De~Roeck,} \affiliation[51]{Antwerp University, B 2610 Wilrijk, Belgium} 

\author[52]{Marcel~Demarteau,} \affiliation[52]{Oak Ridge National Laboratory, 1 Bethel Valley Road, Oak Ridge, TN 37830, USA}

\author[53]{Andrei~Derevianko,} \affiliation[53]{ University of Nevada, Reno, Nevada 89557, USA}

\author[54]{Marco~Di~Clemente,} \affiliation[54]{Italian Space Agency, Via del Politecnico, Rome, 00133, Italy}

\author[55]{Goran~S.~Djordjevic,} \affiliation[55]{University of Nis, Visegradska 33, Nis, Serbia}

\author[16]{Sandro~Donadi,} 

\author[56]{Olivier~Dor\'e,} \affiliation[56]{Jet Propulsion Laboratory, California Institute of Technology, Pasadena, CA 91109, USA}

\author[4]{Peter~Dornan,}

\author[5,*]{Michael~Doser,}

\author[57]{Giannis~Drougakis,} \affiliation[57]{Foundation for Research and Technology-Hellas, 100 N. Plastira Street, Vassilika Vouton, GR 70013, Heraklion, Crete, Greece}

\author[36]{Jacob~Dunningham,}

\author[22]{Sajan~Easo,}

\author[58]{Joshua~Eby,} \affiliation[58]{The University of Tokyo Institutes for Advanced Study, The University of Tokyo, Kashiwa, Chiba 277-8583, Japan}

\author[33]{Gedminas~Elertas,} 

\author[7,5,*,@]{John~Ellis,} \emailAdd{ John.Ellis@cern.ch}

\author[4]{David~Evans,} 

\author[57]{Pandora~Examilioti,}

\author[30]{Pavel~Fadeev,}

\author[59]{Mattia~Fan\`i,} \affiliation[59]{Los Alamos National Laboratory, Los Alamos, New Mexico, USA} 

\author[60]{Farida~Fassi,} \affiliation[60]{Mohammed V University in Rabat, Faculty of Science, 4 Avenue Ibn Battouta, B.P. 1014 RP, Rabat, Morocco}

\author[9]{Marco~Fattori,}

\author[61]{Michael~A.~Fedderke,}   \affiliation[61]{The Johns Hopkins University, 3400 N Charles St, Baltimore, MD 21218, USA}

\author[38]{Daniel~Felea,}

\author[17]{Chen-Hao~Feng,}

\author[22]{Jorge~Ferreras,}

\author[62]{Robert~Flack,} \affiliation[62]{University College London, Gower Street, London WC1E 6BT, UK}

% ...


\maketitle
\section{Preface}
\label{sec:preface}
This document contains a summary of the {\it \bf Community Workshop on Cold Atoms in Space}~\cite{COMMUNITYWORKSHOP} that was held virtually on September 23 and 24, 2021. 
The purpose of this community workshop was to discuss objectives for a cold atom quantum technology development programme coordinated at the Europe-wide level, and to outline a possible community road-map and milestones to demonstrate the readiness of cold atom technologies in space, as proposed in the recommendations of the Voyage 2050 Senior Science Committee (SSC)~\cite{VOYAGE2050}, and in synergy with EU programmes. 

The SSC was set up by the ESA Director of Science to advise on the space science programme for the period 2030--2050, and drew attention to the potential of cold atom technology in fundamental physics and planetary science as well as in navigation, timekeeping and Earth Observation~\cite{VOYAGE2050}. 
The SSC set out a plausible programme of technology development in the Voyage 2050 report that would prepare cold atom payloads for evaluation by the ESA science committees on scientific merit alone, without technical concerns about robustness for the space environment. 
One aim of the workshop in September 2021 was to engage the cold atom community in defining possible science payloads that might be used to establish a recognised pathway towards the use of cold atoms in the ESA science programme.

This community workshop brought together representatives of the cold atom, astrophysics, cosmology, fundamental physics, geodesy and earth observation communities to participate in shaping this development programme. 
It built upon one organised two years ago~\cite{AEDGEWORKSHOP}, which reviewed the landscape of present and prospective cold atom experiments in space. 
Subsequently, several White Papers were submitted~\cite{WP1EA, AEDGE, WP2EA, WP3EA,WP4EA, WP5EA, WP6EA, WP7EA,Sedda_2020} in response to the Voyage 2050 call, which outlined possible ultimate goals and reviewed experiments and technical developments underway that could help pave a way towards these goals.


\begin{figure}[t]
\centering
\vspace{-0.2cm}
\includegraphics[width=0.475\textwidth]{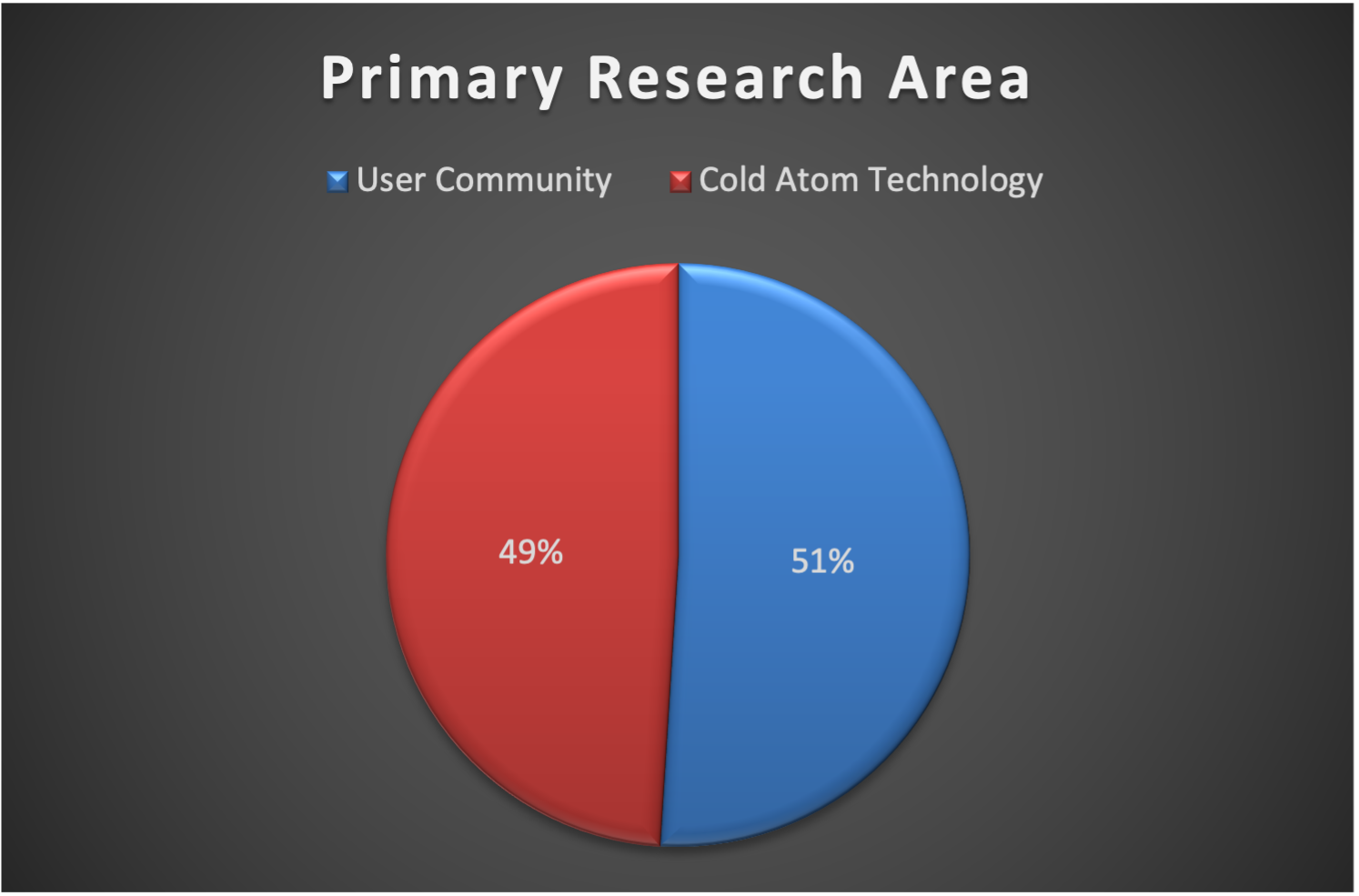}
\hspace{0.25cm}
\includegraphics[width=0.475\textwidth]{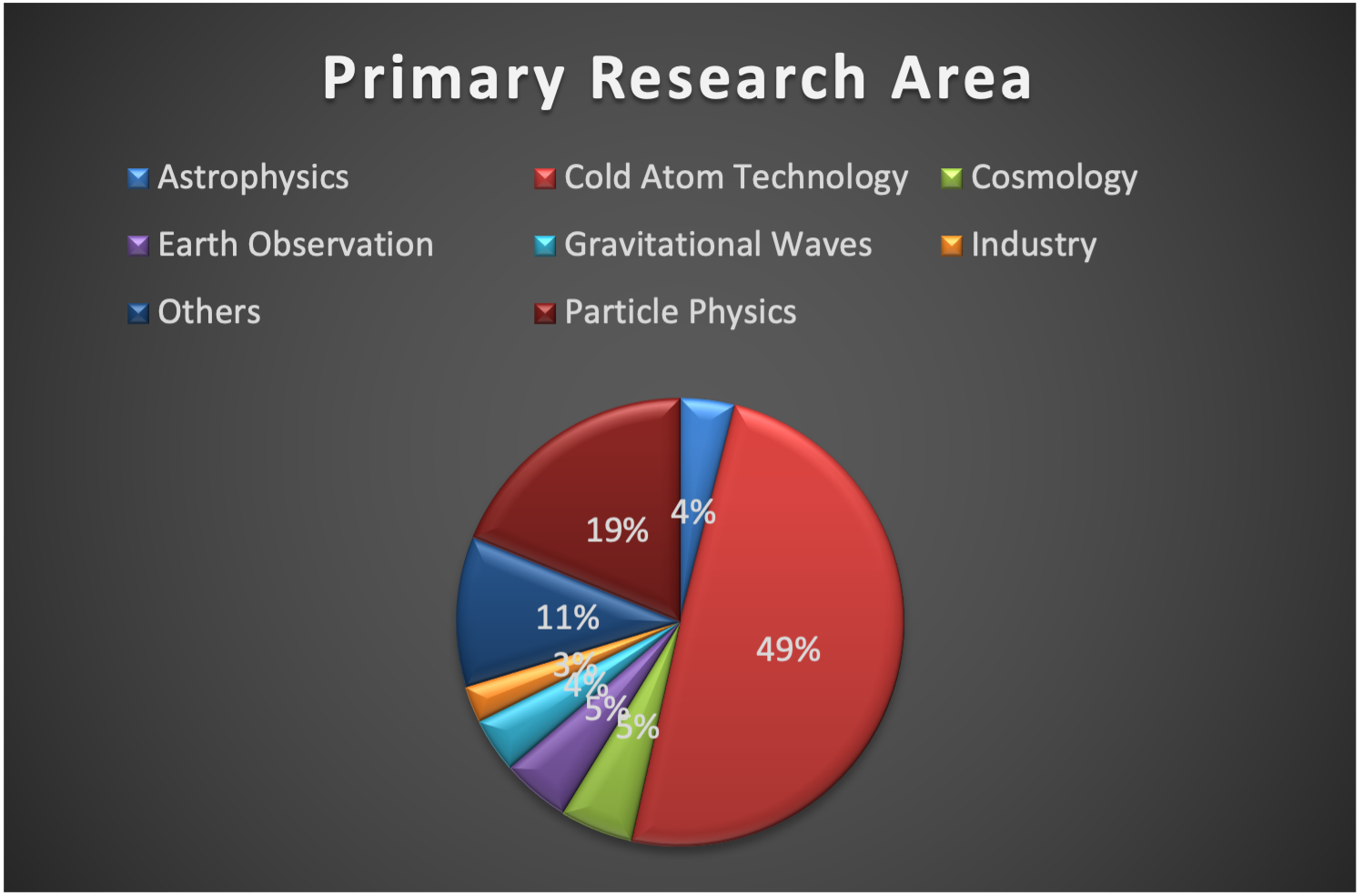}\\
\vspace{0.3cm}
\includegraphics[width=0.475\textwidth]{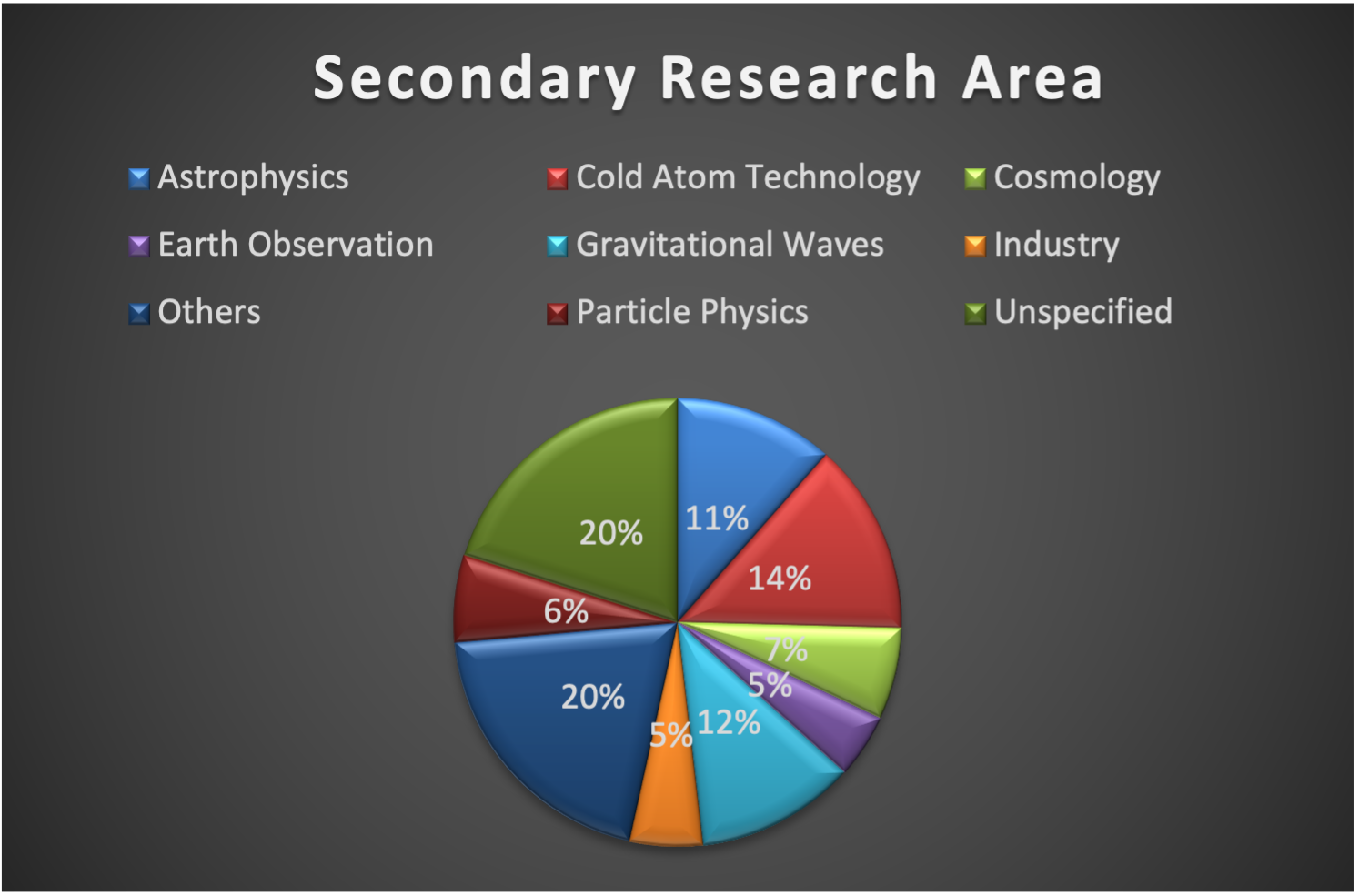}
\hspace{0.3cm}
\includegraphics[width=0.475\textwidth]{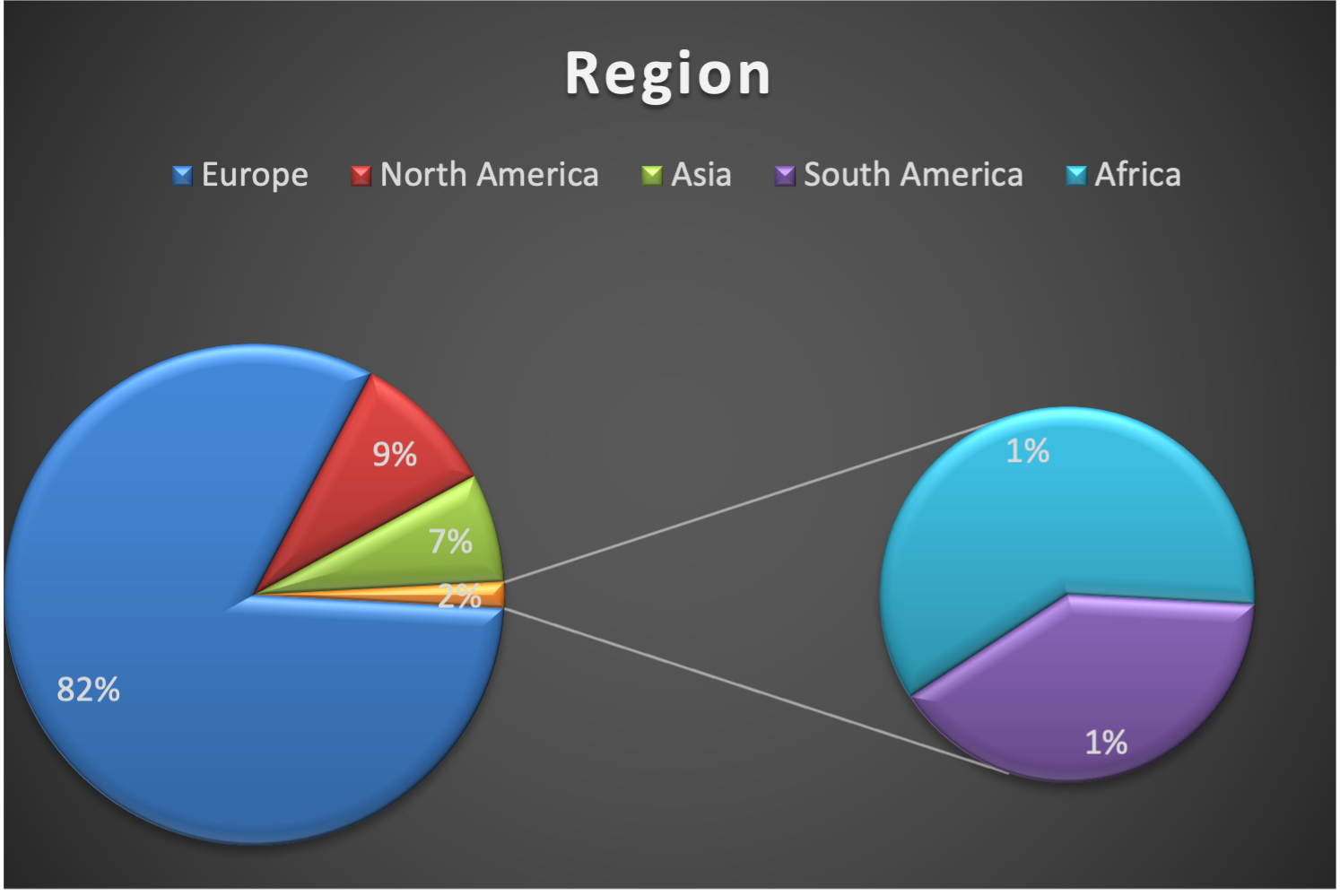}
\caption{\it Statistical analysis of the 331 registered to participate in the Community Workshop on Cold Atoms in Space, held virtually on Sept. 23rd and 24th, 2021, and/or to follow the Community Road-map process.
Upper left: Breakdown of registered participants into the prospective User community and those working on Cold Atom technology.
Upper right: Breakdown of registered participants by self-declared primary research area. Lower left: Breakdown of registered participants by self-declared secondary research area.
Lower right: Breakdown of registered participants by geographical location of institute where they are affiliated.}
\label{fig:mainRS}
\end{figure}

One of the key objectives of this Workshop was to enhance the strength and coherence of a community embracing Cold Atom technology experts as well as prospective Users.
In total, 331 people from 36 countries registered as participants in the Workshop and/or to follow the community roadmap process. 
As seen in the two upper panels of Fig.~\ref{fig:mainRS}, the registered participants comprise essentially equal numbers of declared cold atom experts and prospective Users.
The upper right panel breaks the user community down by their self-declared primary research areas, which range from Earth Observation to cosmology, and the
lower left panel provides another breakdown by their secondary research areas.
As anticipated for a Europe-based event, about 80\% of the registrations are from European countries, as seen in the lower right panel of Fig.~\ref{fig:mainRS},
but there is also significant North American and Asian participation. 

This community provides the backbone for long-term planning and the support needed to see the challenging cold atom missions foreseen in the road-map through to their successful completions.  
The participants display an excellent and diverse mix of expertise, building an outstanding basis for the success of the workshop and the development
of the corresponding road-map, which defines possible interim and long-term scientific goals and outlines technological milestones. 
Section~\ref{sec:intro} is an introduction to our document, Sections~\ref{sec:clocks} to \ref{sec:PH} discuss our main science themes, namely atomic clocks, Earth Observation and fundamental physics, Section~\ref{sec:Tech} discusses the technology developments required before quantum sensors can be deployed
in space, Section~\ref{sec:sum} summarises the discussions during the Workshop, and in Section~\ref{sec:road-map} we outline the corresponding Community road-map.

\section{Introduction}
\label{sec:intro}

Quantum physics was developed primarily in Europe in the first half of the 20th century. 
In the second half, the first “quantum revolution” took place and was the engine of the main technological and societal transformations in recent decades considering, e.g., solid-state electronics and hence all information and computing technologies. 
It also enabled the space era thanks, e.g., to onboard semiconductor technologies (solar cells, avionics, communication systems radars, detectors, etc.). 
Similarly, the first half of the 21st century is being deeply impacted by the second “quantum revolution”, with the possibility of exploiting and controlling quantum phenomena so far not applied outside the laboratory, e.g., macroscopic quantum coherence, superposition, entanglement. 

Atomic quantum sensors are a newly-emerging technology of unparalleled accuracy and precision. 
Spaceborne quantum inertial sensors (e.g., accelerometers, gravimeters, gyroscopes) are today the most advanced sensing technologies that benefit from this revolution, exploiting matter-wave interferometry with Bose-Einstein condensates, using atom clouds cooled below nanoKelvin temperature.
For example, whereas classical accelerometers suffer from high noise at low frequencies, cold atom interferometers (CAI) are highly accurate over the entire frequency range and do not need any external calibration.

In the past twenty years, gravimetry missions have demonstrated a unique capability to monitor major climate-related changes of the Earth directly from space - quantifying the melt of large glaciers and ice sheets, global sea level rise, continental drought, major flooding events, and also the effects of large earthquakes and tsunamis. 
Adding to fundamental knowledge of the Earth, a quantum gravimetry mission for climate will provide essential climate variables (ECVs) of unprecedented quality for groundwater, mass balance of ice sheets and glaciers, heat and mass transport, building upon the successes of previous missions using
classical technology, such as GOCE~\cite{GOCE}, GRACE~\cite{tapley2019contributions} and GRACE-FO~\cite{GRACE-FO}. 
A combination of classical sensors with CAI or, at a later stage, a full quantum sensor will bring the Quantum Mission for Climate to a sensitivity that will open many applications and satisfy user needs~\cite{pail2015observing} with respect to water management and hazard prevention. 
In this connection, we take special note of the adoption of Quantum Technology for Earth Observation by the European Commission, notably (but not exclusively) in the Horizon Europe programme, under the thrust of Commissioner T. Breton, and of the inclusion of Quantum Technology in the ESA Agenda 2025~\cite{Agenda2025}.

Quantum Technology on Earth has revolutionised the measurement of time since the first atomic clocks in the 1950s, and these now provide the fundamental time frame across the globe. 
In space, atomic clocks have widespread applications such as satellite-based navigation systems (GPS, GALILEO). 
Terrestrial clocks based on atomic transitions are now reaching a relative accuracy on the order of $10^{-18}$, a level at which a change of height in the Earth's gravitational field of 1 cm would be detectable as a gravitational redshift. 
This sensitivity brings both challenges and opportunities.
The challenge for terrestrial clocks will be that changes in the local gravitational potential, either by human activity or by alterations in the local water table, will destroy the stability of the clock. 
This issue will certainly drive the siting of such clocks in space, with the implication that space qualification of the quantum technology will be essential for future development. 
The availability of such sensitive technology in space also offers significant opportunities to explore many aspects of fundamental physics.

Mounted on a space platform in a highly eccentric orbit, a sensitive atomic clock would provide an ideal laboratory to test 
 the Einstein Equivalence Principle (EEP) of General Relativity beyond current precision as the spacecraft experiences varying gravitational potentials around the orbit. 
This is a test that is at the heart of General Relativity and all metric theories of gravitation and space-time.
A fundamental aspect of the EEP of General Relativity is the Universality of Free Fall (UFF), which has been tested since the days of Galileo with ever-increasing accuracy. 
Quantum gravimetry using atom interferometers in space will allow pushing tests of UFF to new frontiers, with the potential of unveiling new physics beyond the Standard Model
that modify our theory of gravity, e.g., via fifth forces that exhibit screening mechanisms. 
These experiments represent one of the best ways of exploring the unknown theoretical interface between quantum physics and our best-tested theory of gravity, General Relativity.

The deployment of cold atom technology in space will also enable many other sensitive experiments in fundamental physics, cosmology and astrophysics, such as searches for ultralight dark matter particles, measurements of gravitational waves from the mergers of massive black holes and phenomena in the early Universe, and ultrasensitive probes of quantum mechanics, complementing terrestrial laboratory
experiments such as those at particle colliders.

The commonality of many subsystems between atomic clocks, gravimeters and fundamental physics experiments means that a well-planned programme of technical development should lead to the availability in space-borne missions of all these applications in fundamental science, Earth Observation, time keeping and navigation. 




\section{Atomic Clocks Review}
\label{sec:clocks}
\subsection{Scientific and societal opportunities}
\label{Sec:ClocksSciOpp}
\subsubsection{Fundamental science} \label{sec:clocks_fundamental_science}
High-stability and -accuracy atomic clocks combined with state-of-the-art time and frequency links can be used to measure tiny variations in the space-time metric and test the validity of the Einstein's Equivalence Principle.

As predicted by General Relativity, gravity influences the flow of time. 
When identical clocks experiencing a different gravitational potential are compared by exchanging timing signals, a relative frequency difference proportional to the difference of the gravitational potential at the location of the clocks can be measured. 
The effect, known as gravitational redshift, has been tested in 2018 with a precision of about $2\times10^{-5}$ \cite{Delva2018, Herrmann2018} by using the clocks on-board the Galileo 5 and 6 satellites. 
The ACES (Atomic Clock Ensemble in Space) mission \cite{ACES,cac20,Savalle2019b} will perform an absolute measurement of the redshift effect between the PHARAO clock on-board the International Space Station (ISS) and clocks on Earth, improving this limit by an order of magnitude. 
Optical clock missions on highly elliptical orbits around the Earth or cruising towards the Sun are expected to improve redshift tests by several orders of magnitude and to measure higher-order relativistic effects to high precision.

Local Lorentz Invariance (LLI) postulates the independence of any local test experiment from the velocity of the freely-falling apparatus. 
Optical clocks can be used to provide very stringent test of Lorentz symmetry and the 
Lorentz-Violating Standard Model Extension (SME)~\cite{Colladay:1998fq}. 
Distant Sr optical lattice clocks compared through optical fibre links have been used to constrain the Robertson-Mansouri-Sexl parameter to $1\times10^{-8}$ by searching for daily variations of the relative frequency difference \cite{Delva2017}. 
In \cite{Sanner2019}, two Yb$^+$ clocks confined in two traps with quantization axis aligned along non-parallel directions are compared while the Earth orbits around the Sun. 
The absence of frequency modulations at the level of $1\times10^{-19}$ made possible an improvement in the limits on the Lorentz symmetry violation parameter for electrons. 

Local Position Invariance (LPI) can also be tested by comparing clocks based on different atomic transitions. 
According to LPI, the outcome of any local test experiment is independent of where and when it is performed in the Universe. 
Transition frequencies depend differently on three fundamental constants: the fine structure constant $\alpha$, the electron mass $m_e/\Lambda_{QCD}$ (where $\Lambda_{QCD}$ is the QCD scale parameter), and the quark mass $m_q/\Lambda_{QCD}$. 
Therefore, comparing atomic clocks based on different transitions can be used to constrain the time variation of fundamental constants and their couplings to gravity. 
As an example, the comparison of two $^{171}$Yb$^+$ clocks based on the electric quadrupole and electric octupole transitions and two Cs clocks repeated over several years has recently improved the limits on the time variation of the fine structure constant and of the electron-to-proton mass ratio \cite{Lange2021}. 
At the same time, using the annual variation of the Sun's gravitational potential, it was possible to constrain the coupling of both constants to gravity.   

Atomic clock networks can also be used to place bounds on Topological Dark Matter (TDM) models. 
TDM can be expressed as a scalar field that couples to fundamental constants, thus producing variations in the transition frequencies of atomic clocks at its passage. 
Cross-comparisons between atomic clocks connected in a network over large distances can be used to place bounds on the time variation of three fundamental constants and determine exclusion regions for the effective energy scale (inverse of the coupling strength) of the dark matter field as a function of its Compton wavelength \cite{Roberts2017, Roberts2020, barontini2021qsnet, Barontini:2021mvu}. 
Clock networks providing redundant measurements are a powerful tool to control systematic effects and confirm any detection above the noise threshold.   

The dark matter distribution in the Solar system is critical for the reach of dark matter direct detection experiments.  It is possible that some local over-density of ultralight dark matter could be bound to the Sun.  A clock-comparison satellite mission with two clocks onboard sent to the inner reaches of the solar system was proposed in~\cite{x1} to search for the dark matter halo bound to the Sun, probe natural relaxion~\footnote{The relaxion is a light spin-zero field that dynamically relaxes the Higgs mass with respect to its natural large value~\cite{x2}.} parameter space, and look for the spatial variation of the fundamental constants associated with a change in the gravitation potential. 

Optical clocks have also been proposed for gravitational wave detection \cite{Kolkowitz2016, WP1EA, AEDGE}. 
A pair of clocks in drag-free satellites separated by a long-distance baseline share the interrogation laser via an optical link. 
The clocks act as narrowband detectors of the Doppler shift on the laser frequency due to the relative velocity between the satellites induced by the incoming gravitational wave. 
The atom interrogation sequence on the clock transition can be controlled, enabling precise tuning of the detection window over a wide frequency interval without loss of sensitivity. 
A frequency range between about 10~mHz and 10~Hz can be covered, thus bridging the gap between space-based and terrestrial optical interferometers, as discussed in more detail in Section~\ref{sec:PH}. 

\subsubsection{Metrology} \label{sec:Clocks_Metrology}
The basic `second' in the International System of Units (SI) is the quantity that is fixed with by far the lowest uncertainty of all units. 
This is done by primary frequency standards (laser-cooled Cs fountain clocks) operated at National Metrology Institutes. 
Global time scales rely on the comparison of such high-performance atomic clocks connected in a global network. 
The Bureau International des Poids et Mesures (BIPM) generates International Atomic Time (TAI),
which is based on the cross-comparison of the best primary frequency standards and, more recently, also optical clocks worldwide. Its rate is defined to be close to proper time on the geoid.
Coordinated Universal Time (UTC) is produced by the BIPM and differs from TAI only by an integral number of seconds as published by the BIPM:
it is the only recommended time scale for international reference and the basis of civil time in most countries.

Since optical clocks already outperform the primary frequency standards that operate in the microwave domain (see Section~\ref{Sec:Clocks}), the international metrology community represented by the Comité International des Poids et Mesures (CIPM) and its committees have devised a road-map for the redefinition of the second. 
This documents the high priority and strong commitment of a large community to the development and operation of optical clocks, with high relevance for society. 
Such a redefinition will enable a more accurate and stable international timescale \cite{Grebing2016, yao19}, which is key for precise navigation services via the 
global navigation satellite system (GNSS) network, the synchronization of worldwide exchanges and markets, communication networks, and national defence and security.

The coordination of time requires the permanent comparison and synchronisation of national timescales and clocks. 
With the increasing performance of optical clocks, the demands on the link quality are also  increasing. 
Today's microwave links achieve neither the necessary stability nor the accuracy required by the new optical clocks \cite{Riedel2020}. 
Locally, fibre-optical links can be an alternative \cite{Lisdat2016}, but a global network is not within reach. 
Long-distance time and frequency links enabling frequency comparisons at the level of $1\times 10^{-18}$ are urgently needed. 
Such local networks may even be combined by space clocks, as in the ACES \cite{cac20} or the proposed Space Optical Clock (SOC) \cite{Origlia2016} missions. 
Potentially, a space clock can overcome the limitations on the realization of the SI second and of timescales set by the knowledge of the gravity potential on the ground because the relativistic frequency shift, needed to transform the proper time of the clock to TAI, can be computed more accurately for a space clock than for a ground clock.
Presently, this correction can only be determined with a fractional uncertainty of about $3\times10^{-18}$, equivalent to 3~cm height \cite{Denker2017}, which is already larger than the uncertainty of today's optical clocks (see Section~\ref{Sec:ClocksLabClocks}).

\subsubsection{Earth observation \& geodesy}

In view of climate change and its consequences for society, Earth observation and geodesy are of  increasing importance. There exist highly accurate geometric reference frames based on the GNSS,  Very Long Baseline Interferometry (VLBI), Satellite Laser Ranging (SLR) and Doppler Orbitography and Radiopositioning Integrated by Satellite (DORIS) networks~\cite{IERS, ITRF}.  Physical height reference systems related to the geoid and the flow of water are much less accurate and fall behind the requirements set by the UN resolution for sustainable development~\cite{unr15} by more than an order of magnitude. 


Presently, physical heights are locally derived by spirit levelling tied to reference points such as tide gauges or global observations from satellite missions like CHAMP~\cite{CHAMP}, GOCE~\cite{GOCE}, GRACE~\cite{GRACE}, and GRACE-FO~\cite{GRACE-FO}. 
Although these missions were and are very successful, they lack spatial resolution and require considerable data processing, because the sensors are only sensitive to derivatives of the gravity potential. 
With clocks at a fractional uncertainty level of $10^{-18}$, we now have sensors at hand that are directly sensitive to the gravity potential via the relativistic redshift~\cite{wu2019clock, wu2020towards}. 
Therefore, we have an opportunity to establish a novel technique to realize a height reference system using a network of optical clocks from which the physical height differences at the respective locations can be derived. 

Present clock performance already provides a height resolution better than the current geodetic state-of-the-art \cite{Denker2017}, and is likely to reach the millimetre level within the next decade. 
It is essential to establish links for cross-comparisons of optical clocks that are flexibly accessible and span the globe. 
Satellite-based approaches fulfil these requirements in an ideal fashion. 
While satellite-mediated ground-to-ground links with improved performance compared to, e.g., ACES, will enable a fast development of this field of application.
Clocks operated in space are expected to improve the products from GRACE-like missions, in terms of estimates of potential coefficients, 
arguably in the low-frequency range of the spherical harmonics. 
If a suitable satellite formation is exploited, an improvement in the estimation of time variations in the potential could also be achieved.
Space clocks may ultimately provide an independent, long-term stable and reproducible height reference for decades, even centuries of Earth monitoring.

\subsection{Clocks: state of the art}
\label{Sec:Clocks}
\subsubsection{Laboratory-based clocks}

\begin{figure}[h!]
\centering
\includegraphics[width=0.65\textwidth]{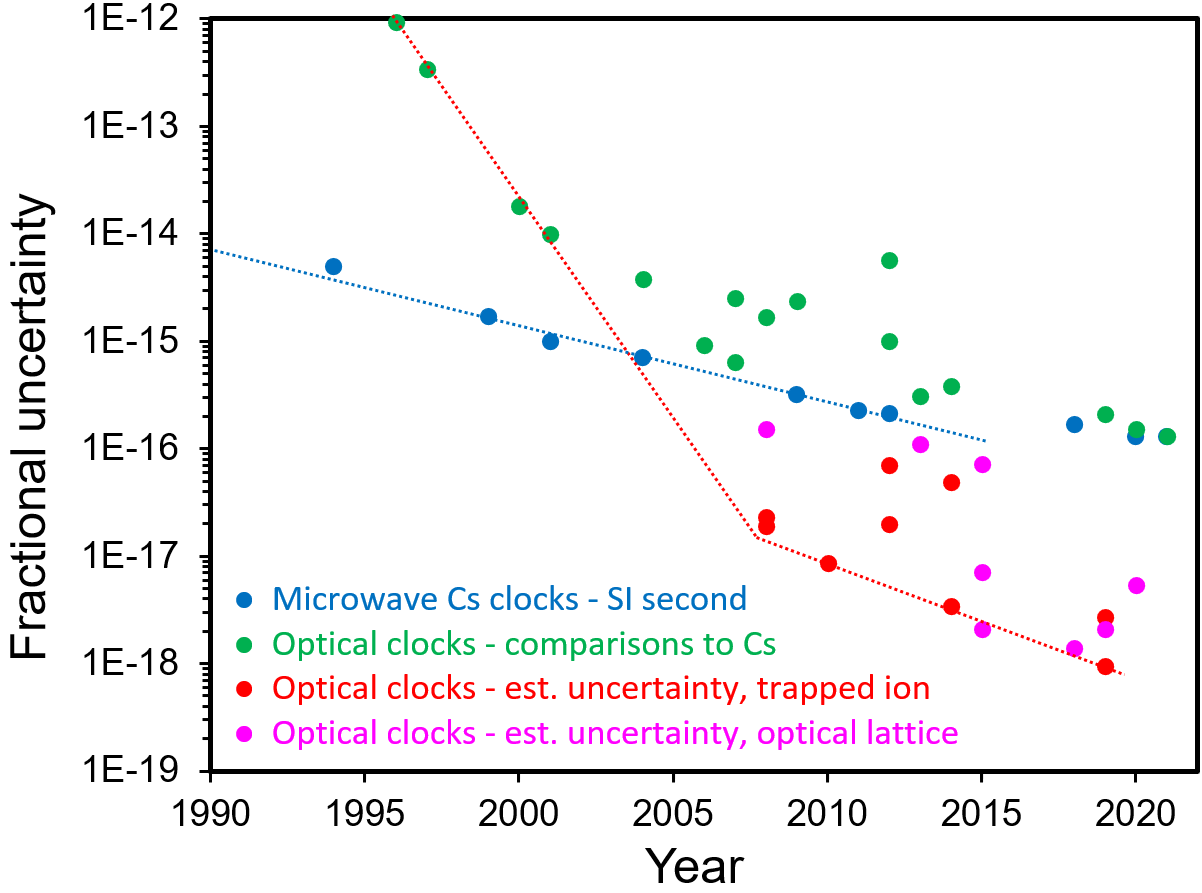}
\caption{Progress in the relative accuracy of atomic clocks \cite{Schnatz1996, Udem1997, Niering2000, Udem2001, Margolis2004, Oskay2006, Boyd2007, Stalnaker2007, Poli2008, Chwalla2009, King2012, McFerran2012, Targat2013, Huntemann2014, McGrew2019, Schwarz2020, Weyers2018, Lange2021, Rosenband2008, Chou2010, Madej2012, Huntemann2012a, Huntemann2014a, Barwood2014, Brewer2019, Sanner2019, Nicholson2015, Ushijima2015, Yamanaka2015, McGrew2018, Bothwell2019, tak20}. 
Cs microwave clocks have steadily improved since the emergence of laser-cooled fountain clocks in the early 1990s \cite{Kasevich1989,Clairon1991,Jefferts2002}, but two distinct types of optical clock currently compete at a fractional frequency uncertainty of approximately $1\times 10^{-18}$: trapped-ion clocks  (Yb$^+$ \cite{Sanner2019}, Al$^+$ \cite{Brewer2019}, Hg$^+$ \cite{Rosenband2008}, Sr$^+$ \cite{Madej2012}, Ca$^+$ \cite{cao17}) and optical lattice clocks (Sr \cite{Bothwell2019}, Yb \cite{McGrew2018}, Hg \cite{Yamanaka2015}, Cd \cite{Yamaguchi2019}).}
\label{fig:clock_accuracy}
\end{figure}

Figure \ref{fig:clock_accuracy} shows the historical progress of state-of-the-art laboratory atomic clocks. 
cesium (Cs) microwave atomic clocks have been the primary standard for the SI second since 1967, which has helped to motivate the development of several generations of Cs clocks with reduced fractional frequency errors. 
However, in recent years, optical atomic clock technology has matured significantly: the best optical atomic clocks now surpass Cs in relative accuracy by a factor of more than 100. 
The field of optical clocks encompasses a diverse range of trapped-ion clocks and optical lattice clocks, each with distinct merits. 
Trapped ion clocks naturally offer exquisite environmental isolation and quantum control, leading to small systematic shifts and high accuracy \cite{Sanner2019,Brewer2019}.
However, optical lattice clocks have the key advantage of using many atoms in parallel, resulting in greater frequency stability and therefore allowing high-precision measurements within a significantly shorter averaging time \cite{Schioppo2017, sch20d}.

\label{Sec:ClocksLabClocks}

In order to verify the $10^{-18}$ precision of the best optical clocks and pursue some scientific opportunities discussed in Section~\ref{Sec:ClocksSciOpp}, it is important to compare different clocks of comparable precision with each other. 
Local measurements of optical clocks within a single laboratory have been used to measure general relativistic redshifts corresponding to vertical displacements at the millimetre scale \cite{Bothwell2021} and to search for possible variations in local physics induced by dark matter \cite{Kennedy2020,Godun2014,Huntemann2014}. 
Distant comparisons between clocks in separate laboratories allow large numbers of independent clocks to be included, but must be mediated by a complex frequency link infrastructure. 
The longest distances are spanned by satellite links, which allow comparison at $10^{-16}$ fractional frequency uncertainty \cite{Riedel2020}. 
However, much higher precision at the $10^{-18}$ level can be carried out over shorter distances using terrestrial links, either with free-space lasers \cite{Giorgetta2013} or telecoms-wavelength lasers sent through optical fibres \cite{Yamaguchi2011,Lisdat2016,Beloy2021}. 
The most extensive optical fibre network is operated between European metrology institutes \cite{Lisdat2016}, across which several clocks have been compared to search for physics beyond the Standard Model (see Section~\ref{sec:clocks_fundamental_science}) \cite{Delva2017,Roberts2019}.

The development of laboratory atomic clocks is fuelled by a broad community that spans universities, industry and several National Metrology Institutes. 
Optical lattice clocks are now particularly widespread, with more than a dozen strontium (Sr) \cite{Bober2015,Bongs2015,Falke2014,Targat2013,Lin2015,Ludlow2006,Poli2014,Strelkin2015,Takamoto2005,Ushijima2015,Yamaguchi2011,Hobson2020,Zheng2021,Akamatsu2014a} and some ytterbium (Yb) \cite{Akamatsu2014a,Barber2006,Pedrozo-penafiel2020,Gao2018,Kim2021} clock laboratories in operation worldwide.~\footnote{Importantly, these optical lattice clocks use the same Sr and Yb technology as proposed for atom-interferometer science missions such as AEDGE~\cite{AEDGE}.}
The commitment of the metrology community to continue investing in optical clocks is highlighted by the CIPM road-map for an optical redefinition of the SI second (see Section~\ref{sec:Clocks_Metrology}). 
The road-map mandates a research programme likely to span at least the next decade, in which several optical clocks will be developed at $1 \times 10^{-18}$ relative accuracy and validated through clock-clock measurements. 
To carry out such measurements, the priority of the optical clock community will be to develop cold atom technology with higher technology readiness levels (TRLs), capable of combining state-of-the-art accuracy with robust, long-term operation---an investment which should have close synergies with a future programme for cold atoms in space.

\subsubsection{Transportable clocks}
Early in the development of optical frequency standards it was recognized that mobile devices (see, e.g., \cite{ker99}) enable applications (see Section~\ref{Sec:ClocksSciOpp}) of clocks that are impractical if the availability of clocks is restricted to only a few laboratories. 
The required engineering to develop delicate laboratory systems into robust mobile devices also opens the door to commercialization and space applications of clocks. 
While there have been several impressive demonstrations of compact optical frequency standards with high performance \cite{ols19, doe19}, we focus here on activities that target a clock performance similar to the state-of-the-art of laboratory setups (see Section~\ref{Sec:ClocksLabClocks}).

To maintain the outstanding frequency stability of optical clocks, ultra-stable interrogation lasers are required. 
For reasons of seismic and thermal insulation, these are typically neither robust nor compact. 
Therefore, the further development of these devices was identified by the community as an important challenge \cite{ arg12, che14a, hae20} and supported, e.g., by ESA activities~\cite{Webster2011, Sanjuan2019, hil21a} and is -- with demonstrated fractional frequency instabilities significantly below $10^{-15}$ -- on a good path. 
As ultra-stable laser systems have numerous applications beyond optical frequency standards, e.g., in atom interferometry, ultra-stable microwave generation, or optical telecommunication, the continued support of these activities is of high importance.

The realization of a fully transportable optical clock requires more lasers and a complex technical infrastructure (see Fig.~\ref{fig:ACES}), and thus poses a larger challenge. 
Nevertheless, several such systems working with neutral atoms \cite{Origlia2018, kol17, ohm21} or single ions \cite{cao17} have been realized, which already outperform the most accurate microwave standards. 
These setups are developed for space applications \cite{Origlia2016}, and have been used in a geodetic context \cite{gro18a, hua20} or to test fundamental aspects of physics \cite{tak20}. 
We therefore conclude that the construction and reliable operation of optical clocks with fractional uncertainties of $1 \times 10^{-17}$ and below and compact dimensions with volumes $\lesssim 1$~m$^3$ is already possible today~\cite{bon15, focos}.

\subsubsection{Free space-time and frequency links}
\label{Sec:ClocksLinks}
Connecting (optical) atomic clocks worldwide lays the basis for applications such as the creation of TAI or a Positioning, Navigation and Timing (PNT) standard, and would also open the route to testing theories of fundamental physics (see Sections~\ref{Sec:ClocksSciOpp} and \ref{Sec:ClocksIntSpace}).

Currently, primary microwave clocks are connected via satellites \cite{Riedel2020} in the microwave domain by the existing GNSS infrastructure \cite{pet21} or dedicated two-way time and frequency transfer (TWTFT) links \cite{kir93,itu15,hua16a}. 
Demonstrated frequency transfer uncertainties of existing microwave links (MWLs) reach into the $10^{-16}$ range after averaging times of days \cite{pet21,dro15} 
and into the $10^{-17}$ range after averaging times of weeks with integer ambiguity resolution~\cite{pet21}, and demonstrated time transfer uncertainties lie in the nanosecond region \cite{hua16a}. 
A new generation of MWL equipment is under development \cite{sch16b,cac20}, which reaches in laboratory tests timing instabilities of $<100$~fs for averaging times $\tau = 10$~s to 2000~s \cite{cac20}, which is equivalent to fractional frequency transfer uncertainties of $<5 \times 10^{-17}$ at 2000~s. 
Similar performances are achieved in the optical domain by Time Transfer by Laser Link (T2L2) \cite{sam14} and the European Laser Timing (ELT) experiment \cite{sch10h} employing time-of-arrival measurements of laser pulses.

A significantly improved uncertainty is achieved by techniques exploiting the optical carrier. 
Optical frequency dissemination using continuous wave laser signals \cite{dje10} reaches fractional frequency transfer instabilities $<5 \times 10^{-19}$ already after 100~s of averaging time in path-length stabilized operation \cite{Dix-Matthews2021}. 
A team at NIST has developed an optical TWTFT (OTWTFT) technique \cite{Giorgetta2013} combining carrier and time-of-flight information, allowing phase-coherent averaging over the signal dropouts that occur inevitably due to atmospheric turbulence \cite{sin18}. 
Using this technique, the NIST team has demonstrated sub-$10^{-18}$ frequency transfer uncertainty and sub-1~fs timing uncertainty at an averaging time of 1000~s in a 3-node network of two concatenated 14~km links \cite{bod20a}. 
Furthermore, they demonstrated OTWTFT to a flying drone with similar performance~\cite{ber19}. 
Despite the proven performance, however, the remaining steps to achieve ground-to-satellite world-wide coverage remain challenging: demonstrate techniques for higher relative speeds between sender and receiver, assess the impact of atmospheric turbulence, signal loss, potential loss of reciprocity on such ground-to-satellite links, and consider inclusion of relativistic effects. 
Recently, a first study addressed this scaling to ground-to-satellite connections \cite{she21} and came to a positive conclusion regarding the feasibility. 
Nevertheless, further experimental evidence gradually approaching the long-term ground-to-satellite goal is required. 
Synergies can be expected with the proposed combination of microwave and optical links in the context of new GNSS constellations \cite{Glaser2020}.


\subsection{International space activities}
\label{Sec:ClocksIntSpace}
Space is the ideal laboratory to test general relativity and alternative theories of gravitation with atomic clocks. 
The large velocities and velocity variations, the access to large variations in the gravitational potential, and the possibility to establish a global network able to compare ground clocks across continents from space provide new opportunities both for fundamental physics research and for applications in other areas of research, such as clock synchronization and timescale distribution, geodesy, Earth observation, navigation, etc., as discussed elsewhere in this report.

\begin{figure}[h!]
\centering
\includegraphics[width=0.80\textwidth]{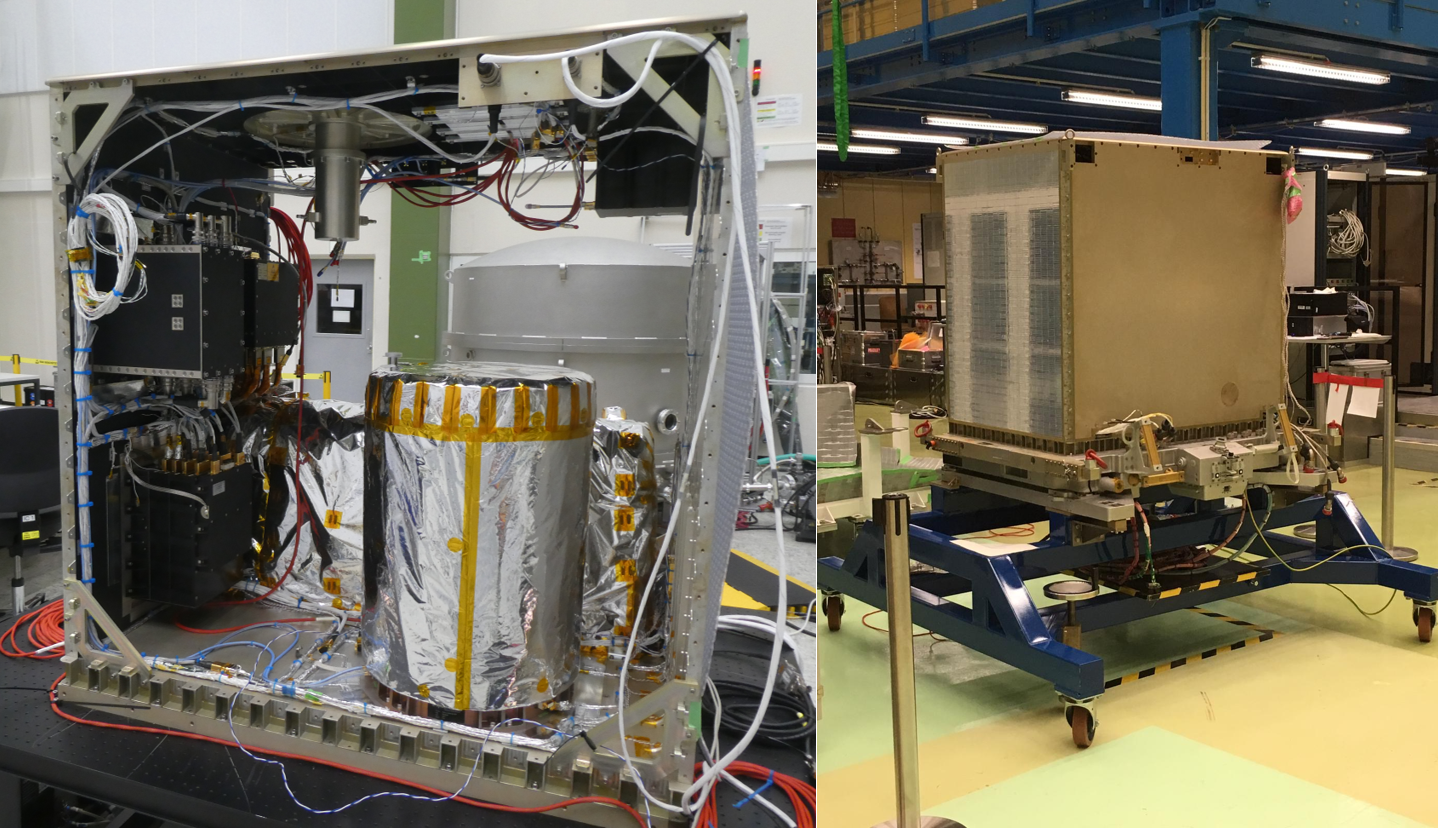}
\caption{(Left) Flight model of the ACES payload during the assembly phase. 
SHM (vertical cylinder) and PHARAO (in the background) are installed on the bottom panel. 
The ACES computer, the PHARAO computer, and the on-board phase comparator are visible on the left panel. 
MWL electronics (not installed yet) and the antennae are accommodated on the top panel. 
(Right) The ACES payload installed on the Columbus External Payload Adapter (CEPA) for interface tests.}
\label{fig:ACES}
\end{figure}

ACES (Atomic Clock Ensemble in Space) \cite{cac20} is an ESA mission designed to operate on the International Space Station. 
The two on-board clocks rely on atomic transitions in the microwave domain. 
The PHARAO clock, a primary frequency standard based on laser cooled Cs atoms, provides the ACES clock signal with a long-term stability and accuracy of $1\times10^{-16}$ in fractional frequency; the active H-maser SHM is the on-board flywheel oscillator that will be used for the characterization of the PHARAO accuracy. 
The ACES clock signal is distributed to ground clocks by using two time and frequency links: MWL is a link in the microwave domain; ELT is an optical link using short laser pulses to exchange timing signals. 
A distributed network of MWL ground terminals will connect the clocks operated in the best research institutes worldwide (SYRTE, PTB, NPL and Wettzell in Europe, NIST and JPL in the US, NICT in Japan) to the ACES clock signal. 
Satellite laser ranging stations will also be connected to the clock network by using the ELT optical link. 
The space-to-ground clock de-synchronization measurement produced by MWL and ELT will be used to perform an absolute measurement of the gravitational redshift in the field of the Earth to $< 2$~ppm, to probe time variations of fundamental constants, and to perform tests of the Lorentz-Violating Standard Model Extension (SME)~\cite{Colladay:1998fq}. 
The possibility of searching for topological dark matter with the ACES network is also being investigated. 
ACES is expected to fly to the ISS in the 2025 time frame. 
The flight model of the ACES payload is shown in Fig~\ref{fig:ACES}.

Significant advances in the development of microwave cold-atom clocks have been achieved in China with the launch and on-orbit operation of the CACES (Cold Atom Clock Experiment in Space) clock based on laser-cooled Rb atoms \cite{Liu2018}. 
The experiment was successfully operated on the Chinese space laboratory Tiangong-2. 
From an analysis of the Ramsey fringes, an estimated short-term stability at the level of $3\times10^{-13}/\sqrt{\rm s}$ was attained under free fall conditions. 
Unfortunately, a full characterization of the clock in space was not possible due to the absence of a stable frequency reference and of a space-to-ground link on board Tiangong-2. 
The clock performance is still to be optimised, but the experiment clearly demonstrates the robustness of the cold-atom clock technology for space.

In the US, the NASA's Jet Propulsion Lab has successfully demonstrated mercury trapped-ion clock technology in space \cite{Burt2021}. 
The Deep Space Atomic Clock (DSAC) payload, consisting of a Hg$^+$ microwave clock and a dedicated GPS receiver, was launched into a 720-km orbit around the Earth in June 2019. 
The space clock was compared to the clocks from the US Naval Observatory, and demonstrated a fractional frequency stability between $3\times10^{-15}$ and $5\times10^{-15}$ at 1 day and $3\times10^{-15}$ after 23 days. 
The short-term stability of the clocks, which is below the GPS measurement system noise, could be estimated as $7\times10^{-13}/\tau^{1/2}$, where $\tau$ is the integration time. 
This technology can be used for navigation, planetary science, and fundamental physics. 

As discussed above, optical clocks can provide an improvement in stability and accuracy of 2 orders of magnitude with respect to microwave clocks. 
Following the impressive progress of atomic clocks based on optical transitions, several initiatives are currently ongoing to advance the required technology to flight readiness. 

Europe is developing key optical clock technology for space, e.g., cooling lasers, the clock laser, a high-finesse reference cavity, a clock control unit to stabilize the laser frequency on the atomic transition, and the lattice laser. 
A design study for a Sr clock physics package has been completed. 
Compact and transportable ground-based prototypes for a Sr optical lattice clock \cite{Origlia2018} and a Sr ion clock \cite{Barwood2017} are being characterized. 
Free-space coherent optical links reaching a fractional frequency uncertainty of $1\times10^{-19}$ in a few days of measurement time are under development \cite{Dix-Matthews2021}. 
In parallel, the I-SOC Pathfinder platform has been proposed as the ACES follow-on mission. 
I-SOC Pathfinder is pushing further the microwave and optical link technology \cite{Prochazka2018, Prochazka2020} developed for ACES to continue operating a worldwide network of optical clocks on the ground to test fundamental laws of physics, to develop applications in geodesy and time \& frequency transfer, and to demonstrate key time and frequency link technologies that are essential for all future atomic clock missions in space. The I-SOC Pathfinder mission could be operated on the ISS, but other platforms as geostationary satellites could be even preferable to enhance common view time slots between selected ground stations.

In the US, the FOCOS (Fundamental physics with an Optical Clock Orbiting in Space) mission concept has been proposed~\cite{focos}. 
FOCOS relies on a Yb optical lattice clock with $1\times10^{-18}$ stability and accuracy on a highly elliptical orbit around the Earth. 
A coherent optical link is used to compare the space clock to ground clocks for general relativity tests and timing applications. 
FOCOS will also serve as a pathfinder for future atom interferometry missions to test the Equivalence Principle, clock constellations in space to hunt for dark matter \cite{Roberts2017, Roberts2020}, and gravitational wave observatories \cite{Kolkowitz2016}.

In parallel, CACES follow-on experiments based on optical clock technology are under development in China.

Finally, major efforts and resources are being invested worldwide to improve the atomic clocks of the Global Navigation Satellite System (GNSS). 
Currently, available technology relies on the passive H-maser, the Rb atomic frequency standard, and the Cs beam frequency standard \cite{Steigenberger2017, Vannicola2010}. 
Clocks for navigation satellites have reduced stability (in the $10^{-15}-10^{-14}$ range for fractional frequency), but offer a more compact design with low mass (3 to 20~kg), low power consumption (30 to 70~W), and long lifetime on orbit. 
Alternative technologies are under study for the next generation of atomic clocks for navigation. 
Among them, it is worth mentioning the mercury ion clock technology \cite{Burt2021}, the pulsed optically pumped Rb clock \cite{Shen2020}, the Rb optical atomic clock \cite{Phelps2018}, and the iodine frequency reference~\cite{doe19, Schkolnik2017Jokarus} that will soon be tested in the COMPASSO experiment on the ISS platform Bartolomeo \cite{DLRCOMPASSO}. 
These developments are maturing key technology for space (see Section~\ref{sec:Tech}) making it possible not only to deploy compact atomic clocks for global positioning and navigation, but also high-performance atomic clocks for fundamental physics and geodesy (see Section~\ref{sec:EO}).     

\subsection{Recommendation: Road-map to space clocks}


As discussed above and in Section \ref{sec:PH}, there are many possible configurations of space-clock missions. 
Here we envisage the possibility of three missions, undertaken in stages:

\begin{itemize}
    \item Complete ACES and launch it to the ISS with utmost urgency.
    \item Implement I-SOC Pathfinder (or an equivalent) on the ISS as an ACES follow-on mission. 
    The payload, containing an active H-maser, a microwave link, and a laser-pulse optical link, is designed for comparison of ground-based optical clocks to $10^{-18}$ fractional frequency precision in 1 day. 
    This would have applications in fundamental physics discovery, proof-of-concept optical timescales, and geodesy.
    \item Launch a dedicated satellite in a highly elliptical orbit containing a strontium optical lattice clock with a $1\times10^{-18}$ systematic uncertainty and $1\times10^{-16}/\sqrt{\tau}$ instability, with a coherent optical link to ground. 
    The goals of such a mission are similar to the ones proposed in FOCOS \cite{focos}; strong cooperation would be very beneficial. 
    A mission of this type will enable more precise comparisons across a wider network of ground clocks, and direct searches for new physics, including stringent tests of general relativity.
\end{itemize}

Within this road-map, we highlight an urgent need to develop coherent free-space optical links capable of clock comparisons at the $10^{-18}$ level in less than 1 day. 
Once optical links are in place, a critical element of the road-map is then the qualification of a strontium optical lattice clock for operation in space~\footnote{Ytterbium could be used in place of strontium, as proposed in FOCOS~\cite{focos}. 
The atoms share similar complexity and capability: either would be suitable for an atomic clock mission, or for an atom-interferometer science mission of the type proposed in AEDGE~\cite{AEDGE}.}. 
To achieve this, a technology development programme must be undertaken for several components of the clock: optical resonators to stabilize lasers to a noise floor of $1\times10^{-16}$ in fractional frequency; laser sources at six different wavelengths to cool, optically confine and interrogate strontium atoms; compact physics packages with a controlled black-body radiation environment; and compact frequency combs. 
Further details of the required technologies are outlined in Section~\ref{sec:Tech}. 

The possibility of establishing collaborations between Europe, US, and China on major future scientific missions should be investigated.~\footnote{For example, in the context of the US National Academies Decadal Survey on Biological and Physical Sciences Research in Space 2023-2032~\cite{Decadal}.}
Further, synergies should be exploited between the atomic clock missions discussed here and the requirements for the AEDGE mission~\cite{AEDGE} discussed in Section~\ref{sec:PH}, for which the same cold-strontium technology is envisaged.


\section{Quantum Gravimetry for Earth Observation Review}
\label{sec:EO}
\subsection{The observation of mass change and Earth Observation requirements}

\subsubsection{Earth sciences and gravity field observations}

The Earth sciences need gravity-field observations from satellites for understanding the Earth's structure and monitoring its changes related to geodynamics and climate change, as well as many other needs of society related to its glacio- and hydrosphere. 
Satellite earth observations (EO) enable the observation and monitoring of the Earth's density distribution and mass variations, and provide a significant contribution to the determination of many Essential Climate Variables (ECVs) as defined by the Global Climate Observing System (GCOS)~\cite{GCOS}. 
ECVs monitor phenomena that are changing the world we live in, such as climate change, changing water resources, flooding, melting of ice masses, global sea level rise and atmospheric changes. 
Better knowledge of such phenomena is bound to lead to significant societal benefits via, e.g., the operational prediction of floods and droughts, monitoring and prediction of sea level rise, and in applications regarding water management.

For these reasons, several initiatives and studies have been launched in the past years at the international level to foster continued observation and monitoring of mass and mass transport phenomena. 
In 2015 the International Union of Geodesy and Geophysics (IUGG) issued a resolution on ``Future Satellite Gravity and Magnetic Mission Constellations”~\cite{IUGG2015} and launched an international multidisciplinary study on science and user needs for the observation of mass transport to understand global change and to benefit society. 
In the resulting report it was stated that ``… a satellite gravity infrastructure is needed with increased space-time sampling capability, higher accuracy and sustained observations”~\cite{IUGGPail2015}, see also~\cite{Pail}. 
Moreover, in 2019 the International Association of Geodesy (IAG) started the ``Novel Sensors and Quantum Technology for Geodesy" (QuGe) initiative; in its framework, three working groups have been defined, one specifically devoted to ``Quantum gravimetry in space and on ground”~\cite{Pereira} and a second one targeting ``Relativistic geodesy with clocks''~\cite{Petit} (See Section~\ref{Sec:ClocksSciOpp}).

\subsubsection{Past, present and planned gravimetry missions}

\begin{figure}[h!]
\centering
\includegraphics[width=\textwidth]{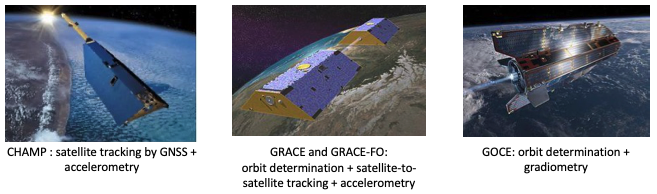}
\vspace{-0.5cm}
\caption{\it The CHAMP~\cite{CHAMP}, GRACE~\cite{GRACE} and GOCE~\cite{GOCE} satellites and mission concepts.}
\label{CHAMP}
\end{figure}

In the past twenty years, satellite gravity missions have helped form a well-organized user community tracking the Earth mass movements and to study environmental changes on a global scale using data from satellite observations (see Fig.~\ref{CHAMP} and Table~\ref{tab:models}). 

The general principle of gravity missions is based on precise tracking of satellite position (in free fall) and inter-satellite ranging, combined with the determination of non-gravitational accelerations by means of accelerometers on board the satellites. 
The technology exploited so far for gravimetry missions is represented by electrostatic accelerometers (EA), as in the GFZ CHAMP mission that flew from 2000 to 2010~\cite{CHAMP}, where the accelerometer provides observations that represent the surface forces acting on the satellite, i.e., all non-gravitational accelerations (drag, solar and Earth radiation pressure), so that the Earth gravity field can be obtained from purely gravitational orbit perturbations (observed by satellite tracking). 
Based on the same EA technology but measuring orbit perturbations by satellite-to-satellite tracking, the NASA/DLR GRACE mission that flew from 2002 to 2017~\cite{GRACE} and the GRACE-FO mission launched in 2018~\cite{GRACE-FO} have provided and are providing routine measurements of the spatial variations of the Earth gravity field in space and in time on a monthly basis. 
These missions are based on the concept of flying a pair of low-Earth-orbiting satellites, precisely tracked using the Global Navigation Satellite System (GNSS) and an inter-satellite microwave ranging system, with the accelerometers enabling the measurement of non-gravitational forces. 
This allows for long-term monitoring of the gravity field and its time variations. 

However, the 20-year-long time series of observations needs to be prolonged: this is why several space agencies are working on follow-up missions. 
ESA plans a Next-Generation Gravity Mission – NGGM~\cite{NGGM}, scheduled for launch in 2028 and also known as MAGIC (Mass-change and Geosciences International Constellation), in cooperation with the corresponding NASA  MCDO mass change satellite initiative ~\cite{MassChange}. 
The MAGIC mission configuration will be based on two pairs of twin satellites: one pair will fly on a quasi-polar orbit, the other one on an orbit with inclination of about 67$^{\circ}$~\cite{MAGIC}, with laser ranging between each pair. 
This will allow for a reduction in the revisit time, providing higher temporal and spatial resolution.

Based on a different measurement concept, gravimetry can also be performed by acquiring observations from a gradiometer, measuring gravity gradients inside the satellite. 
This was done in the very successful GOCE mission flying from 2009 to 2013~\cite{GOCE}, which exploited gradiometry for a unique mapping of the static gravity field, providing models with unprecedented accuracy for a range of geophysical and oceanographic applications (e.g., sea-level currents, a reference system for global height systems, and background data for geophysics and understanding the Earth interior).

Table~\ref{tab:models} summarizes the status of the gravity missions based on classical technology, the accuracies attained so far and the prospects for the near future.

\begin{table}[htb]
   \centering
\hspace{-1cm}
{\small    \begin{tabular}{|c|c|c|c|c|}
\hline
    & CHAMP & GRACE/GRACE-FO & NGGM & GOCE \\
    & 2000 - 2010 & 2002 - ongoing & Launch scheduled 2028 & 2009 - 2013 \\
        \hline \hline
     Measurement type  & & \multicolumn{2}{c|}{Monitoring gravity field time variations} & Static gravity field \\
      \hline
     EA accuracy & $\sim 10^{-10}$~m/s$^2$ & $\sim 10^{-11}$~m/s$^2$ & $\sim 10^{-11}$~m/s$^2$ & $\sim 10^{-12}$~m/s$^2$  \\
     \hline
    Geoid & $\sim 10$~cm & $\sim 10$~cm & $\sim 1$~mm @ 500~km & $\sim 1$~cm \\    
    undulations  & $@ 350~km$  & $@ 175~km$    & every 3 days & $@ 100~km$ \\
     &  & & $\sim 1$~mm @ 150~km & \\
      &   &   & every 10 days &  \\
    \hline
     Gravity  & $\sim 0.02$~mGal &  $\sim 1$~mGal& & $\sim 1$~mGal  \\  
    anomalies & $@ 1000~km$ &  $@ 175~km$  & & $@ 100~km$\\
    \hline 
  \end{tabular}}
   \caption{\it Accuracies in the determination of the gravity field by ``classical" measurements (including the planned ESA-NASA NGGM/MAGIC mission~\cite{MAGIC}).}
    \label{tab:models}
\end{table}

Although the classical gravity field missions have been highly successful, they  have not satisfied all the user needs of science and society. 
These have been summarized  in various international reports, in the form of tables such as those provided in~\cite{Pail2}, see Table~\ref{tab:models2}.\\

\begin{table}[htb]
   \centering
{    \begin{tabular}{|c|c|c|c|c|}
\multicolumn{5}{c}{\it Threshold requirements}\\
\hline
 Spatial   & \multicolumn{2}{c|}{Equivalent water height} & \multicolumn{2}{c|}{Geoid} \\
 resolution &  Monthly field & Long-term trend & Monthly field & Long-term trend \\
        \hline
400~km  & 5~mm & 0.5~mm/yr & 50~$\mu$m & 5~$\mu$m/yr \\      
200~km  & 10~cm & 1~cm/yr & 0.5~mm & 0.05~mm/yr \\      
150~km  & 50~cm & 5~cm/yr & 1~mm & 0.1~mm/yr \\      
100~km  & 5~m & 0.5~m/yr & 10~mm & 1~mm/yr \\      
    \hline 
    \multicolumn{5}{c}{\it Target objectives}\\
\hline
 Spatial   & \multicolumn{2}{c|}{Equivalent water height} & \multicolumn{2}{c|}{Geoid} \\
 resolution &  Monthly field & Long-term trend & Monthly field & Long-term trend \\
        \hline
400~km  & 0.5~mm & 0.05~mm/yr & 5~$\mu$m & 0.5~$\mu$m/yr \\      
200~km  & 1~cm & 0.1~cm/yr & 0.05~mm & 5~$\mu$m/yr \\      
150~km  & 5~cm & 0.5~cm/yr & 0.1~mm & 0.01~mm/yr \\      
100~km  & 0.5~m & 0.05~m/yr & 1~mm & 0.1~mm/yr \\      
    \hline 
  \end{tabular}}
\caption{\it Consolidated science and users' requirements for earth observation, as reported in~\cite{Pail2}.}
  \label{tab:models2}
\end{table}

\subsubsection{Earth observation requirements}

\begin{figure}[h!]
\centering
\includegraphics[width=0.5\textwidth]{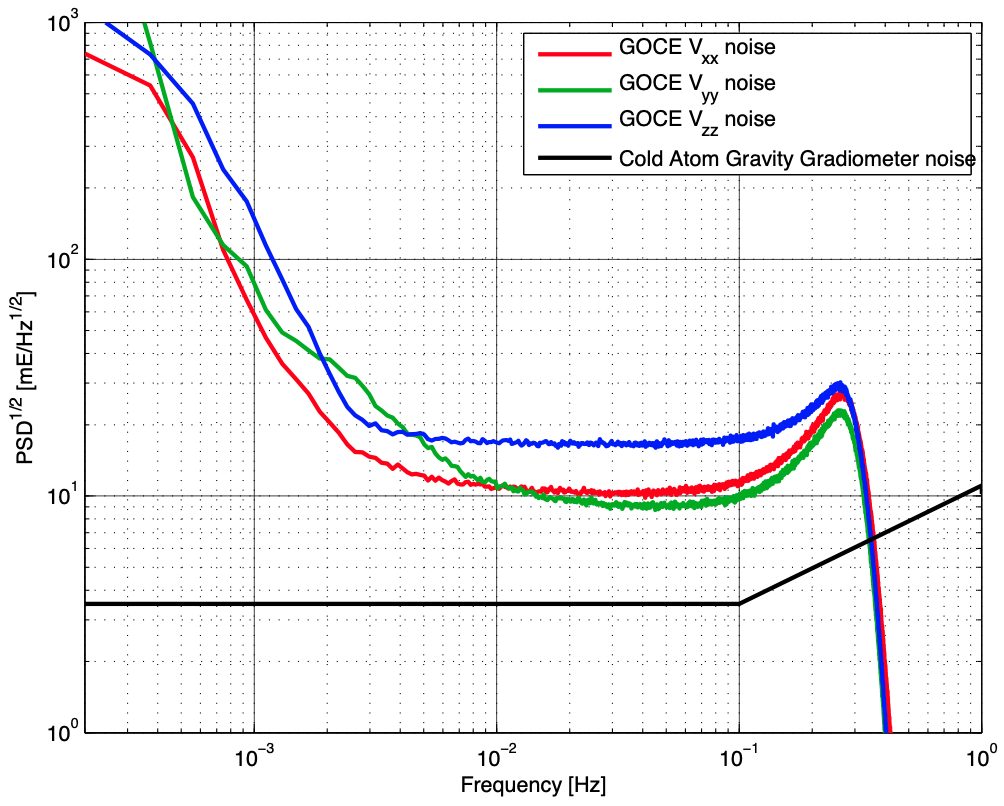}
\caption{\it Comparison between the noise spectra in GOCE gradiometers and in a prospective cold atom gradiometer, illustrating the latter's {potentially} reduced level at low frequencies. 
Figure taken from ~\cite{Carraz2014}, {see~\cite{trimeche2019concept} for a recent discussion.}} 
\label{fig:Carraz}
\end{figure}

The numbers in Table~\ref{tab:models2} point towards mission requirements that deliver measurements with higher sensitivity,
greater accuracy, and more long-term stability. 
In summary, the needs are:\\
$\bullet$ Higher spatial resolution (implying lower orbits, 300-350 km) for detection of gravity changes due to movements of mass in the Earth system. 
However, it must be remarked that no space mission will be able to map the higher-frequency details of the gravity field, due to the atmospheric limitations of the orbit height. 
Therefore, a full detailed mapping of the spatial gravity field variations down to a few km resolution must be supplemented by airborne and ground gravity measurements;\\
$\bullet$ Shorter revisit times, which require flying a satellite constellation, e.g., double pairs such as in NGGM/MAGIC. 
The improved temporal resolution would be crucial for operational service applications such as near real-time flood tracking. 
A shorter revisit time will also aid in better determination of tidal effects, which must currently be modelled, and represent a limiting factor in the accuracy of current missions;\\
$\bullet$ Greater accuracy in the measurements, by exploiting new technologies such as laser interferometry and cold atom accelerometers, with accelerometers accurate to better than $\sim 10^{-10} - 10^{-11}$~m/s$^2$ (measurement range of $\pm 10^{-4}$~m/s$^2$) and gradiometers accurate to $\sim 10^{-12}$~m$^2$/s$^2$/$\sqrt{\rm Hz}$ for a GOCE-like mission, over a 
larger spectral measurement band, due to the spectral behaviour of the quantum accelerometers at low frequencies,
compared with the spectral behaviour of electrostatic accelerometers, as illustrated in Fig.~\ref{fig:Carraz}, taken from~\cite{Carraz2014}.
These improvements in instrument performance and low-frequency stability will become important for satellite constellations;\\
$\bullet$ Extension of the observation time series, which is essential for long-term monitoring of mass transport and variations, and especially for understanding the separation of natural and anthropogenic forcing.

\subsection{Quantum Sensors in the Context of Earth Observation}

\subsubsection{Classical and quantum sensors for gravimetry}

Quantum sensors based on atom interferometry are extraordinarily sensitive to external forces~\cite{Geiger2020}, reaching an accuracy of $40$ nm/s$^2$ for 
gravimeters~\cite{Karcher2018}, $8\times 10^{-8}$ 1/s$^2$ for gradiometers~\cite{Chiow2016,Asenbaum2016} and $70$ nrad/s for gyroscopes~\cite{Berg2015a,Savoie2018}.
In Table~\ref{tab:models3} some of the most relevant features~\cite{trimeche2019concept, douch2018simulation} of quantum accelerometers and the classical electrostatic accelerometers used so far for space accelerometry are compared. 
On the ground, the best quantum accelerometers are operating at sensitivities of about
$5\cdot10^{-8}$~m/s$^2$/Hz$^{1/2}$~\cite{Karcher2018} for an interrogation time of about $100-200$ ms.  
A space quantum accelerometer is expected to reach sensitivities in the low $10^{-12}$ m/s$^2$/Hz$^{1/2}$ when stretching the interrogation times to $20$ s, similar to the very best electrostatic accelerometers, such as used for the GOCE~\cite{GOCE} mission, but in a wider measurement band extending down to lower frequencies
as illustrated in Fig.~\ref{fig:Carraz}.~\footnote{For the
capabilities of hybrid electrostatic-quantum interferometers, see~\cite{2013ApPhL.102n4107B, 2018ApPhB.124..181B, 2019AdSpR..63.3235A}.}
Absence of drifts is a consequence of the absolute character of quantum sensors, with stable scale factors determined by the wavelength of the laser beam-splitters and the duration of the measurement, and the possibility of evaluating accurately systematic effects. 
On the other hand, they are so far limited to single axis measurements, and have a much higher Size, Weight and Power (SWaP) budget. 
However, whilst the technology is currently less mature, it is being demonstrated in a number of national and international projects as outlined in Section~\ref{sec:Tech}.

\begin{table}[htb]
   \centering
{\small    \begin{tabular}{|c|c|c|}
\hline
    & Atomic & Electrostatic  \\
    & accelerometer & accelerometer  \\
    \hline \hline
Sensitivity    & $ 4 \times 10^{-8}$~m/s$^2$/Hz$^{1/2}$ on ground & $3    \times 10^{-12}$~m/s$^2$/Hz$^{1/2}$ \\
 & (projection for space at $10^{-12}$~m/s$^2$/Hz$^{1/2}$ & (demonstrated) \\
 & for interrogations of more than $20\,\mathrm{s}$) &     \\ 
 \hline
Measurement bandwidth    & $\leq 0.1$ Hz & [0.005-0.1] Hz \\
 \hline
     Scale factor  & Absolute & Calibration required \\
      \hline
     Stability & No drift & Drift  \\
     \hline
    Measurement & Single axis & Three axes \\    
capability     &  &  \\
    \hline
Proof mass motion & Residual velocities $\to$ Coriolis acceleration &  \\  
    \hline 
SWaP & High & Low \\
    \hline 
TRL & Intermediate & High \\
    \hline 
  \end{tabular}}
   \caption{\it Comparison of classical and quantum sensors.}
    \label{tab:models3}
\end{table}


\subsubsection{Potential gain in Earth observation by quantum gravimeters}

\begin{figure}[h!]
\centering
\includegraphics[width=\textwidth]{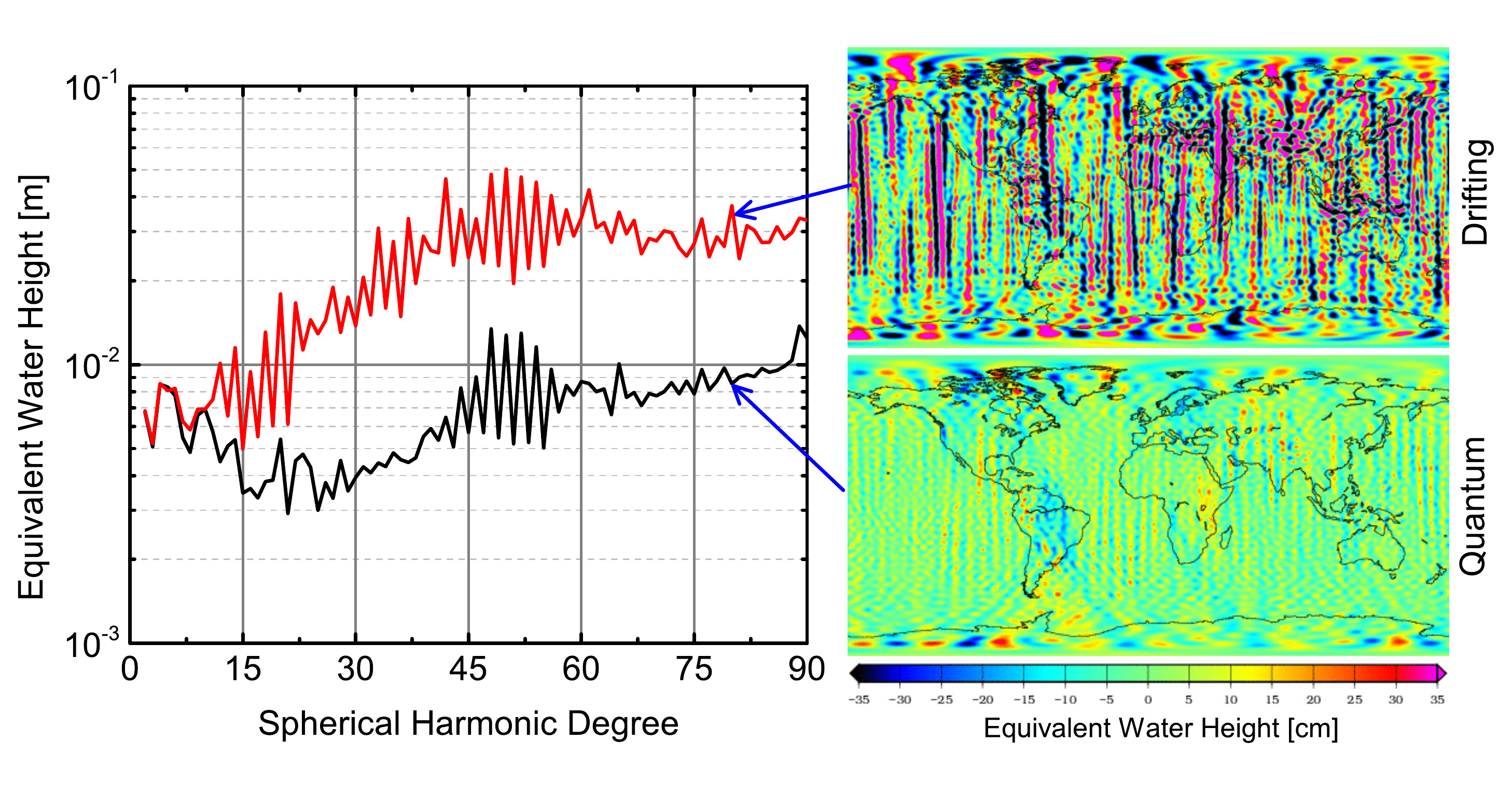}
\vspace{-0.5cm}
\caption{\it Spectra of gravity field recovery in equivalent water height obtainable with an atom interferometer and an electrostatic accelerometer, shown as black and red lines, respectively, in the left panel and colour-coded in the lower and upper maps in the right panels. 
Figure taken from ~\cite{Leveque2021}.}
\label{GRICE}
\end{figure}

The promise of atomic accelerometers for providing better long-term stability, i.e., smaller measurement noise at the lowest frequencies, below 10\,mHz, will enable the reconstruction of the Earth gravity field to be improved~\cite{muller2020using, trimeche2019concept, douch2018simulation} over many spherical harmonic degrees, as seen in Fig.~\ref{GRICE}~\cite{Leveque2021}. 
This shows the uncertainties in the gravity-field recovery as a function of spherical harmonic degree, evaluated in equivalent water height, considering quantum (black) or drifting (red) accelerometers in the left panel, and colour-coded in the right panels. 
The computations were carried out without any empirical periodic parameter adjustment in the gravity-field reconstruction.

The potential gains in Earth observation obtainable using quantum sensors are illustrated in Fig.~\ref{Gain}. 
The ellipses represent the required measurement resolutions for the indicated scientific objectives, with the spatial resolution on the horizontal axis and the temporal resolution on the vertical axis. 
Also shown are the sensitivity curves of the ``classical" CHAMP~\cite{CHAMP}, GRACE~\cite{GRACE} and GOCE~\cite{GOCE} missions and the prospective sensitivity of a possible quantum gravimetry mission employing atom interferometry.


\subsubsection{Potential gain in Earth observation by chronometric levelling}

As already mentioned in Sec.~\ref{Sec:ClocksSciOpp}, a combination of ground and space clocks together with high performance time and frequency dissemination capabilities in particular via satellites can benefit the stabilization and long-term validation of physical height networks. Though this approach is today less developed than quantum gravimeters, it is highly appealing~\cite{wu2019clock, wu2020towards} because clocks offer access to a new observable in Earth observation, namely to gravity potential differences in addition to the established determination of its derivatives. Time and frequency dissemination via satellite supported by space clocks can thus be a valuable ingredient~\cite{muller2018high} in the establishment of an improved height reference system as required by the United Nations initiative on the global geodetic reference frame for sustainable development~\cite{unr15}.   

\subsection{Quantum space gravimetry pathfinder mission\label{sec:QSG-pathfinder}}

\subsubsection{Concepts for a quantum gravimetry pathfinder mission}

For the European Union, deploying a quantum space gravimetry (QSG) pathfinder mission within this decade is a strategic priority to ensure 
technological non-dependence and leadership in this field and to pave the way towards an EU/ESA QSG mission within the next decade~\cite{Domps}.~\footnote{To this end, the European Commission has issued a call to enhance the TRLs of components for quantum gravimeters in space, e.g., components for cold atom interferometry (including Bose-Einstein Condensates)~\cite{EU}.}
In this regard, the pathfinder will represent a fundamental technological step towards the feasibility of such a mission.
As such, the pathfinder mission will be important in showing the fitness of cold atom interferometry for the purpose of gravity sensing, even if this first mission will not provide observations allowing for an improvement in the gravity field recovery.
We note that the challenges inherent to the launch and operation of a dedicated quantum pathfinder mission have at an early stage led ESA to consider the possibility of embarking the pathfinder mission on one of the MAGIC satellite pairs, to be launched around 2028-30. This mission is at the moment at the stage of a phase A study, but a pathfinder add-on is currently deemed unrealistic from a risk/payload point of view.  

\begin{figure}[h!]
\centering
\includegraphics[width=0.7\textwidth]{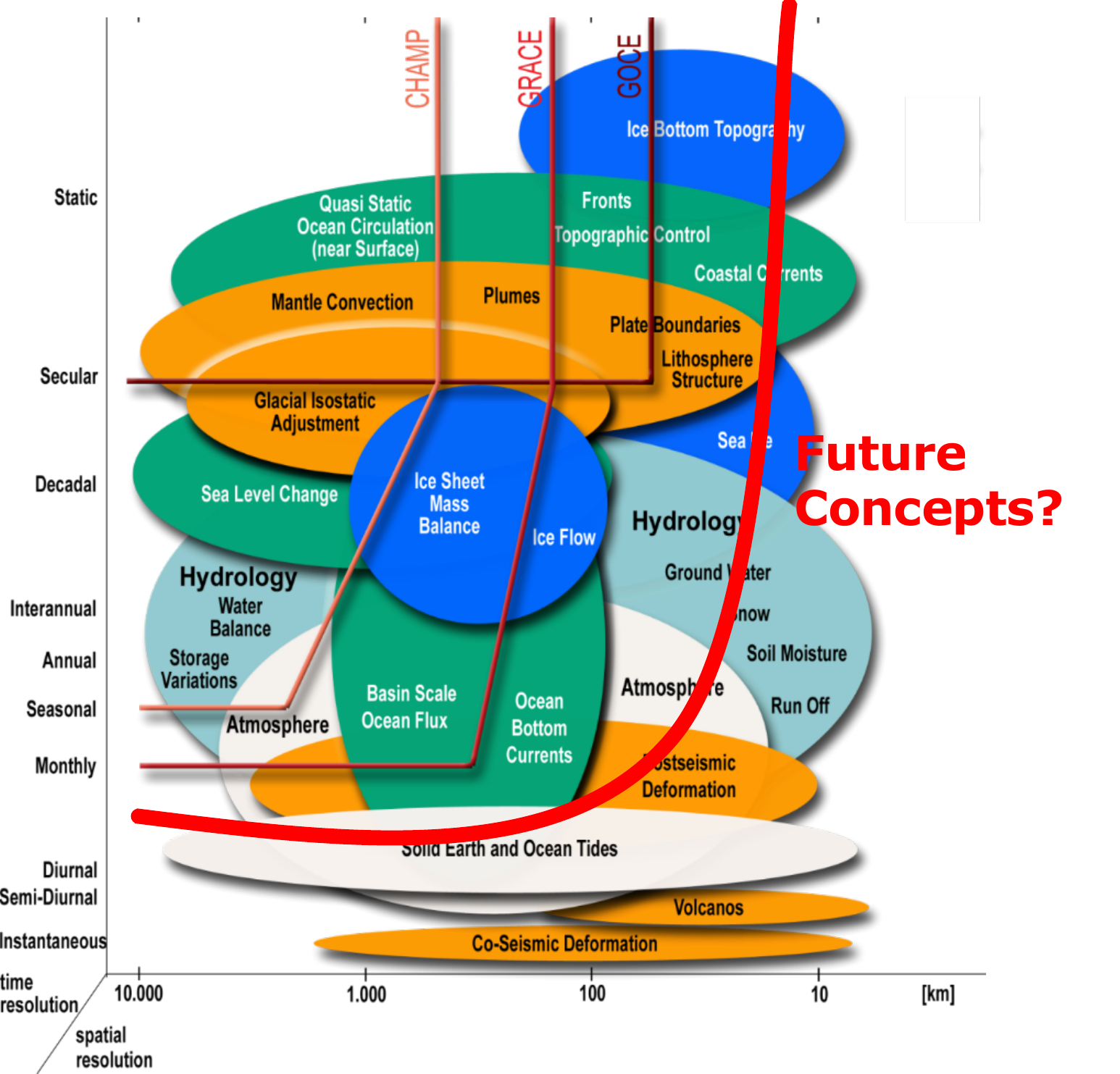}
\vspace{-0.5cm}
\caption{\it The spatial and temporal resolutions required for the indicated scientific objectives, compared with the sensitivity curves of the ``classical" CHAMP~\cite{CHAMP}, GRACE~\cite{GRACE} and GOCE~\cite{GOCE} missions and the prospective sensitivity of a possible quantum gravimetry mission. 
Figure adapted from~\cite{Visser2016}.}
\label{Gain}
\end{figure}

\subsubsection{Profile for a quantum gravimetry pathfinder mission}

The main goal of the pathfinder mission should be to demonstrate the maturity of the cold atom technology to operate in space. 
It should also go beyond the present-day performance of ground atom accelerometers (few $10^{-8}$~m/s$^2$/$\sqrt{\rm Hz}$) by one to two orders of magnitude thanks to the long interrogation times (several seconds) in microgravity. 
It will also help to demonstrate the technical maturity of key components of cold atom sensors in space, such as the long operation times or the rotation compensation (see Section \ref{sec:Tech} for further details). 
Furthermore, the pathfinder mission will have a strategic importance also for geodesists who are looking forward to analyse observations and obtain meaningful geodetic results from the next (fully-fledged) quantum gravimetry mission. 
The pathfinder will in any case provide interesting observations and results useful for the recovery of the gravity field, even though a clear improvement will be available to end-users in geodesy and geophysics only from the following quantum gravimetry mission. 
In fact, this pathfinder mission will allow preparing for missions with larger interrogation times (more than 10s) at resolutions of $\sim 10^{-11}-10^{-12}$~m/s$^2$/$\sqrt{\rm Hz}$ suitable for the wide user community.

The payload is expected to have a mass of a few hundred kg and would require a few hundred W to operate. 
In order for the quantum sensor to perform optimally, the platform needs to be designed to meet clear constraints (centre of gravity, rotation compensation, etc.). 
To operate with optimal performance, the orbit, altitude, flight modes and position  within the platform should be chosen to ensure a successful operation of the quantum sensor.\\

\subsection{Recommendation: Road-map to quantum Earth Observation in space} 

As has been outlined in Section~\ref{sec:QSG-pathfinder}, several possible scenarios can be envisaged for future Earth Observation missions based on quantum technology. 
Here we summarize the main possibilities, their  drawbacks and strengths::

$\bullet$ In parallel with the conventional planned MAGIC gravimetry mission~\cite{MAGIC}, it is necessary to update the quantum instrument specifications and requirements. 
However, embarking the quantum sensor as a passenger on an SST geodesy mission poses tremendous technical challenges and carries significant technological and programmatic risks for both aspects of the mission (classical and quantum sensors).

$\bullet$ Hence the prevailing outcome of the discussions among scientific experts during the workshop is a recommendation to launch a pathfinder mission within this decade with a performance of up to $10^{-10}$~m/s$^2$/$\sqrt{\rm Hz}$ on a dedicated platform. 
Such a mission would balance the need to have a test of the quantum technology in space and the level of expectation from the quantum gravimetry pathfinder mission. 
It would also be a clear milestone for other communities, such as the fundamental physics one.

$\bullet$ The success of MAGIC and the Pathfinder mission will then enable the implementation of a full-fledged quantum space gravimetry mission to be launched to follow MAGIC. 
The definition of the mission scenario and the instrument baseline will be based on lessons learned from MAGIC and the Pathfinder mission.






\section{Atomic Sensors for Fundamental Science Review}
\label{sec:PH}
\subsection{Scientific opportunities}
\label{sec:fun_sc_opp}

The promise of Cold Atom technologies for making precise experimental probes of topics in fundamental science such as general relativity, cosmology, quantum mechanics and the search for new physics beyond the Standard Model has been recognised in many terrestrial and space projects.
To cite just a few examples: in the US the MAGIS 100 m atom interferometer is under construction at Fermilab~\cite{Abe2021} and NASA has operated the Cold Atom Lab (CAL) Bose-Einstein condensate (BEC) experiment successfully for several years on the ISS~\cite{CALISS, Aveline_2020}; in Europe the ELGAR project~\cite{ELGAR} has been proposed; initial funding has been provided for a suite of experiments applying Quantum Technology for fundamental physics in the UK, including the terrestrial AION atom interferometer~\cite{Badurina2019}; in France the MIGA atom interferometer is under construction~\cite{Canuel:2017rrp}; in Germany there is the VLBAI programme~\cite{VLBAI} and a series of BEC experiments in microgravity using MAIUS sounding rockets~\cite{MAIUS, Lachmann_2021}; the STE-QUEST experiment has been proposed~\cite{Schubert2013, STE-QUEST}, following the success of the MICROSCOPE experiment~\cite{Touboul2017, MICROSCOPE} that made a pioneering test of the Equivalence Principle in space; and in China the terrestrial ZAIGA atom interferometer~\cite{ZAIGA} is under construction and quantum correlations have been verified by the Micius satellite experiment~\cite{QUESS} over distances exceeding a thousand km.

The deployment of cold atom technologies in space offers unique 
research opportunities
in the fields of fundamental physics, cosmology and astrophysics, as
represented in several White Papers submitted to the ESA Voyage 2050
call for mission concepts~\cite{WP1EA,WP2EA,WP3EA,WP4EA,WP5EA,WP6EA,WP7EA}.
We focus in the following on two of these mission concepts.

\begin{figure}[h!]
\hspace{0.5cm}
\centering
\includegraphics[width=0.9\textwidth]{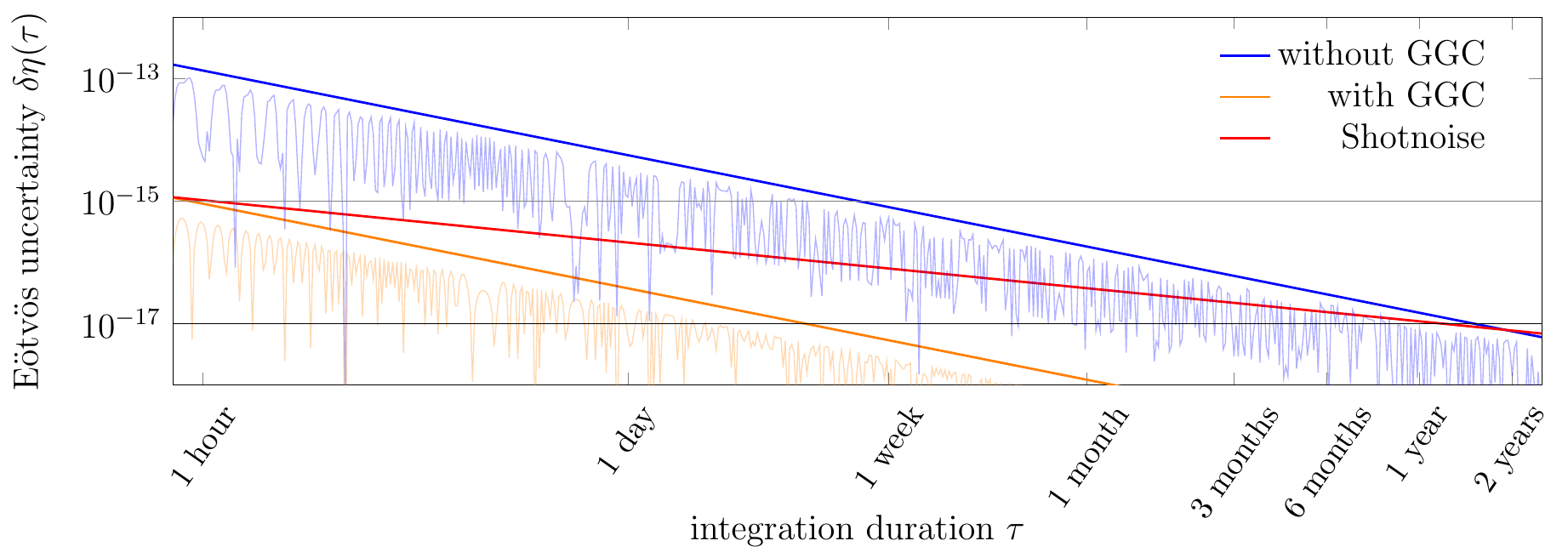}
\caption{\it Averaging of systematic uncertainties due to gravity gradients in a UFF test with Rb and K quantum sensors. 
Gravity Gradient Compensation (GGC) significantly reduces the systematic contributions, such that the residual differential acceleration may be attenuated to an unprecedented degree through signal demodulation (orange curve). This not only allows for requirements on the source preparation that are greatly reduced compared to other mission proposals as STE-QUEST~\cite{STE-QUEST}, but also paves the way for more ambitious mission scenarios targeting $\delta \eta \leq 10^{-17}$ in shot-noise limited operation (red curve). In contrast, even though the systematics are integrated down thanks to demodulation, the measurement would be limited by systematics without GGC  (blue curve). The figure is taken from~\cite{Loriani}.}
\label{fig:GGC_int}
\end{figure}

One is based on the previous STE-QUEST proposal~\cite{STE-QUEST}, and proposes a double atom interferometer with rubidium and potassium ``test masses" in quantum superposition to test the universality of free fall (UFF).
It assumes a single satellite in a 700~km circular orbit, and applies recent developments on gravity gradient control by offsetting laser frequencies, which enables the atom positioning requirements to be relaxed by a factor $>100$~\cite{WP2EA, Loriani}.
This offers the possibility of probing the UFF, i.e., the equivalence principle, with an unparalleled precision ${\cal O}(10^{-17})$ after 18 months in orbit~\cite{Loriani}, see Fig.~\ref{fig:GGC_int}.

\begin{figure}[h!]
\vspace{0.5cm}
\centering
\includegraphics[width=0.95\textwidth]{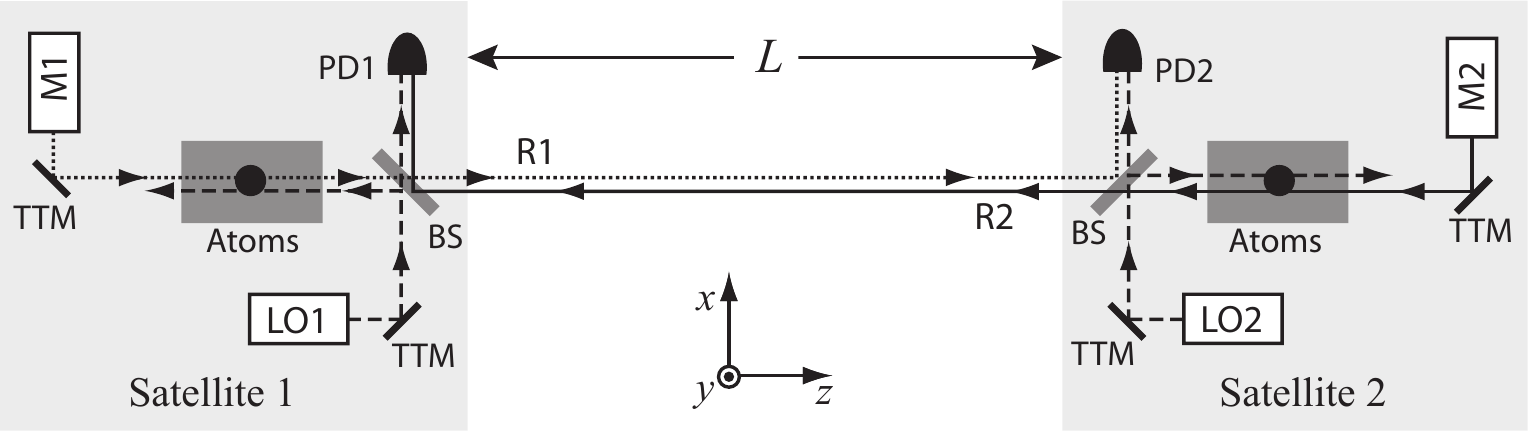}
\caption{\it Possible scheme for a satellite atom interferometer experiment~\cite{AEDGE}.  It has two master laser beams M1 and M2 (dotted and solid lines) and two reference beams  (R1 and R2) there are two local oscillator lasers LO1 and LO2 (dashed  lines) phase-locked with R2 and R1, respectively. Photodetectors (PD1 and PD2) measure the heterodyne beatnote between the reference beams R2 and R1 and the corresponding local lasers LO1 and LO2, respectively, providing feedback for the laser link. Non-polarizing beam splitters are denoted by BS, and tip-tilt mirrors used for controlling the directions of the laser beams are denoted by TTM. Small offsets between overlapping laser beams have been introduced for clarity. The figure is taken from~\cite{Graham:2017pmn}.}
\label{sat-concept}
\end{figure}

The other is the AEDGE concept for a satellite atom interferometer using
strontium~\cite{AEDGE}, illustrated in Fig.~\ref{sat-concept}, which is similar in principle to the previous SAGE concept for a satellite atom interferometer experiment~\cite{Tino:2019tkb}. In the baseline AEDGE configuration~\cite{AEDGE} the atom clouds were assumed to be located inside the spacecraft and have sizes $\sim 1$~m. In addition to this version, here we also consider in the following the possibility that the clouds are
outside the spacecraft and have longitudinal sizes $\sim 100$~m that enable a much higher number of momentum transfers and increase sensitivity, a concept called AEDGE+.~\footnote{See also the Workshop talk by Nan Yu~\cite{NanYu}.}
AEDGE can search for waves of ultralight dark matter (ULDM)
particles with masses between ${\cal O}(10^{-21})$ and ${\cal O}(10^{-15})$~eV and measure
gravitational waves in a frequency range between ${\cal O}(1)$ and ${\cal O}(10^{-2})$~Hz~\cite{AEDGE},
intermediate between the ranges of frequency where the sensitivities of the terrestrial
laser interferometers LIGO~\cite{LIGO}, Virgo~\cite{Virgo}, KAGRA~\cite{KAGRA}, ET~\cite{ET} and CE~\cite{CE} are maximal (${\cal O}(1)$ to ${\cal O}(10^{2})$~Hz)
and the optimal frequencies 
of the space-borne laser interferometers LISA~\cite{LISA}, TianQin~\cite{TaiQin} and Taiji~\cite{Taiji}  (${\cal O}(10^{-2})$~Hz).

In the following we present 
some representative examples of the capabilities of these cold atom concepts for probing
fundamental physics, cosmology and astrophysics.

\subsubsection{Tests of the universality of free fall}

The Einstein Equivalence Principle (EEP) is the foundation of all theories of gravitation that describe it as a geometrical phenomenon, i.e., a curvature of space-time. Indeed, the universal coupling to all mass-energy that is implicit in the EEP is necessary for all metric theories of gravitation, including general relativity among many others. As such, the EEP is one of the most foundational building blocks of modern physics. Nonetheless, many modified theories of gravity that go beyond the Standard Model and general relativity and/or account for dark matter/energy entail some violation of the EEP. Examples include scalar-tensor theories of gravity, especially those subject to screening mechanisms
such as chameleon and symmetron models, in which additional scalar degrees of freedom can give rise to long-range fifth forces that cause deviations from general relativity that depend on the local environments and the configurations of experimental tests. Atom interferometry experiments already provide stringent constraints on certain screened fifth-force models (see, e.g.,~\cite{Elder:2016yxm, Burrage:2016bwy, Jaffe:2016fsh, Sabulsky:2018jma}), and further searches for fifth forces have been identified as a priority (see, e.g.,~\cite{WP2EA} 
and Section~\ref{sec:DM} for more examples).

The best known aspect of EEP is the universality of free fall (UFF), i.e., the weak equivalence principle (WEP).~\footnote{The EEP corresponds to the combination of the UFF/WEP with local position invariance and Lorentz invariance. See~\cite{2019PhRvL.123w1102B} for a test of Lorentz invariance using MICROSCOPE data.} The history of experimental tests of UFF/WEP goes back as far as the Renaissance, and probably beyond. A simple phenomenological figure of merit for all UFF/WEP tests is the E\"otv\"os ratio $\eta_{AB}$ for two test objects
$A$ and $B$ and a specified source mass of the gravitational field:
\begin{equation}\label{etaAB}
\eta_{AB} = 2\,\frac{a_A - a_B}{a_A + a_B}\,,
\end{equation}
where $a_i$ ($i=A,B$) is the gravitational acceleration of the object $i$ with respect
to the source mass. Note that for a given experiment the data can be
interpreted with respect to different source masses (see, {e.g.},
Ref.~\cite{Schlamminger:2007ht}) with correspondingly different results for
$\eta_{AB}$.

Whilst $\eta_{AB}$ is a useful tool for comparing different experiments, it cannot account for the diversity of possible underlying theories, e.g., different types of couplings depending on the source and test objects, or couplings to space-time varying background fields other than local gravity, {e.g.}, \cite{Tasson2011,Hees:2018fpg}. Thus, not only best performance in terms of the E\"otv\"os ratio is required, but also a large diversity of test objects and source masses. 

Table \ref{tab:StateArt} presents the state of the art in UFF/WEP tests, separated into different classes as a function of the type of test-masses employed. In particular, we distinguish between tests using macroscopic test masses and atom-interferometry (AI) tests~\footnote{See also~\cite{2015PhRvA..92b3626B}.} that use matter waves in a quantum superposition, possibly condensed to quantum degenerate states (Bose-Einstein Condensates) with coherence lengths $\geq \mu$m. The ``game-changing'' results of the MICROSCOPE mission demonstrate the potential of going into a quiet and well-controlled space environment, with potentially ``infinite'' free fall times, as exemplified by the
prospective sensitivity of the STE-QUEST-like space-borne cold atom interferometer mission displayed in
Table \ref{tab:StateArt}. 
\begin{table*}[t]
\hspace{-5mm}
\begin{center}
\hspace{-5mm}
\begin{tabular}{ccccc}
\hline 
{\bf Class} & {\bf Elements} & {\bf $\eta$} & {\bf Year [ref]} & {\bf Comments} \\
\hline\hline
\multirow{4}{*}{Classical} & Be - Ti & $2\times10^{-13}$ & 2008 \cite{Schlamminger:2007ht} & Torsion balance \\
& Pt - Ti & $1\times10^{-14}$ & 2017 \cite{Touboul2017} & MICROSCOPE first results \\
& Pt - Ti & ($10^{-15}$) & 2019+ & MICROSCOPE full data \\
\hline
\multirow{3}{*}{Hybrid} & $^{133}$Cs - CC & $7\times10^{-9}$ & 2001 \cite{Peters2001} & Atom Interferometry\\ 
& $^{87}$Rb - CC & $7\times10^{-9}$ & 2010 \cite{Merlet2010} & and macroscopic corner cube \\
\hline
\multirow{5}{*}{Quantum} & $^{39}$K - $^{87}$Rb & $5\times10^{-7}$ & 2014 \cite{Schlippert2014} & different elements \\
& $^{87}$Sr - $^{88}$Sr & $2\times10^{-7}$ & 2014 \cite{Tarallo2014} & same element, fermion vs. boson  \\
& $^{85}$Rb - $^{87}$Rb & $3\times10^{-8}$ & 2015 \cite{Zhou2015} & same element, different isotopes  \\
& $^{85}$Rb - $^{87}$Rb & $3.8\times 10^{-12}$ & 2020 \cite{Asenbaum2020} & \multirow{2}{*}{$\geq$ 10 m towers} \\
& $^{85}$Rb - $^{87}$Rb & ($10^{-13}$) & 2020+ \cite{Overstreet2018} & \\
& $^{170}$Yb - $^{87}$Rb & ($10^{-13}$) & 2020+ \cite{Hartwig2015} & \\
& $^{41}$K - $^{87}$Rb & $10^{-17}$ & 2035+ & STE-QUEST-like mission \\
\hline
Antimatter & $\overline{\rm H}$ - H & ($10^{-2}$) & 2020+ \cite{Perez2012} & under construction at CERN \\
\hline
\end{tabular}
\caption{\it Compilation of UFF/WEP tests. Numbers in brackets are results expected in the near future, and we also show the performance of an STE-QUEST-like atom-interferometry mission in the context of this road-map. Further information about this road-map milestone are provided in Section~\ref{sec:road-map}. In addition, lunar laser ranging probes the UFF at a level $\sim 2 \times 10^{-14}$ for the Earth and Moon~\cite{universe7020034}.}\label{tab:StateArt}
\end{center}
\end{table*}

\subsubsection{Ultralight dark matter detection} \label{sec:DM}

\begin{figure}[h!]
\centering
\vspace{-0.2cm}
\includegraphics[width=0.475\textwidth]{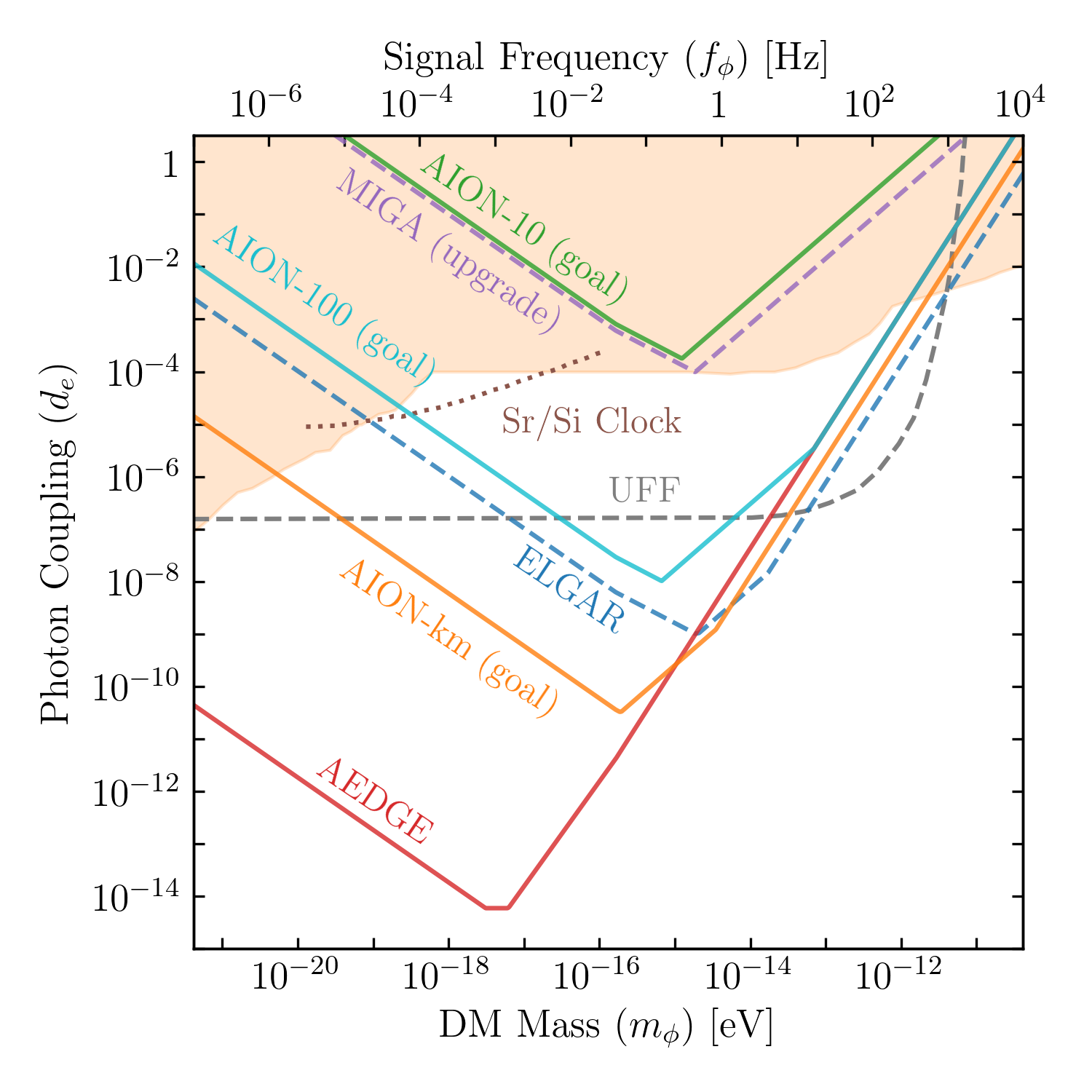}
\includegraphics[width=0.475\textwidth]{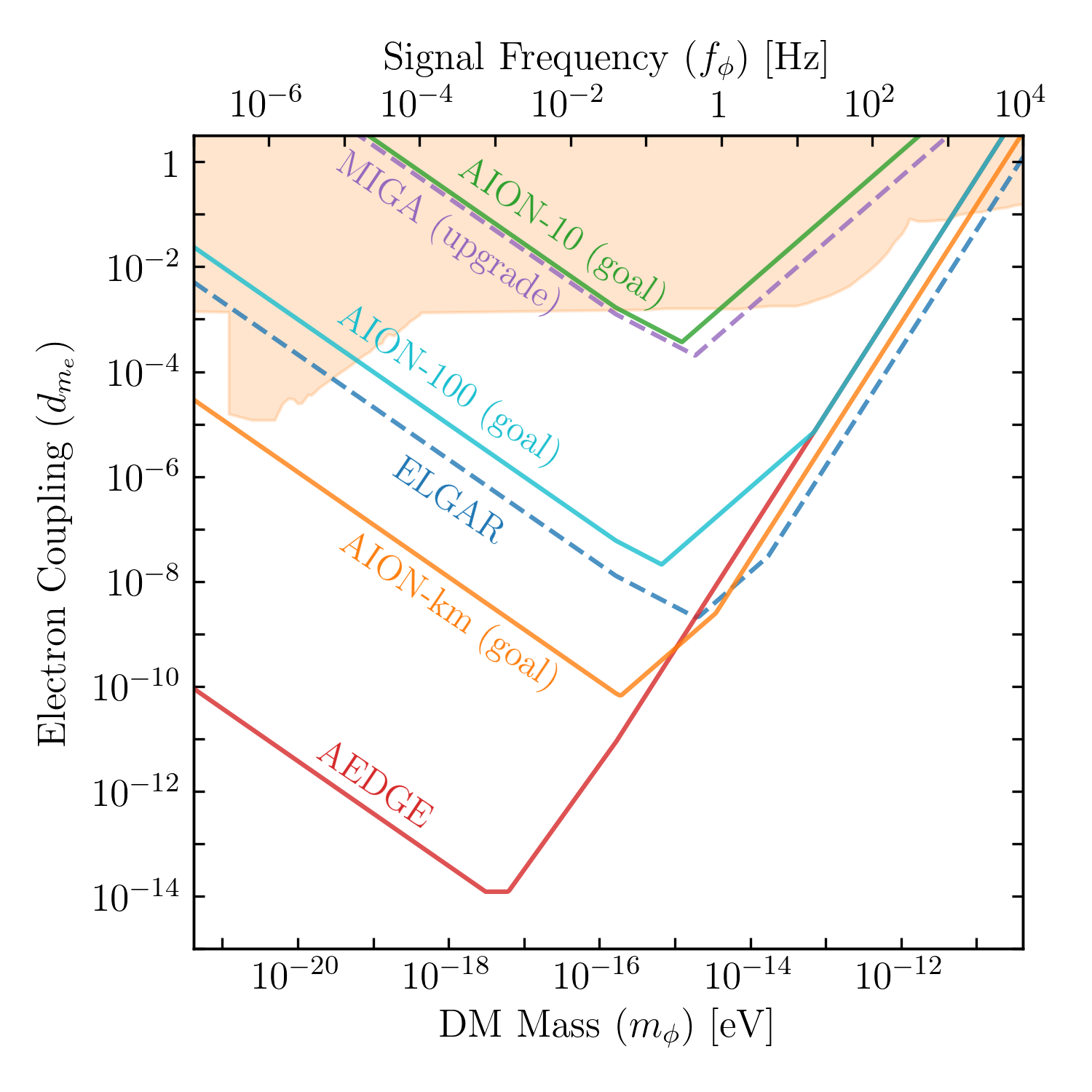}
\vspace{-0.5cm}
\caption{\it Sensitivities to the possible couplings of scalar ULDM to
photons (left panel) and to electrons (right panel) of the terrestrial
AION~\cite{Badurina2019}, MIGA~\cite{Canuel:2017rrp} and ELGAR~\cite{ELGAR} experiments, and of the AEDGE space-borne concept~\cite{AEDGE}. Also
shown is a combination of the current constraints from MICROSCOPE~\cite{MICROSCOPE} and
terrestrial torsion balance and atomic clock experiments and (in the
left panel) the sensitivity of a prospective measurement of the universality 
of free fall (UFF) with a precision of $10^{-17}$~\cite{WP2EA}.}
\label{DM}
\end{figure}

Figure~\ref{DM} shows examples of the present and prospective sensitivities of searches for linear couplings of ultralight scalar dark matter to photons (left panel)
and electrons (right panel).~\footnote{One may also consider quadratic couplings, which would be screened in terrestrial experiments if the quadratic couplings are positive, reducing the experimental sensitivity~\cite{Hees:2018fpg} (see also~\cite{Stadnik:2020bfk}). A space-based experiment such as AEDGE is less affected by this screening, and hence maintains sensitivity at larger masses, as discussed in~\cite{Badurina2019, AEDGE}. See~\cite{geraci2016sensitivity, Figueroa_2021} for other suggestions how atom interferometers, gravimeters and accelerometers could be used to search for alternative dark matter candidates.}
Such dark matter, because of its non-universal coupling to the fields of the standard model, is also one example of a violation of the Einstein equivalence principle and the universality of free fall, UFF. The shaded regions are excluded by current experiments including MICROSCOPE~\cite{MICROSCOPE} and atomic clocks~\cite{Kennedy2020}.~\footnote{See also the results from the WReSL experiment on
the couplings of ultralight scalar dark matter to photons and electrons oscillating at higher frequencies~\cite{Antypas_2021},
and recent experimental bounds on the couplings to the up, down, and strange quarks and to gluons obtained using molecular spectroscopy~\cite{oswald2021search}.} 
We also show the prospective sensitivities of 
rubidium-based terrestrial interferometers (MIGA~\cite{Canuel:2017rrp} and ELGAR~\cite{ELGAR}) and strontium-based terrestrial and space-borne atom interferometers
(AION~\cite{Badurina2019} and AEDGE~\cite{AEDGE}, respectively). MIGA, ELGAR and the 100 m and km
versions of AION  offer significantly greater sensitivity 
than current experiments (torsion balances, atomic clocks and MICROSCOPE~\cite{MICROSCOPE}) to couplings of scalar ULDM with masses $\lesssim 10^{-12}$~eV to both
photons and electrons. A sensitivity of $\sim 10^{-7}$
to the ULDM-photon coupling in this mass range 
could be obtained with a prospective cold-atom interferometer UFF probe
with a precision of $10^{-17}$ as discussed in the context of this road-map \cite{WP2EA}, using Rb and K quantum probes (see also Table \ref{tab:StateArt}). As seen in Fig.~\ref{DM},
the AEDGE strontium-based space-borne
atom interferometer concept~\cite{AEDGE} could offer the greatest sensitivity to
both the ULDM-photon and -electron couplings for masses between
$10^{-15}$ and $10^{-21}$~eV, with a maximum sensitivity 
$\sim 10^{-14}$ for masses between $10^{-17}$ and $10^{-18}$~eV.

\subsubsection{Probes of general relativity}

The measurements of gravitational waves using an atom interferometer in space offer unique prospects for probing modifications of general relativity. 
For example, AEDGE measurements of the inspiral of merging black holes with a combined mass of $10^4$ solar masses at a redshift  $z \sim 1$ would be sensitive to the possible appearance of a  graviton mass $\lesssim 10^{-26}$~eV, over three orders of magnitude below the current upper limit from LIGO and Virgo~\cite{EV}.
These measurements could also be used to search for possible violations of Lorentz invariance in the propagation of gravitational waves, with a sensitivity complementary to the searches by LIGO/Virgo and for gravitational {\v C}erenkov radiation~\cite{EV}.

More classical GR tests using modern optical clocks are measurements of the gravitational Shapiro delay down to $10^{-8}$~\cite{Ashby2010} or tests of the gravitational red-shift at $10^{-9}$ as in the FOCOS mission~\cite{focos}, proposed recently in the context of the NASA decadal survey. In both cases, these would provide 3-4 orders of magnitude improvements on best current knowledge. 

There is also a proposal to probe models of dark energy by deploying a smart constellation of four satellites in an elliptic orbit around the Sun and making orientation-independent measurements of the differential accelerations  between each pair of satellites, using as  test masses atomic clouds far away from the spacecraft in open-space vacuum (see the Workshop presentation by Nan Yu~\cite{NanYu}).
It has also been suggested to deploy an atomic clock 
at a distance ${\cal O}(150)$~AU to probe the low-acceleration frontier of
gravity and the local distribution of dark matter~\cite{WP5EA}. 
Another
suggestion is to detect the gravito-magnetic field of the galactic dark halo by locating atomic clocks at Sun-Earth Lagrange points and measuring the time-of-flight asymmetries between electromagnetic  signals travelling in opposite directions, which would be generated partly by the angular momentum of the Sun and partly by the angular  momentum of the dark halo~\cite{WP4EA}.

\subsubsection{Quantum mechanics}

It has
been proposed to test quantum correlations 
over astronomical distances~\cite{WP3EA,WP6EA}, e.g., between the Earth and the Moon
or Mars, or between LISA spacecraft~\footnote{See also~\cite{x4} for a description of the proposed Deep Space Quantum Link mission, whose goals include long-range teleportation, 
tests of gravitational coupling to quantum states, and advanced tests of quantum nonlocality.}. The Micius measurements~\cite{QUESS}
already demonstrate that quantum correlations extend over 1200~km and
that the apparent effective correlation speed exceeds $10^7$~c, and Earth-Moon
experiments could improve these sensitivities by factors 
$\sim 2 \times 10^4$~\cite{WP6EA}.

It has also been proposed to test 
wavefunction collapse and models predicting the violation of the quantum
superposition principle~\cite{Bassi2003,Bassi2013} by monitoring the 
expansion of a cloud of cold atoms~\cite{BILARDELLO2016764}. 
Current results already impose relevant constraints on the correlation length and rate parameters of the Continuous Spontaneous Localisation (CSL) model: 
see~\cite{BILARDELLO2016764} for a detailed discussion.

\subsubsection{Cosmology}

\begin{figure}[h!]
\centering
\vspace{-0.2cm}
\includegraphics[width=0.475\textwidth]{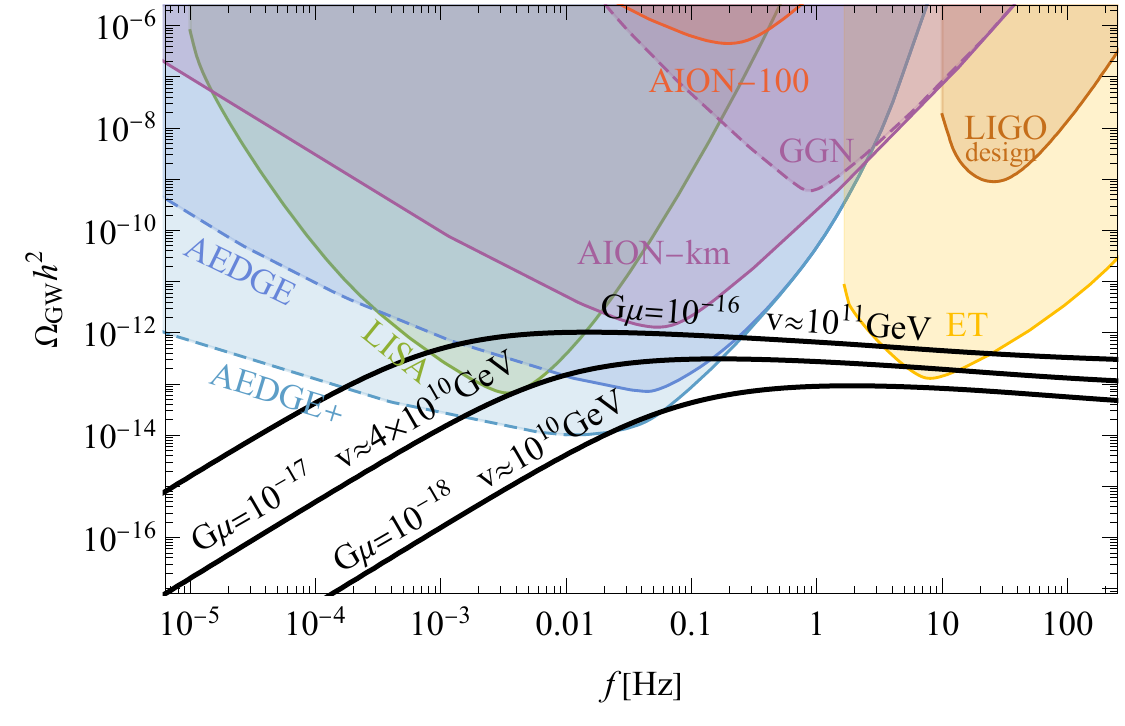}
\includegraphics[width=0.475\textwidth]{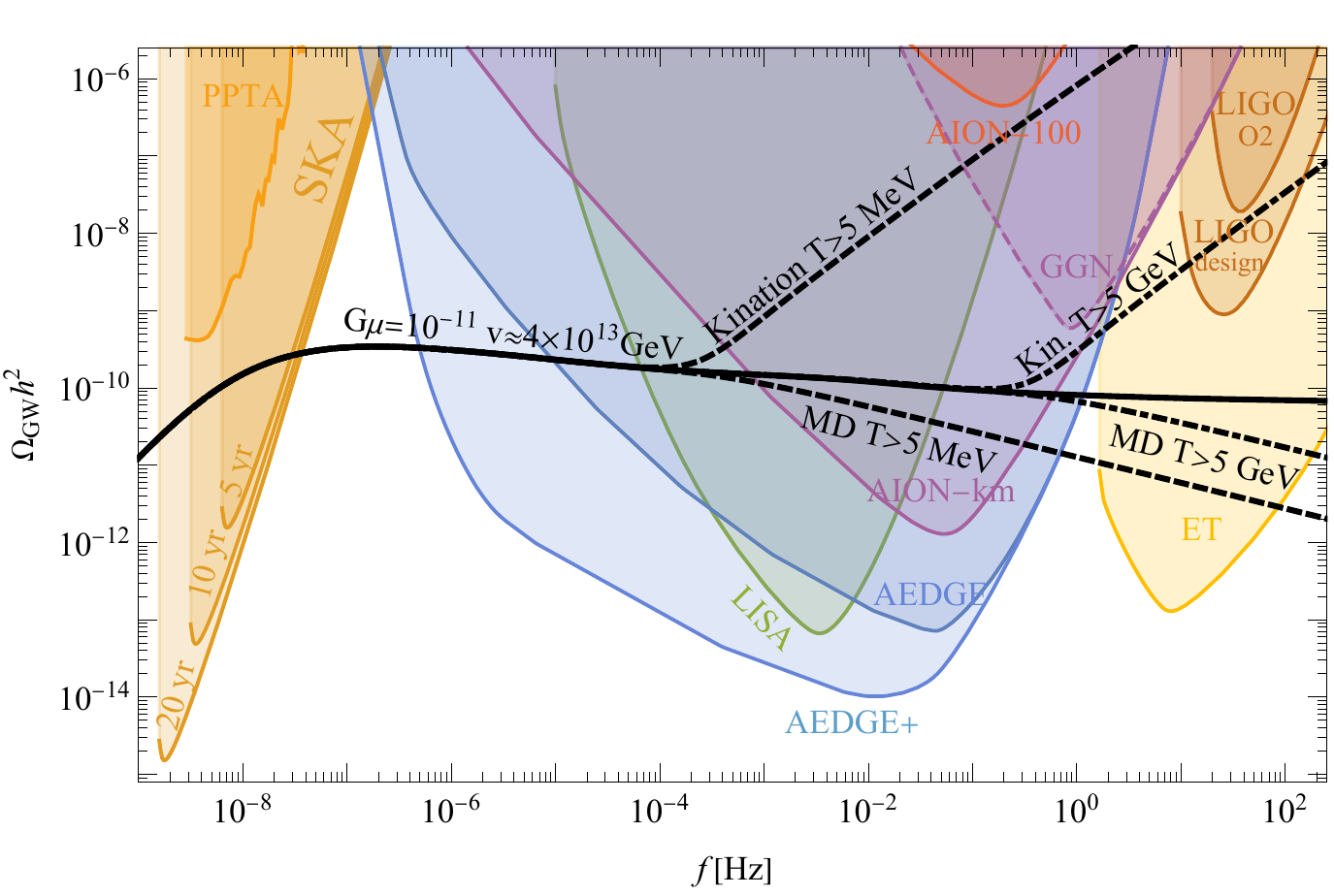}
\vspace{-0.5cm}
\caption{\it Left panel: Sensitivities to the GWs from a
network of cosmic strings with tension $G \mu$ of AION-100 and -km, 
AEDGE~\cite{AEDGE} and AEDGE+ (a version of AEDGE that would use an $\sim 100$m atomic cloud outside the spacecraft~\cite{RS}), LIGO, ET and LISA.
Right panel: Possible effects on the spectrum of GWs from cosmic strings
with tension $G\mu = 10^{-11}$ of an epoch of matter domination (MD) or
of kination~\cite{RS}.}
\label{string}
\end{figure}

AEDGE measurements also offer new opportunities in cosmology, such as unparalleled sensitivity to possible emissions from collapsing loops of cosmic strings in a network with tension $G \mu \gtrsim {\cal O}(10^{-18})$, more than 3 orders of magnitude below the current limit from the third Advanced LIGO–Virgo observing run~\cite{LIGOScientific:2021nrg} and an order of magnitude beyond the reach of LISA, as seen in the left panel of Fig.~\ref{string}~\cite{AEDGE}.
Such measurements could also be sensitive to effects in the early Universe that cause it to deviate from the conventional expectation of adiabatic expansion, such as a period of matter dominance or kination, as seen in the right panel of Fig.~\ref{string}~\cite{RS}.~\footnote{The string tension
$G \mu = 10^{-11}$ used in this plot is the maximum consistent~\cite{Ellis:2020ena,Blasi:2020mfx}  with current data from pulsar timing arrays~\cite{NG, PPTA, EPTA, IPTA},
which report strong evidence for a spectrally-similar low-frequency stochastic process.} 
There are also some models that yield primordial gravitational waves that could be detected by AEDGE, see, e.g., Fig.~12 in~\cite{Campeti}.

\begin{figure}[h!]
\centering
\vspace{-0.2cm}
\includegraphics[width=0.38\textwidth]{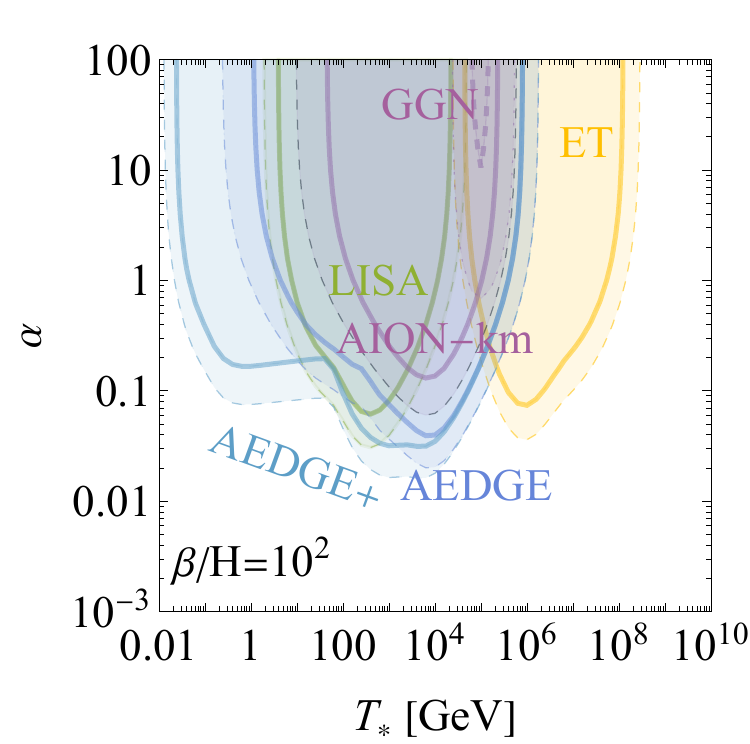}
\includegraphics[width=0.475\textwidth]{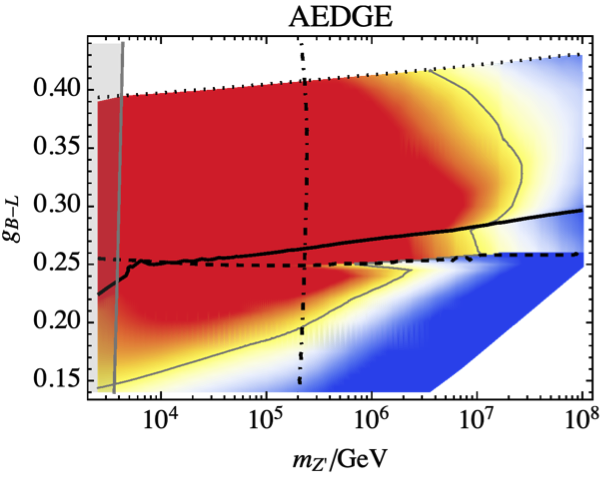}
\vspace{-0.5cm}
\caption{\it 
Left panel: Sensitivities of AION-100 and -km, AEDGE, AEDGE+ and LISA in the $(T_*, \alpha)$ plane
to GWs from generic first-order transitions with a transition rate $\beta/H = 100$~\cite{FOPhT}. Right panel: The expected signal-to-noise ratio (SNR)
with AEDGE for observing the stochastic GW background from a
U(1)$_{B-L}$-breaking phase transition. In the red regions 
SNR $> 1000$ and the solid gray contour corresponds to SNR $= 10$~\cite{FOPhT}.}
\label{PT}
\end{figure}

Another cosmological opportunity is the search for a stochastic background of gravitational waves generated by a first-order phase transition in the early Universe, either in the electroweak sector of the Standard Model or in some extension that includes a new interaction generated by a massive boson beyond the reach of present and proposed collider experiments.
The left panel of Fig.~\ref{PT} compares the sensitivities in the $(T_*, \alpha)$ plane (where $T_*$ denotes the critical temperature and $\alpha$ the strength of the transition) of the indicated experiments to GWs from a generic phase transition with a transition rate $\beta/H = 10^2$, and the right panel shows the signal-to-noise ratio (SNR) in the $(m_{Z'}, g_{B-L})$ plane expected from AEDGE measurements of the stochastic GW background from a U(1)$_{B-L}$-breaking phase transition~\cite{FOPhT}.
In the red regions SNR $> 1000$ and the solid gray contour corresponds to SNR $= 10$~\cite{FOPhT}.


\subsubsection{Astrophysics}

\begin{figure}[h!]
\centering
\vspace{-0.2cm}
\includegraphics[width=0.6\textwidth]{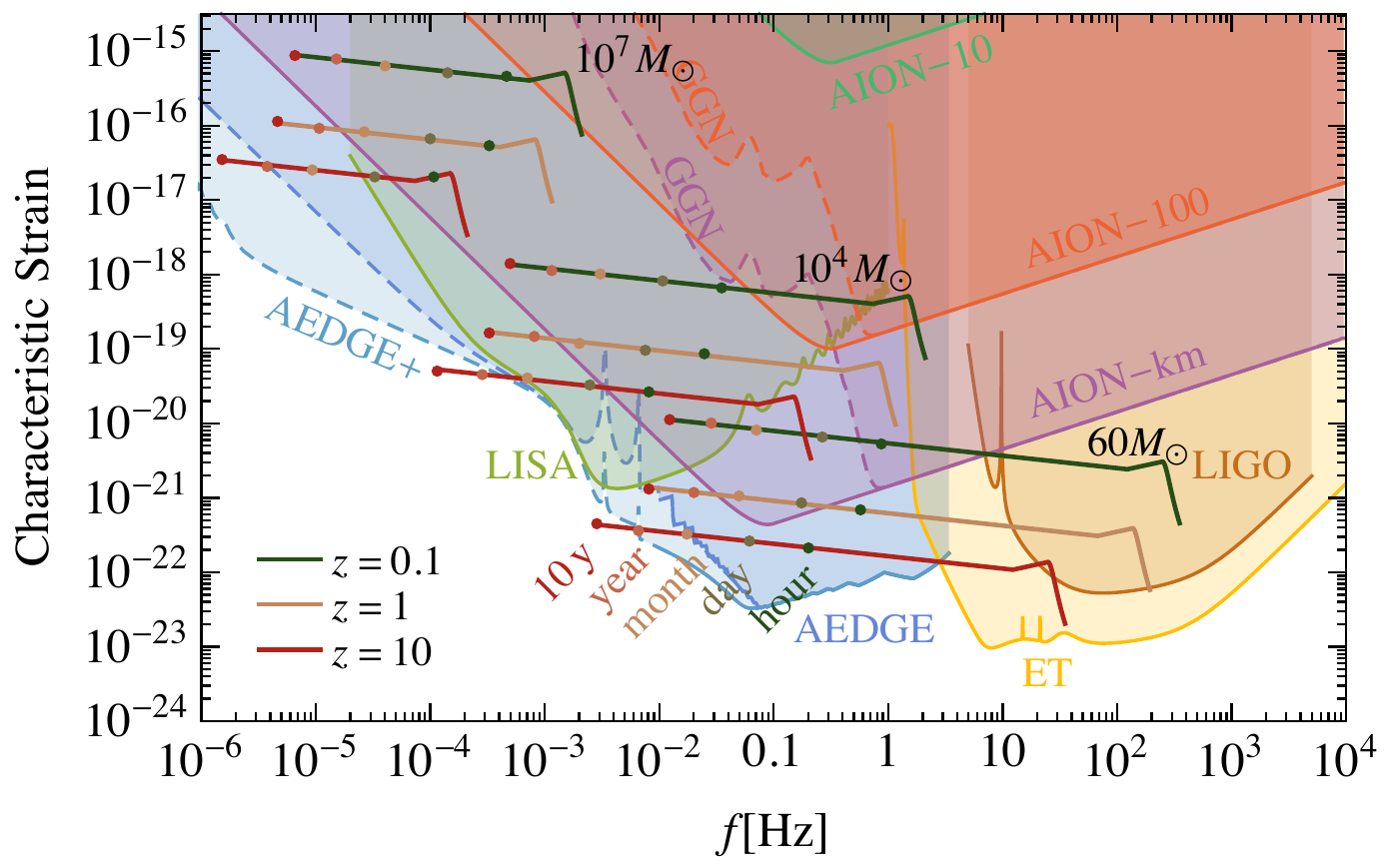}
\vspace{-0.5cm}
\caption{\it Strain sensitivities of AION-10, -100 and -km, AEDGE and AEDGE+, compared
with those of LIGO, LISA and ET and the signals expected from mergers of equal-mass
binaries whose masses are $60, 10^4$ and $10^7$ solar masses.  The assumed redshifts are
$z = 0.1, 1$ and $10$, as indicated. Also shown are the remaining times during inspiral before the
final mergers~\cite{AEDGE,RS}.}
\label{IMBH}
\end{figure}

\begin{figure}
\centering 
\includegraphics[width=0.475\textwidth]{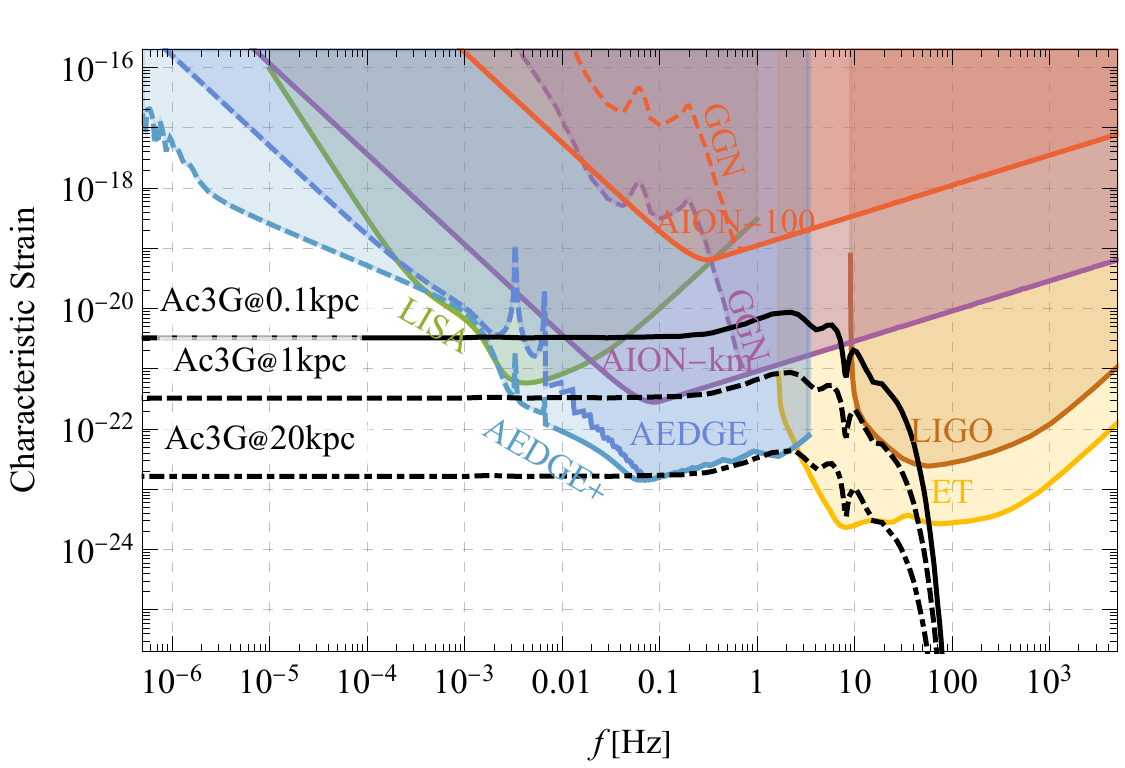}
\includegraphics[width=0.475\textwidth]{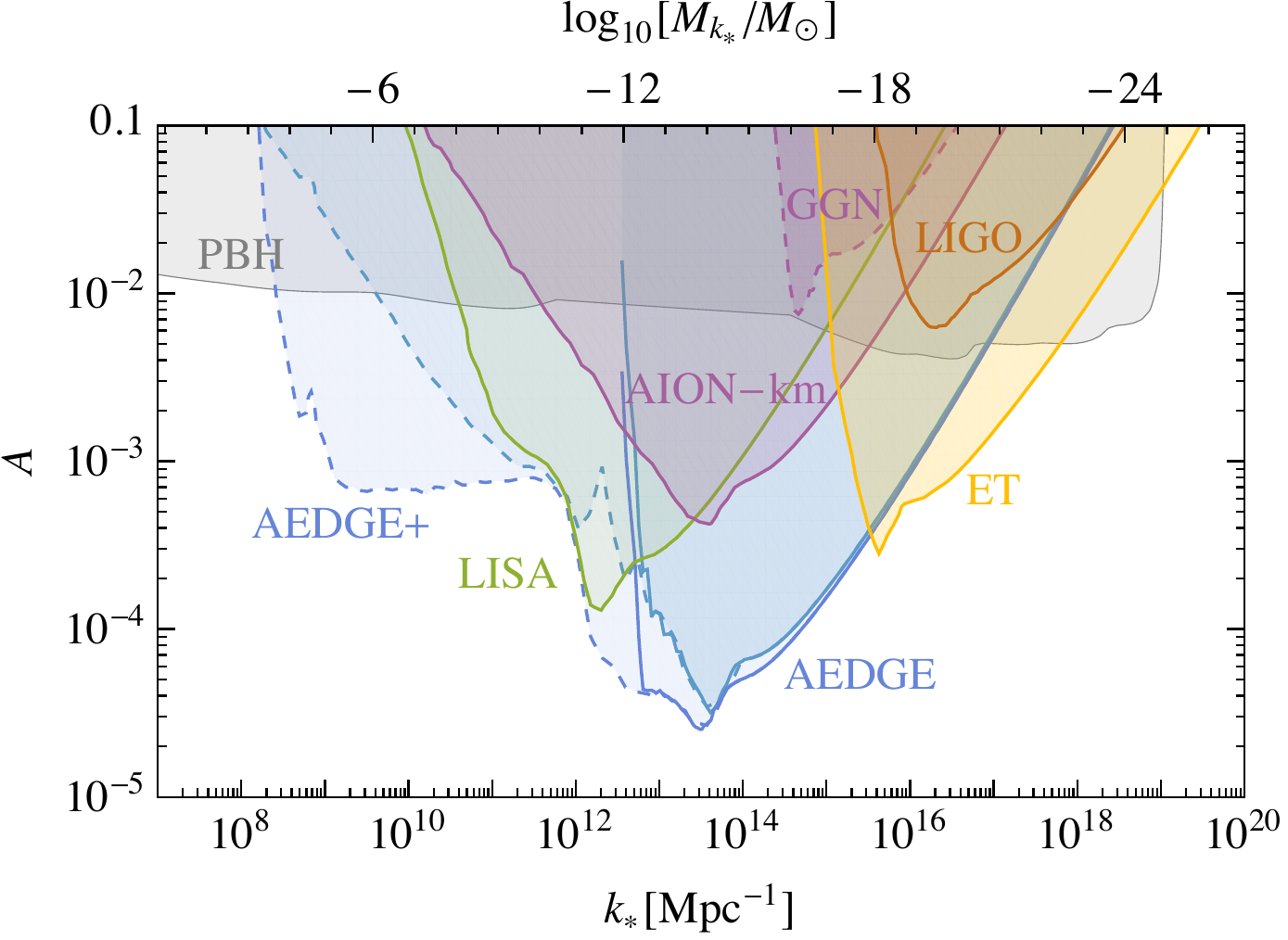}
\vspace{-3mm}
\caption{\it Left panel: The strain sensitivities of AEDGE and other
experiments compared with the neutrino
gravitational memory signal expected from a core-collapse supernova calculated using model Ac3G in~\cite{Mukhopadhyay:2021zbt}.
Right panel: Sensitivities to the amplitude $A$ of a delta function peak in the curvature power spectrum at scale $k_*$ of AION-km, 
AEDGE and AEDGE+, LIGO, ET and LISA. The gray region is excluded by constrains on primordial black holes. The top axis indicates the horizon mass that corresponds 
approximately to the mass of the formed primordial black holes~\cite{RS}.}
\label{astro}
\end{figure}

Opportunities in astrophysics are also opened up by these gravitational-wave measurements, including the observation of mergers of 
intermediate-mass black holes (IMBHs) that could reveal how the super-massive black holes at the centres of many galaxies were assembled~\cite{AEDGE}, 
as illustrated in Fig.~\ref{IMBH}.~\footnote{We note also that timing measurements using atomic clocks located on asteroids have been proposed as a
way to measure gravitational waves in the
$\mu$Hz range~\cite{Fedderke:2021kuy}, in the gap between LISA and pulsar timing arrays.}
As also seen in Fig.~\ref{IMBH}, there are good prospects of synergies obtained by networking detectors working in different frequency ranges, e.g., LISA might measure the initial inspiral stage of IMBH mergers whose final stages would be measured by AEDGE, and AEDGE measurements of the initial inspiral stages of mergers of lower-mass black holes could be used to predict the direction and timing of their final stages, providing advance warning for multimessenger observations.
Also, as shown in the left panel of Fig.~\ref{astro}, AEDGE could observe~\cite{RS} the gravitational memory effect due to neutrinos emitted from a collapsing supernova in our galaxy~\cite{Mukhopadhyay:2021zbt} and, as shown in the right panel of Fig.~\ref{astro}, AEDGE would also be uniquely sensitive to possible features in the spectrum of cosmological density fluctuations that could lead to a population of primordial black holes with a density orders of magnitude below the current astrophysical limits~\cite{RS}.~\footnote{We comment in passing that radio astronomy would gain greatly from extending the Event Horizon Telescope~\cite{EventHorizonTelescope:2019dse} 
concept to a space mission
with a much longer baseline, which would benefit from using atomic clocks for synchronization.}.

\subsection{Connection to Technology Development Section}
\label{sec:fun_tec}
\subsubsection{Optical Clocks}
As outlined in Section~\ref{Sec:Clocks}, several current activities exist to advance the capabilities for optical clocks in space. 
Depending on the type, environment, and targeted timescale of these clocks, the achievable stability varies. 
Similar to the capabilities, the requirements for the above discussed experiments differ.
LISA, similar to navigational applications and earth observation, require high short-term stability~\cite{LISA, Schuldt2017},
with the stability of the reference directly impacting the quality of the intended measurement.
On the other hand, space-borne tests of special relativity, as proposed in~\cite{GuerlebeckBoost} for instance, or made using the signals of misaligned Galileo satellites~\cite{Herrmann2018}, require a high stability at orbit time.

As described in Section~\ref{Sec:ClocksLabClocks}, the stabilization concept has to be chosen for a specific purpose. Optical frequency references based on \textit{optical resonators} usually show the best performance on short time scales, where fluctuations in the resonator length do not impact the stability as severely. 
If integrating over longer times, these fluctuations, caused by thermal or mechanical stress, impact the achievable stability. To increase the long-term stability, efforts have been undertaken to reduce the impact of outside effects, such as by the choice of spacer material~\cite{Ludlow2007} and length of spacer, enclosure in thermal shields~\cite{Sanjuan2019}, or cryogenic environments~\cite{Seel1997}.
The latter is especially interesting in ground-based experimental systems, such as for instance in~\cite{Tobar2010}.

Absolute frequency references based on the spectroscopy of atoms or molecules, on the other hand, benefit from prolonged integration times.
In general, the width of the probed line determines the achievable stability. 
For such references, usually a trade-off has to be made between the required stability and the available SWaP budget. 
As described in Section~\ref{Sec:ClocksIntSpace}, different concepts have been considered for space operation. We note in particular that space operation, especially outside the International Space Station, requires full automation and high reliability of the system.

As outlined above, fundamental experiments require quiet environments to measure the desired effects. 
While this does not always necessitate operation in orbit, space-qualified optical frequency references are a key technology to enable the research.
Section~\ref{Sec:ClocksIntSpace} details current and planned missions for operation in orbit, demonstrating operation at a stability below $10^{-13}$ at orbit time.
To improve on the current best measurements performed on the ground, the frequency stability of any space-borne frequency reference needs to be below $1 \cdot 10^{-15}$ at orbit time. 
Based on the current developments and demonstrated optical references in space, this appears to be feasible for both short-term (in the range of seconds to minutes) and long-term operation (in the range of hours).

\subsubsection{Optical Links}
Optical links are required to fulfill the scientific goals of various planned and executed missions.
Those missions can be divided in such examples with inter-satellite links, such as LISA~\cite{LISA} and AEDGE~\cite{AEDGE}, and those with ground-to-satellite links, such as Micius~\cite{QUESS}. 
As outlined in Section~\ref{Sec:ClocksLinks}, currently mainly microwave links have been established, which do not satisfy the requirements for fundamental science missions.

As an obvious example, entangled optical photons, as required for Micius, can only be transferred by optical ground to satellite links.
The success of Micius was a key step towards space-based fundamental quantum entanglement experiments.
Future quantum entanglement experiments could be envisaged using two space-borne platforms to eliminate the atmospheric impact on the optical link.

Bidirectional ground-to-satellite optical links are a necessity for many applications, such as the proposed Kepler constellation~\cite{Glaser2020}. 
Additionally, they could play a part in comparisons of optical frequency references operating in different gravitational potentials, i.e., on the ground and in space.

Precision in optical inter-satellite links is crucial to the scientific goals of missions employing long-range laser ranging, such as LISA~\cite{LISA}, where picometre level changes over the separation of the satellites of the order of $5 \cdot 10^{9}$ metre will be detectable. 
In this case, the frequency of the deployed laser enables more precise measurements than achievable with a microwave link.

Finally, gravitational wave detection measurements with cold atom sensors require a pair of atom interferometers on two satellites irradiated by the same laser beams to achieve the necessary coupling. 
Current proposals foresee a linkage over $4 \cdot 10^{7}$ metre to reach the strain sensitivity displayed in Fig.~\ref{IMBH}.
The achievable precision described in Section~\ref{Sec:ClocksLinks}, underline the feasibility of missions such as LISA and AEDGE.

Outside of space based experiments, also ground-based projects, such as ELGAR~\cite{ELGAR}, require coherent long-distance free-space laser beams.
Since the necessary length is short compared to the optical links discussed above and the accessibility of the system, the involved technology is currently at hand. 

\subsubsection{Atom Interferometry}

In the sections above, laser stabilization and information transfer over optical links has been discussed. 
The execution of future spaced-based experiments for fundamental science exploitation, such as 
STE-QUEST~\cite{STE-QUEST} and AEDGE~\cite{AEDGE}, requires high-precision atom interferometers, with different interferometer schemes. Whereas, for a test of the weak equivalence principle two different masses (here atomic species) need to be observed in the same place, the detection of gravitational waves and other astrophysical phenomena require elaborate schemes with a single atomic species and optical links between the atom interferometers, interrogated by the same laser.

As described in~\cite{STE-QUEST}, weak equivalence principle experiments are made using two different masses, whose free-fall behaviours are observed and compared. 
Improvements on the current experimental sensitivity are possible, in principle, using any pair of different atomic species. 
The visibility of any deviation from the equivalence principle prediction increases with the interrogation time.  
As an example, according to current estimates, experiments with $^{85}$Rb and $^{87}$Rb require interrogation times of $10$\,s or more in interleaved operation. 
This enables measurements of differential accelerations better than $10^{-13}\,\mathrm{m\,s}^{-2}$, necessary for the targeted precision in the  E\"otv\"os parameter, see Table~\ref{tab:StateArt}.  
These types of experiments require two species interferometers on a single satellite, which also allows for the increased free-fall time.

The detection of gravitational waves, on the other hand, strongly profits from large distances between two atom interferometers in a gradiometric configuration. 
Such experiments require optical links over long periods with two connected satellites in Earth orbit.
The specific interferometer concept requires an atomic species with a clock transition at optical frequencies, e.g., strontium.
The free-fall times of the atoms anticipated for these experiments are substantially longer than those discussed above. 
For example, to enable gravitational wave detection as outlined in Section~\ref{sec:fun_sc_opp}, free-fall times in the order of $600$\,s are required to enable characteristic strain measurements down to $10^{-23}$ at about $80\,\mathrm{mHz}$.

Atom interferometry in microgravity has been performed for several years. 
Experiments have been performed in a drop tower (QUANTUS)~\cite{vanZoest2010, Muentinga2013}, on parabolic flights (I.C.E.)~\cite{Nyman2006}, on sounding rockets (MAIUS)~\cite{MAIUS}, and on the ISS (CAL)~\cite{CALISS}. 
The planned Bose Einstein Condensate and Cold Atom Laboratory (BECCAL) mission~\cite{Frye2021} is the next-generation ultra-cold atom laboratory, including high precision atom interferometry, and is being prepared for deployment on the ISS.
Important challenges in miniaturization and automation have been addressed in the development of CAL and BECCAL.
{We note in addition that pathfinders for the key underlying optical technologies such as FOKUS~\cite{Lezius}, KALEXUS~\cite{Dinkelaker} and JOKARUS~\cite{PhysRevApplied.11.054068} have already demonstrated reliable operation.}
The next step with the proven technology is leaving the International Space Station and integrating an atom interferometer into a satellite.
With that development, space-borne tests of the weak equivalence principle, as proposed in~\cite{STE-QUEST}, would come within reach.

AEDGE represents the following step in advancing quantum technologies for deployment in space. 
It requires two optically-linked atom interferometers, increasing the distance and thereby sensitivity of gravitational wave detectors with respect to ground-based measurements. 
The measurement is enabled by employing strontium as opposed to rubidium or potassium. 
This leads to an additional complexity in the system, such as availability of miniaturized laser systems, electronics, and optics.

Based on the present technology in space-based rubidium and potassium systems~\cite{MAIUS, CALISS, Frye2021} and ground-based strontium systems~\cite{Zhang2016, Rudolph2020}, the developments required for the next steps appear feasible.

In this connection, we note that the experimental landscape of atom interferometry projects for fundamental science exploitation has expanded significantly in recent years, with several terrestrial experiments, based on different cold atom technologies, currently under construction, planned or proposed. 

As mentioned above, four large-scale prototype projects are funded and currently under construction, i.e., MAGIS~\cite{Abe2021} in the US, MIGA~\cite{Canuel:2017rrp} in France, ZAIGA~\cite{ZAIGA} in China and  AION~\cite{Badurina2019} in the UK. 
These will demonstrate the feasibility of atom interferometry at macroscopic scales, paving the way for terrestrial km-scale experiments as the next steps. 
There are proposals to build one or several km-scale detectors, including AION-km at the STFC Boulby facility in the UK,  MAGIA-advanced and ELGAR~\cite{ELGAR} in Europe, MAGIS-km at the Sanford Underground Research facility (SURF) in the US, and advanced ZAIGA in China. 
It is foreseen that by about 2035 one or more km-scale detectors will have entered operation. 
These km-scale experiments would not only be able to explore sensitively ultralight dark matter and the mid-frequency band of gravitational waves, but would also serve as the ultimate technology readiness demonstrators for space-based missions like STE-QUEST~\cite{STE-QUEST} and
AEDGE~\cite{AEDGE} that would reach the ultimate sensitivity for exploring the fundamental physics goals outlined in this section.
     
The perspectives for large-scale atom interferometer projects are very rich today, with a central focus on establishing the readiness of cold atom technology for use in  experiments to explore fundamental science. 
Theses terrestrial pathfinders are advancing further the relevant technologies, closing gaps and addressing fundamental physics questions.

However, in order to deploy effectively strontium or ytterbium on any microgravity platform, in addition to these terrestrial developments a space-borne pathfinder mission or technology demonstrator is key.
The development of critical components benefits from synergies between all systems. 
As such, laser modules developed for clock deployment could be used as the basis for interferometric missions. 
Similar synergies are present for vacuum generation, frequency stabilization, and low noise electronics between the various systems. 
With a technology demonstrator for a space-bourne optical lattice system, many applied and fundamental missions could be supported and further developments triggered.

\subsection{Recommendation: Road-map for fundamental physics in space} 
To summarize Section~\ref{sec:fun_tec}, which details the requirements to enable the scientific opportunities from Section~\ref{sec:fun_sc_opp}, the following recommendations for developments can be made: 

\begin{itemize}
    \item Build upon the ongoing large-scale terrestrial atom interferometer projects for fundamental science exploitation, such as MAGIS~\cite{Abe2021} in the US, MIGA~\cite{Canuel:2017rrp} in France, ZAIGA~\cite{ZAIGA} in China and  AION~\cite{Badurina2019} in the UK to construct one or more of the proposed km-scale successors, which will enable technology development for space-based missions like STE-QUEST~\cite{STE-QUEST} and AEDGE~\cite{AEDGE}.
    The ELGAR experiment~\cite{ELGAR} would also supplement existing and future, possibly satellite-based, scientific measurements by targeting relevant gravitational wave frequencies. 
    \item Perform fundamental tests in ground based microgravity facilities, such as the drop tower, the Einstein Elevator, parabolic flights, and sounding rockets, to develop technology and support scientific findings. 
    \item Prepare a satellite mission as the next step in cold and condensed atom technology in space with a two-species interferometer based on the available rubidium and potassium sources. 
    \item Prepare optical frequency references for operation in space to test General Relativity and probe modified
    gravity theories, 
    as well as support space missions such as LISA. 
    This is in accordance with the recommendations in Section~\ref{sec:clocks}. 
    \item Develop components for deployment in space, with a focus on those with synergies for different missions, for instance optical preparation, laser modules, vacuum generation, and magnetic field control.
    \item Advance the development of optical lattice systems on the ground, in ground-based microgravity platforms, and in space-based pathfinder missions to enable future gravitational wave detection missions. 
    This includes miniaturization of subsystems and proof-of-principle missions or studies for individual components. 
\end{itemize}

\section{Technology Development, Space Qualification and Pathfinders}
\label{sec:Tech}


\subsection{Requirements for cold atoms in space/Specifications for space borne quantum sensors}

\subsubsection{Atomic clock mission} 
We advocate a single dedicated satellite in a highly elliptical orbit, containing a strontium optical lattice clock with a $1\times10^{-18}$ systematic uncertainty and $1\times10^{-16}/\sqrt{\tau}$ instability, and including a coherent optical link for comparisons to ground-based clocks. 
Such a mission would enable a wide network of ground-based clocks to be compared to each other with $10^{-18}$ precision in 1 day, which would have applications in fundamental physics discovery, proof-of-concept optical timescales, and geodesy. 
Furthermore, the clock in elliptical orbit would enable direct searches for new physics, including stringent tests of general relativity (see Sections \ref{sec:clocks} and \ref{sec:PH}). 
To support such a mission, improved optical links must be developed for clock comparisons at $1\times10^{-18}$. 
Efforts to realize such links are currently underway in the ACES and I-SOC Pathfinder ESA programmes. 
However, the bulk of the challenge of a space-clock mission will be to develop a space-qualified strontium optical lattice clock. 
Present strontium clock technology occupies at least a few m$^3$ and consists of several complex, delicate components (see Section~\ref{sec:TRL}) which must all be brought up through the TRL scale. 

\subsubsection{Earth Observation mission}
As outlined in Section~\ref{sec:EO}, quantum sensors offer the perspective of enhancing missions for Earth Observation by embarking them either in a gradiometric configuration on a single satellite for a GOCE-like mission concept, or as accelerometers combined with laser links between satellite pairs for a GRACE-like mission concept in an earth orbit with an altitude of few hundred kilometers and nadir pointing mode.
Similar to previous missions in Earth Observation, the targeted mission duration is several years.
Sensitivities of $^{87}$Rb atom interferometers to accelerations or differential accelerations would be in the range of $10^{-10}$ to $10^{-12}\,\mathrm{m\,s}^{-2}\,\mathrm{Hz}^{-1/2}$ at low frequencies, complementing current sensor technology.
The SWaP budget of a current quantum sensor may include a few hundred kilograms of payload and a few hundred Watts of power,
which need to be reduced. Moreover, the payload could be designed to link
multiple quantum sensors on a single satellite by interrogating them with the same beam splitter laser for a GOCE-like mission, 
or they could be implemented as accelerometers for drag correction in a GRACE-like mission.

\subsubsection{Fundamental Physics}
As discussed in Section~\ref{sec:fun_tec}, key technologies in three areas are required for fundamental physics tests: optical clocks, optical links, and atom interferometers. 
The requirements for optical frequency references are described in the paragraph above.
In case of optical links, the requirements depend on the specific mission, and range from single photon transmission in case of entanglement missions to long distance coherent laser light transmission in case of gravitational wave detection.
For atom interferometers, the requirements are naturally more restrictive the higher the desired precision, which has implications for the atomic flux, preparation of the atoms, coherent manipulation, and interrogation times. 
The test of the universality of free fall outlined in Section~\ref{sec:fun_tec} targets a measurement of the Eötvös ratio at the level of $\leq 10^{-17}$, and requires atom interferometers using two different kinds of atomic species simultaneously (e.g., $^{85}$Rb and $^{87}$Rb or $^{41}$K and $^{87}$Rb) with interrogation times of over $10$\,s to reach a projected sensitivity to differential accelerations of $10^{-13}\,\mathrm{m\,s}^{-2}$ at a cycle time of $10\,\mathrm{s}$ in an interleaved operation.
For the case of the proposed space-borne gravitational detector, the scheme is based on strontium atoms, requiring an increased atomic flux, large momentum transfers, and interrogation times of $600$\,s to enable characteristic strain measurements down to $10^{-23}$ at about $80\,\mathrm{mHz}$. 
These parameters place requirements, e.g., on the atomic source as well as the dimensions of the central elements, such as the vacuum chamber, affecting the overall design of the sensor head.

Fundamental physics experiments require a variety of different orbital scenarios. For example,
entanglement experiments require large distances to close additional loopholes and 
test the validity of quantum mechanics, hence a geostationary or lunar satellite would be of interest.
Clock redshift and local position invariance tests benefit from orbits with large variations in the
gravitational potential, whereas
local Lorentz invariance tests benefit from short orbital times with high orbital velocities and velocity variations.
A prime candidate for such a mission is a satellite in low Earth orbit, 
while nevertheless avoiding vibrations caused by drag in the atmosphere.
Orbital heights of $600$\,km with orbital times in the order of $90$~min appear appropriate. 
A similar orbit may be chosen for EEP tests, 
motivated by the effect of the gravitational field and the absence of vibrations due to drag.
The satellite should have an optical link to an optical frequency reference on the ground.
The proposed gravitational-wave experiment described here
would require two satellites in medium Earth orbit with a longer-baseline optical link 
in a calm environment to reduce gravitational noise.

In all cases, the proposed mission duration spans multiple years.

\subsection{Technology Development Path and Milestones}

As indicated in the previous Section, the requirements for Cold Atoms in space basically call for three types of instrument to be developed: Atomic Clocks, Atom Interferometers and Optical Links.

In order to have those instruments introduced and accepted into a space mission, a solid development and qualification approach should be established. 
This is expected to be based on existing and well-proven approaches currently applied in space projects, that need to be tailored to the specific technologies and trends to reach a suitable balance between risk and mission objectives. 
Depending on the technologies and objectives, the approach could include in-orbit demonstrators or pathfinder missions. 
As a guideline, and irrespective of the type of mission (in-orbit demonstrator, pathfinder, ...) a generic development approach for such instruments would typically include the following steps.

First, the scientific/mission objectives are defined, e.g., a test of the universality of free fall or a strain measurement to a certain level, together with a high-level baseline mission scenario including, e.g., the orbit, in order to derive the expected preliminary mission lifetime and environment (mechanical, thermal, magnetic field, radiation, …).
In parallel, the technical requirements for the instrument(s) should be defined (functional, performance, operational, volume/mass/power, interfaces, …).
This is based on both a flow-down of mission requirements (top-down) and the review of existing ground instruments and/or experiments (bottom-up). 
The outcome of this first step would be the issue of the consolidated mission definition and technical requirements, e.g., a certain sensitivity to differential accelerations or phase shifts induced by a gravitational wave.

Once these requirements have been reviewed and agreed by the community, each instrument can follow its own development path.
This includes first the definition of the instrument architecture and its external interfaces with the spacecraft and with other instruments.
Secondly, the instrument architecture definition is further refined into subsystems, modules or units, e.g.,
the vacuum system including peripheral optics, the laser system, or the control electronics.
For each of these elements, interfaces with upper levels are defined, together with technical requirements based on a flow-down from the upper level.
The granularity of the instrument architecture definition depends on the type and complexity of the instrument.
The outcome of this step is the issue of the instrument architecture definition and technical requirements for the subsystems.

The next step is the development of the subsystems, modules or units, whose approach is tailored to the specific element and the maturity of its underlying technology. It is usual practice to start the development at Breadboard level to demonstrate the basic functionalities and performance, and to develop further to an Elegant Breadboard and/or Engineering Model (EM). An Engineering Model is fully representative of the Flight Model in terms of form, fit and function, but does not require the use of qualified high-reliability parts. 

Once all subsystems have demonstrated compliance with their technical requirements, they are integrated into the instrument, which is in turn validated and verified according to an agreed method. It is likely that instruments based on Cold Atom technologies will not be able to reach their full performance on-ground, and therefore appropriate verification methods will have to be defined (e.g., based on a combination of test and analysis/extrapolation through modelling, and/or microgravity test facilities such as a drop tower, Einstein Elevator, or zero-g Airbus flight). In the event of successful validation and verification at instrument level according to EM standards, the instrument will have reached TRL6.

From there, the instrument will follow a qualification phase. 
For complex instruments like the ones we are considering here, this will most probably involve a Qualification Model (QM) that is fully representative of the Flight Model in terms of build standard (using qualified high-reliability parts), and that will be subject to a qualification test campaign according to agreed qualification levels and duration. 
This qualification step also includes the qualification of all lower-level elements, including their materials, parts and processes, in accordance with the requirements applicable to the mission (e.g., due to launch loads, the operational environment. etc.). 
After successful completion of the qualification phase, the instrument will have reached TRL7, and the manufacturing of the Flight Model is released.

\subsection{Technology Readiness Level} \label{sec:TRL}

\subsubsection{Atomic clock mission} 
Once optical links are in place, several technologies must be developed with improved TRL in order to launch a strontium optical lattice clock into space: optical resonators to stabilize lasers to a noise floor of $1\times10^{-16}$ in fractional frequency, laser sources at six different wavelengths to cool, optically confine and interrogate strontium atoms, compact physics packages with a controlled black-body radiation environment, and compact frequency combs.

\subsubsection{Earth Observation}
Key concepts have been demonstrated, including gravimeters~\cite{Hu2013PRA} with sensitivity $4.2~\cdot~10^{-8}\mathrm{m/s}^{2}/\mathrm{Hz}^{1/2}$ and gradiometers~\cite{Chiow2016,Asenbaum2016} with sensitivity $3\cdot 10^{-8}\mathrm{s}^{-2}\,\mathrm{Hz}^{-1/2}$, both operating with $^{87}$Rb atoms, as well as matter-wave collimation of BECs and BEC interferometers in a drop tower and onboard of a sounding rocket, utilising atom-chip technology for fast and robust production of $^{87}$Rb BECs.
The latter relied on adaptation and developments of the physics package, the laser system, and the control electronics to realise compact and robust setups.
Further developments are required for operation in the specific conditions imposed in satellite missions on a component level, which is partially ongoing, but also for demonstrating the desired performance which relies on the extended free fall times in a microgravity environment.

\subsubsection{Fundamental Physics}
As outlined above, different scenarios for testing the limits of quantum mechanics with cold and condensed atoms exist. 
As such, the TRL differs widely for the involved components. 
At this time, payload for experiments on cold Rb atoms and BECs as well as quantum entanglement have been operated on satellites or the ISS. 
To achieve the targeted sensitivities for the discussed missions, further developments for components, such as laser systems, vacuum technology, radiation-hard electronics and autonomous operation are necessary to accommodate, for instance, high-precision Rb/K or Sr interferometers. 
Specific payloads, such as a setup for BEC experiments and interferometry, or components, such as frequency combs, have been successfully operated on sounding rockets and therefore feature a higher TRL than other parts. 
Moreover, several components 
require developments to cope with vibrational loads during launch, SWaP budget limitations, and other environmental conditions. 

\subsection{Technology evolution}
Operating quantum sensors based on cold atoms on a satellite implies a completely new technology going to space.
Between various mission scenarios utilising atomic clocks or atom interferometers for earth observation or fundamental physics several general building blocks are shared.
They consist of a physics package with a vacuum system surrounded by optics, coils and other peripherals, a laser system with laser sources and optical benches for light distribution and switching, an electronics system with various controllers, e.g., laser drivers, and a computer for executing sequences, collecting, storing and evaluating data.
This leads to potential synergies in technology developments.
Within these subsystems, it is crucial to identify critical components and to start their development without delay for a mission within this decade.
While different missions may require modified components imposing a new verification, this does not imply a completely new development, as the concepts, their capabilities and the approach for verification are known.

Mission-specific technology developments that may focus on performance, miniaturisation, robustness, lifetime or other relevant topics, as required, will typically follow a stepwise approach.
On a conceptual level, ground-based facilities, although incompatible with deployment in space, can serve to test and verify experimental procedures, sequences, and concepts for a future mission.
Initially, the system needs to be defined based on the necessary functionalities.
Top-level examples are, e.g., an atom interferometer based on Rb or Sr, a Sr optical lattice clock, optical frequency dissemination based on fibres, free space, and ground-space-ground communication with mission-specific performance requirements.
The next step is to identify suitable subsystems (e.g., a laser system or vacuum system/science chamber) with respect to performance, to define components (e.g., a laser head or an atom chip) and develop either as required.
This may include reliability and partial environmental tests on a subsystem level, depending on the estimated critically.
The subsystems have then to be integrated, and subjected to end-to-end verification and performance tests.
Subsequently, the full ground system is to be implemented and tested, including reliability and partial environmental tests.
Facilities such as the microgravity simulator in Bordeaux, the drop tower in Bremen, or the Einstein Elevator in Hannover offer the possibility for operating a payload or parts thereof in up to a few seconds of microgravity.
Additional options for such tests are early flight tests, as enabled by a zero-g Airbus flight or a sounding rocket.
Gravity, the available microgravity time of the aforementioned facilities, or special (e.g., environmental) constraints of a mission may lead to the recommendation of a pathfinder mission with opportunities sponsored by ESA, the EU or national agencies.
Finally, after the development and verification steps, the definition and planning of a full mission concludes the technology evolution.

\subsection{Development milestones}
Preceding or in parallel to the technology developments, several scientifically-justifiable milestones can be defined to assess the maturity on a conceptual level.
These are related to the demonstration of basic functionalities and concepts, including feasibility studies.
A starting point is the definition of the mission concept and the scientific motivation, be it for earth observation, fundamental physics or other purposes.
Critical concepts can then be tested in ground-based setups (e.g., trapping and cooling atoms with novel atom's chips/gratings), in certain cases with reduced performance (e.g., due to reduced free fall time), but still demonstrating their basic feasibility.
This may include a special scheme for an atom interferometer up to a large-scale device for gravitational wave detection as in MIGA, MAGIS, ELGAR or AION.
Depending on the possibility for extrapolation, the confidence in the modelling and understanding (e.g., due to different behaviour in the absence of gravitational sag, non-moving atoms), deploying a compact test setup in a microgravity facility such as the drop tower can demonstrate source performance, interrogation procedures, and detection of the atoms, accompanied by refined modelling, without the need for components qualified for a satellite mission.
On the side of the mission concept, the orbit needs to be defined and evaluated accordingly, including the estimated implications on the payload and performance.
The latter requires a simulator for the atom interferometer, clock or other relevant payload element to be developed.
It can subsequently provide the modelling of the measurement output, including dependencies on internal and external disturbances, compared to the desired signal.
Within a dedicated pathfinder mission on a satellite, demonstrating, e.g., a high-contrast atom interferometer with extended free-fall times and long-term operation, and performing statistical and systematic studies, provide further milestones for assessing the maturity of sensors.

\subsubsection{In-orbit validation}
To date, following the successful operation of quantum sensors in laboratories, key technologies and concepts have been demonstrated, e.g., in dedicated microgravity experiments on BEC generation and interferometry with rubidium atoms.
Limited microgravity time on the available platforms prevented long-term operation, extensive statistics and a detailed systematic analysis in this special environment.

A dedicated satellite platform for a pathfinder avoids conflicts of programmatic or technical constraints due to interface or other requirements of either part of the payload in joint mission concepts.
Other currently existing or planned payloads have to provide multiple functionalities and are consequently neither dedicated nor optimised for quantum sensors.
Despite considerable efforts for miniaturisation and robustness for accommodation in the available microgravity platforms, a satellite mission will impose additional constraints and requirements on the payload and the operation of the quantum sensor.
This step towards a full-blown mission utilising quantum sensors motivates a timely pathfinder mission.

Several examples of pathfinder missions have been mentioned in earlier sections.
Here we discuss as an  example some aspects of a prospective mission deploying a quantum sensor with rubidium BECs on a satellite.
Such a system offers the opportunity to achieve multiple goals.
It enables the technology demonstration of a BEC source, beam splitters, detection, remote control, and autonomous execution of sequences on a satellite up to the uninterrupted operation of an atom interferometer over weeks to months, validating the maturity of the individual key components and functionalities.
Furthermore, it can serve for testing and validating onboard data evaluation and autonomous optimisation of parameters for the source, interferometer, and other manipulations of the atoms, reducing the need for user intervention.
Depending on the satellite bus and orbit, the quantum sensor is subjected to specific disturbances, necessitating mitigation strategies such as, e.g., rotation compensation with a higher dynamic range than for stationary setups on Earth.
Testing such techniques for free fall times of several seconds is crucial for future high-performance sensors.
Finally, only a pathfinder mission can provide a detailed performance evaluation for a satellite-based quantum sensor due to the persistent microgravity, enabling the accumulation of extensive statistics in the satellite-specific environment.

The higher level of maturity shown by the technology demonstration in such a pathfinder mission would have a direct impact on proposals for an Earth Observation mission based on Rb BEC interferometers or a fundamental physics mission testing the universality of free fall with Rb/K dual-species BEC interferometers.
Future missions with optical lattice clocks would also benefit from a pathfinder mission with a Rb BEC interferometer for validating vacuum technology, parts of the control electronics, the experiment control computer, onboard data evaluation, and autonomous operation.
Introducing other atomic species for atom interferometry such as Sr implies more similarities with the outlined pathfinder than a clock-type mission due to additional shared concepts, but will in addition require dedicated development activities for the atomic source and laser systems.

\section{Workshop Summary}
\label{sec:sum}

As discussed in Section~\ref{sec:intro}, the discovery of quantum mechanics in the first part of the
20th century made possible many of the technological innovations developed in the second half of the century.
We are now witnessing the breakout from the laboratory of many quantum phenomena not yet applied,
which offer sensor technologies of unparalleled accuracy for timing, accelerometry, gravimeters, etc.. 
The potential of this second quantum revolution has been recognized by the Senior Science Committee 
advising ESA on its Voyage 2050 programme, which has recommended an intensive programme of R\&D to 
prepare quantum sensor technologies for deployment in space. 
Many of the cutting-edge developments of
quantum technologies currently taking place in laboratories around Europe and elsewhere were
discussed in this Workshop. 
Realizing their full potential in space-borne applications of immediate value
to society as well as to fundamental science will require a community effort to outline high-level
objectives and a road-map for achieving them that optimizes the synergies between different mission concepts.

Atomic clocks have already attained an accuracy of $10^{-18}$, and the ACES mission is in an
advanced state of preparation for deployment on the ISS. Atomic accelerometers have already
exhibited a precision of $4 \times 10^{-8}$~m/s$^2/\sqrt{\rm Hz}$ on the ground, and offer
drift-free stability extending to frequencies below 10~mHz. There has been a series of
of Bose-Einstein Condensate (BEC) experiments in microgravity using MAIUS sounding rockets,
the CAL BEC experiment has operated successfully for several years on the ISS, the MICROSCOPE 
experiment has tested the Weak Equivalence Principle (WEP) in space, and the MAGIS, MIGA,
ZAIGA and AION atom interferometer experiments to look for ultralight dark matter and pave the way 
for future measurements of gravitational waves are under construction. 

We present below a first draft for this community road-map, based on the cold atom achievements so far, the ongoing
research, and the high-level objectives for the future. 
The draft road-map is centred around three topics:
atomic clocks, quantum accelerometry, and atom interferometry, and is oriented towards the 
following objectives: next-generation standards for time standards and navigation, next-generation
Earth observation and its potential for monitoring mass and climate change, and fundamental science including
tests of relativity, searches for dark matter and novel measurements of gravitational waves. 
We stress the existence of an ongoing technology development programme including terrestrial and space-borne
pathfinder projects, and the need for follow-on pathfinder experiments on Earth and in space.

Milestones in the road-map towards space clocks discussed in Section~\ref{sec:clocks} include the completion of ACES~\cite{ACES} and its deployment on the ISS,
to be followed by a follow-on mission such as I-SOC Pathfinder~\cite{Prochazka2018, Prochazka2020}.  
Its objectives would include a comparison of ground-based optical clocks with a precision of $10^{-18}$ over a day (see Fig.~\ref{fig:clock_accuracy}), which would have applications in fundamental physics
experiments as establishing a proof-of-concept for setting timescales, and in geodesy. 
This should be followed by a dedicated  satellite in  a  highly  elliptical  orbit  containing  a  strontium  optical  lattice clock  with  similar precision and a  coherent
optical link to ground, with goals similar to those of FOCOS~\cite{focos}. 
Such a mission would enable  more precise comparisons across a wider network of ground clocks, and stringent tests of general relativity. 
The development of coherent free-space  optical links will be key, to be accompanied by the  qualification  of  a  strontium  optical  lattice  clock  for operation in space, which will require
a technology development programme for several clock components. 
There are multiple synergies between these atomic clock missions and the requirements for a fundamental science mission such as AEDGE~\cite{AEDGE} based on strontium atom interferometry.
 
As discussed in Section~\ref{sec:EO}, quantum accelerometry has exciting potential (see Figs.~\ref{GRICE} and \ref{Gain}), and we envisage two
atom accelerometer missions aimed at realizing it. 
For the technical reasons discussed in Section~\ref{sec:EO}, deploying a quantum sensor as a passenger on a conventional geodesy mission 
such as MAGIC~\cite{MAGIC} would pose severe technical challenges, implying significant technological and programmatic risks for both the classical and quantum aspects of such a joint mission.
For this reason, the discussions among scientific experts during the workshop led to a recommendation to launch a separate
quantum pathfinder mission within this decade on a dedicated platform, with a target performance of $10^{-10}$~m/s$^2$/$\sqrt{\rm Hz}$. 
Such a mission would combine the need for a test of the quantum technology in space with optimizing the results to be expected 
from a subsequent quantum gravimetry pathfinder mission. 
It would also serve as a milestone for other communities, such as that interested in applications of cold atoms to probes of fundamental physics. 
The success of MAGIC and the quantum pathfinder mission  would pave the way for a full-fledged quantum space gravimetry mission to follow on from MAGIC, whose definition would be based on the experience gained with MAGIC and the pathfinder mission.

We outlined in Section~\ref{sec:PH} the requirements for enabling the opportunities for exploring fundamental science, some of which are illustrated in Figs.~\ref{DM} and \ref{IMBH}.
The first step is to construct and operate the ongoing large-scale terrestrial atom interferometer projects for fundamental science, such  as  MAGIS~\cite{Abe2021}  in  the  US,  MIGA~\cite{Canuel:2017rrp}  in  France, ZAIGA~\cite{ZAIGA} in China and AION~\cite{Badurina2019}, to be followed by one or more of the proposed km-scale experiments such as ELGAR~\cite{ELGAR} and the successors to MAGIS, ZAIGA and AION, which  will  serve  as  ultimate  conceptual  technology  readiness  demonstrators  for  a  space-based mission such as AEDGE~\cite{AEDGE}. 
In parallel, there should be a satellite mission demonstrating cold and condensed atom technology in space, building on the experience with CAL~\cite{CALISS} and MAIUS~\cite{MAIUS} and using a two-species interferometer (see~\cite{STE-QUEST}), based on the available rubidium and potassium sources. 
It will also be necessary to prepare  optical frequency  references for  operation  in space (see also Section~\ref{sec:clocks}), which will also support space missions such as LISA~\cite{LISA}. 
This will require advancing strontium  development  on  the  ground,  in  ground-based  microgravity  platforms, and in space-based pathfinder missions, including individual components, the miniaturization of subsystems 
and proof-of-principle missions.

\begin{figure}[ht]
\centering
\includegraphics[width=\textwidth]{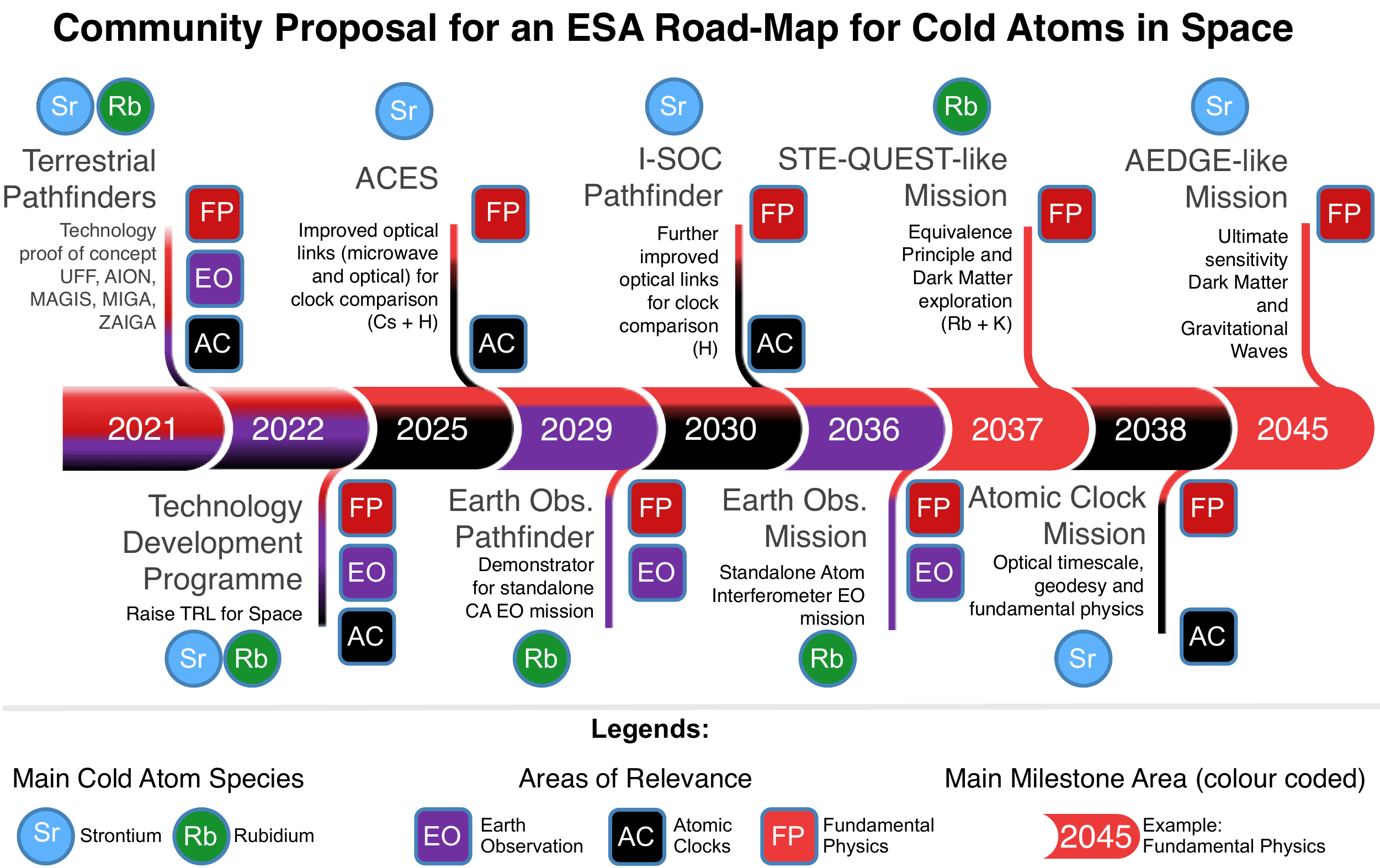}
\vspace{-0.5cm}
\caption{\it A first draft of the community road-map proposed for discussion with ESA. The figure shows the main milestones of the road-map, where the colour code indicates the relevance for the three main areas of Earth Observation (purple), Atomic Clocks (black) and Fundamental Physics (red). The main cold atom species targeted by the milestones are indicated with circles coloured blue for strontium and green for rubidium. These do not necessarily indicate that a milestone is based on this Cold Atom species, but may also indicate that this milestone will contribute to its space readiness. The years indicated in each milestone represent the starting date of the milestone activity or the launch date of a mission, assuming preparatory work took place before. The further the timeline extends into the future, the more uncertain the milestone starting date becomes. Further details about each of the milestones and their interdependences are given in Section~\ref{sec:road-map}.}
\label{fig:road-map}
\end{figure}

\section{Community Proposal for {an ESA} Road-Map for Cold Atoms in Space}

\label{sec:road-map}


We present in this Section a draft for a possible road-map for the development and
deployment of cold atoms in space, formulated following the discussions during the 
Community Workshop that are summarised in the previous Sections of this document,
which we propose for discussion with ESA, European national space agencies and interested
potential partners. The proposed road-map is mainly driven by high-level scientific applications
and includes synergetically the areas of Earth Observation and Fundamental Physics, as well as applications of space-borne Atomic Clocks. The status and prospects in these three areas are reviewed in Sections~\ref{sec:EO}, ~\ref{sec:PH}, and ~\ref{sec:clocks} of the workshop summary, respectively, requirements for technology development, space qualification and pathfinders are
reviewed in Section~\ref{sec:Tech}, and the main points of these Sections are summarised in Section~\ref{sec:sum}.


Fig.~\ref{fig:road-map} presents a high-level timeline of the main milestones of the road-map,
where dates indicate the start of an activity or the launch of a space mission. A detailed discussion of the individual milestones and their interdependences is provided in Section~\ref{milestone-TP} to ~\ref{milestone-AEDGE}. Although the road-map focuses on possible
ESA missions and their milestones, we also mention potential synergies with other missions and options to integrate the ESA programme into a broader international context. {We note that the timeline is conservative, in the sense that it could be shortened in the event of substantial international collaboration, and that uncertainties are larger for later milestones in the road-map.}


\subsection{Terrestrial Pathfinders: Underway in 2021}
\label{milestone-TP}
{
Scientific experiments in terrestrial environments and additional technology development, including relevant environmental testing, such as vibration or radiation hardness, are crucial to the definition, planning, and successful deployment of the space-borne missions discussed in the following road-map. 
Several of these developments are currently ongoing in different institutes. 
While some institutes focus on individual subsystems, such as more compact and robust laser systems, other efforts are underway to operate large-scale experiments, such as MAGIS-100 (Sr) in the US, MIGA (Rb)  in France, AION-10 (Sr) in the UK, VLBAI (Rb and Yt) in Germany, and ZAIGA (Rb) in China, which are funded and currently under construction. 
These are prototypes for planned apparatuses spanning several km, such as ELGAR in Europe, SURF in the USA, and expansions of AION in the UK and ZAIGA in China. 
These systems could explore systematically mid-frequency range gravitational waves and would complement existing and future gravitational wave antennae, both in space and on the ground. 
In addition, these large-scale projects could perform research in other areas, e.g., to search for oltralight bosonic dark matter that is weakly coupled to Standard Model particles, and to test the Equivalence Principle.
These terrestrial Cold Atom pathfinder projects will complement the other pathfinders listed in our road-map and space-borne cold atom experiments such as CACES, MAIUS, CAL and BECCAL, as well as the successful pathfinders for key underlying optical technologies such as FOKUS, KALEXUS and JOKARUS.}

\subsection{Development Programme Launch: 2022}
\label{milestone-DP}
In parallel with the construction, commissioning and operation
of these terrestrial pathfinder experiments, it will be necessary to
start without delay a programme to raise the TRLs of the associated
technologies, demonstrating their robustness and reliability,
and reducing their size, weight and power requirements, so that they
become realistic candidates for deployment in space-borne experiments.
In addition to the sensor technologies used directly in the experimental
apparatus, such as atomic clocks, accelerometers and interferometers,
it will be necessary to raise the TRLs of ancillary technologies such as
optical links.
This technology development programme will be essential for the
deployment in space of atomic clocks, applications to Earth Observation
and experiments in fundamental physics.
Technologies using rubidium are currently more mature than those using strontium,
but both will require substantial development effort.

In addition to a development programme building systematically on the ongoing terrestrial pathfinders outlined in Section~\ref{milestone-TP}, a pathway that has proven to be successful in increasing TRLs for space operation is provided by terrestrial microgravity environments. 
These include, for instance, the Einstein Elevator in Hanover, the drop tower in Bremen, parabolic flights operated by Novespace, and sounding rocket missions from Kiruna. 
Operation within these environments requires the experimental setups to ensure miniaturization of the setup and perform environmental tests. 
This enables aspects of fundamental research to be performed and technology necessary for space operation to be developed. 
The scientific output supplements and focuses planned space-based missions and can bridge gaps until space-based missions become a reality. 
Prominent examples of such experiments are the I.C.E experiment in parabolic flights, the QUANTUS and PRIMUS experiments in the drop tower, and the MAIUS and JOKARUS campaigns in sounding rocket flights.
Other important stepping-stones for fundamental research and technology development for further space-based experiments are the cold atom experiments on the ISS, namely CAL and BECCAL. 
With CAL being currently mounted to the ISS and BECCAL being prepared for deployment within this decade, they pave the way towards more complex and specialized experiments and increase the TRLs of individual components as well as complete setups. 
Establishing this Development Programme will be essential for the success of the following Road-Map milestones outlined in Sections~\ref{milestone-ACES} - \ref{milestone-AEDGE}.

\subsection{ACES: 2025}
\label{milestone-ACES}

The ACES experiment is currently being prepared for launch by 2025.
It will measure the gravitational redshift between the PHARAO clock on-board the ISS
and clocks on Earth, improving on current measurements of the redshift effect
by an order of magnitude. In addition to providing the first in-orbit demonstration of
the operation of a cold atom clock using cesium, as well as the operation of a hydrogen maser,
it will also pioneer the deployment of improved optical links for atomic clocks in space.
ACES will serve as a pathfinder for future projects deploying atomic clocks in space (Section~\ref{milestone-AC})
as well as providing important new tests of general relativity, paving the way for future
cold atom experiments on fundamental physics in space (Sections~\ref{milestone-STE-QUEST} and \ref{milestone-AEDGE}).
The experience gained with ACES will be particularly relevant for subsequent experiments
using strontium clocks in space.

\subsection{Earth Observation Pathfinder Mission: 2029}
\label{milestone-EO}

As discussed in the main text, gravimeters and accelerometers based on cold atom technology
show great promise for future Earth Observation missions, in view of the high precision,
stability and low-frequency performance they offer. However, we judge that it would be
premature to plan a standalone full Earth Observation mission using cold atoms, and that a prior
pathfinder mission will be required. For the reasons discussed in the text, we consider that
it not be advantageous to combine this quantum pathfinder mission with an Earth Observation
mission using classical technology.
While relevant primarily for a subsequent Earth Observation mission (Section~\ref{milestone-EO2}), this pathfinder mission will also pave the way for
subsequent fundamental science missions using atomic clocks. 
It would use rubidium atoms, whose terrestrial development is relatively mature, and its
operational experience would be relevant to an STE-Quest-like fundamental physics mission (Section~\ref{milestone-STE-QUEST}), which
would also use rubidium.

\subsection{I-SOC Pathfinder: 2030}
\label{milestone-I-SOC}
I-SOC Pathfinder would push further the microwave and optical link technologies being
developed for ACES (Section~\ref{milestone-ACES}), with a view to continue the operation of a worldwide network of 
optical clocks on the ground to test fundamental laws of physics, to develop 
applications in geodesy and time \& frequency transfer, and to demonstrate 
key technologies for future atomic clock missions in space.
The main objective of I-SOC Pathfinder will be to increase the versatility of atomic clocks in space,
acting as a pathfinder for a subsequent mission to exploit the full capabilities
of atomic clocks, also for applications in fundamental physics based on strontium.
Its operation will support other technology developments by operating optical links from the space station to ground.

\subsection{Earth Observation Mission: 2036}
\label{milestone-EO2}

This would be a standalone Earth Observation mission to deliver the prospective improvements
in spatial and temporal resolution
over classical Earth Observation missions such as GRACE and GOCE that are illustrated in
Fig.~\ref{Gain}. The full definition of the mission will be informed by the lessons learned 
from the next-generation classical MAGIC mission and the Pathfinder mission outlined in
Section~\ref{milestone-EO}.
Although the primary purpose of this mission would be Earth Observation, the technical
developments it requires will also benefit the STE-QUEST-like fundamental physics mission outlined in
Section~\ref{milestone-STE-QUEST}, which will also use rubidium.

\subsection{STE-QUEST-like M-Class Mission: 2037} 
\label{milestone-STE-QUEST}

Building on experience with the successful MICROSCOPE mission and development work
undertaken for the previous STE-QUEST proposal, this mission would deploy a double 
atom interferometer with rubidium and potassium ``test masses" in quantum superposition to test the
Einstein equivalence Principle (universality of free fall, UFF) and search for ultralight dark matter.
It would use a single satellite in a 700~km circular orbit, and offers the possibility of probing the
Einstein Equivalence Principle, with a precision ${\cal O}(10^{-17})$, as seen in Fig.~\ref{DM}.
This is essentially a fundamental physics mission that would, however, take advantage of the synergy with the Rb-based Earth Observation missions described in Sections~\ref{milestone-EO} and \ref{milestone-EO2} as well as experiments performed on the ISS in the CAL and BECCAL facilities and those performed in terrestrial microgravity facilities outlined in Section~\ref{milestone-TP}, e.g., QUANTUS, PRIMUS, MAIUS and I.C.E.

{Programmatically, this mission profile, supported by the comprehsive and ambitious development programme outlined in Section~\ref{milestone-DP}, would fit well in the scope of an ESA M-class mission. The time-frame of the call announced recently by ESA's Science Directorate~\cite{ESA-F-M-Call} would be fully in line with the timeline proposed here of a STE-QUEST-like mission launching in 2037, and there is widespread support in the cold atom community for the proposal of such a mission profile in
response to this ESA call.}

\subsection{Atomic Clock Mission: 2038}
\label{milestone-AC}

This mission would translate the high precision of the most accurate atomic clocks shown in
Fig.~\ref{fig:clock_accuracy} into a global time standard that would take metrology to the
next level, with corresponding advantages for navigation, geodesy and fundamental physics.
It would use  a  dedicated  satellite  in  a  highly  elliptical  orbit  containing  a  
strontium  optical  lattice clock  with  a  $10^{-18}$ systematic  uncertainty  and  
$10^{-16}/\sqrt{\tau}$ stability,  and  a  coherent optical link to ground. In addition
to enable  more precise comparisons across a wider network of ground clocks, it would
provide direct searches for new physics including stringent tests of general relativity
than possible with previous missions. 
The technology development for a Sr optical lattice clock is ongoing, and is expected to continue and bring to flight maturity the critical elements of the instrument, e.g., lasers, reference cavity and atomic source. The optical clock mission will build upon the operational experience previously gained with ACES (Section~\ref{milestone-ACES})
and I-SOC Pathfinder (Section~\ref{milestone-I-SOC}) and on a continuously improving network of ground-based optical clocks connected to the space clock via a coherent optical link. {We remark that the  time frame envisaged in Fig.~\ref{fig:road-map} could be substantially shortened by cooperation with other space agencies that show a strong interest in the development of optical clocks for space~\cite{focos}, which might allow such a mission to be realised on a substantially shorter time-scale.}

\subsection{AEDGE-like L-Class Mission: 2045} 
\label{milestone-AEDGE}

This would be a fundamental science mission based on atom interferometry
using strontium. It would provide the ultimate sensitivity to ultralight
dark matter, as seen in Fig.~\ref{DM}, and gravitational waves in the
deciHz band intermediate between the maximum sensitivities of 
LIGO/Virgo/KAGRA and other terrestrial laser interferometers and the
space-borne laser interferometer LISA, as seen in Fig.~\ref{IMBH}.
{Thus it would complement the capabilities of these other detectors and offer
interesting synergies when operating as part of an international network.}
The configurations considered assume two satellites in medium Earth orbit
separated by $\sim 40,000$~km using atom clouds that might be either
inside or possibly outside the satellites.
It would be based upon developments pioneered by many of the pathfinders 
described above including the terrestrial atom interferometers now under
construction (Section~\ref{milestone-TP}), ACES (Section~\ref{milestone-ACES}), I-SOC Pathfinder (Section~\ref{milestone-I-SOC}) and the proposed dedicated atomic 
clock mission (Section~\ref{milestone-AC}). It would also have many elements in common with missions using
rubidium atoms such as the dedicated Earth Observation and STE-QUEST-like
missions (Sections~\ref{milestone-EO}, \ref{milestone-EO2} and \ref{milestone-STE-QUEST}),
such as techniques to
minimize the size, weight and power requirements for atomic clocks,
as well as optical links.


\subsection{Road-map summary} 

{We have highlighted in the Section the important synergies and interdependences between fundamental science and Earth Observation missions based on cold atom technology, as well as the key roles played by terrestrial pathfinder experiments and the importance of a vigorous  development programme. We note that are strong, diverse, and well-funded terrestrial pathfinder programmes for deploying cold atom technology already well underway in many countries. ESA leadership of a technical development programme for cold atoms in space, building on the advances made by the terrestrial pathfinders to raise the TRLs of the associated
technologies for space readiness, would provide the backbone for future cold atom missions.}

{This development programme will complement ACES and the Earth Observation and I-SOC pathfinder missions, 
which will pave the way for ultimate Earth Observation and atomic clock missions targeting next-generation geodesy, 
time-keeping and fundamental physics to launch around 2036 and 2038. 
The technical developments in cold atom technology and ancillary technologies
such as optical links will also prepare the ground for an STE-QUEST-like M-Class mission around 2037 to probe
general relativity via tests of the equivalence principle (universality of free fall) and search for ultralight dark matter.
An STE-QUEST-like M-class mission around 2037 is being proposed by cold atom community members in response to the recent call 
by ESA's Science Directorate~\cite{ESA-F-M-Call}, and it enjoys widespread and strong support in cold atom community. The ultimate sensitivity to dark matter, along with unique capabilities for
measuring gravitational waves, would be provided by an AEDGE-like L-class mission in the 2040s.}

{We propose this road-map for discussion by the  interested  cold  atom,  Earth Observation, fundamental physics
and other prospective scientific user communities, together with ESA and national space and research funding agencies.}

\bibliographystyle{JHEP}
\bibliography{AION,library}

\end{document}